\documentclass[10pt,journal,compsoc]{IEEEtran}
\usepackage{pgfplots}
\usepackage{prettyref}
\usepackage{tikz}
\usepackage{xcolor,colortbl}
\usepackage{multirow}
\usepackage[ruled, linesnumbered, vlined, commentsnumbered]{algorithm2e}
\usepackage[section]{placeins}
\usepackage{booktabs}
\usepackage{threeparttable}
\usepackage{makecell}
\usepackage{graphicx}
\usepackage{csquotes}
\usepackage{subcaption}
\usepackage{standalone}
\usepackage[skip=0.5\baselineskip]{caption}
\usepackage{wrapfig}
\usepackage{amsmath}
\usetikzlibrary{pgfplots.statistics}
\usepackage{pifont}
\usepackage{balance}
\usepackage[utf8]{inputenc}
\usepackage[T1]{fontenc}
\usepackage{microtype}
\usepackage{amsthm}
\usepackage{amssymb}
\usepackage[normalem]{ulem}
\usepackage{enumitem}
\usepackage{hyperref}
\usepackage{xfp}
\usepackage{arydshln}
\usepackage{wasysym}
\usepackage[most,listings,skins,breakable]{tcolorbox}

\usepackage[ruled, linesnumbered, vlined, commentsnumbered]{algorithm2e}
\newtcolorbox{quotebox}{colback=steel!10,boxrule=0.4pt,colframe=black,fonttitle=\bfseries,top=2pt,bottom=2pt}

\makeatletter
\def\adl@drawiv#1#2#3{%
        \hskip.5\tabcolsep
        \xleaders#3{#2.5\@tempdimb #1{1}#2.5\@tempdimb}%
                #2\z@ plus1fil minus1fil\relax
        \hskip.5\tabcolsep}
\newcommand{\cdashlinelr}[1]{%
  \noalign{\vskip\aboverulesep
           \global\let\@dashdrawstore\adl@draw
           \global\let\adl@draw\adl@drawiv}
  \cdashline{#1}
  \noalign{\global\let\adl@draw\@dashdrawstore
           \vskip\belowrulesep}}
\makeatother

\mathchardef\mhyphen="2D
\newcommand{\vect}[1]{\boldsymbol{#1}}
\DeclareMathAlphabet\mathbfcal{OMS}{cmsy}{b}{n}

\newbox\aMark
\setbox\aMark\hbox{\begin{pgfpicture}\textcolor{red}{\pgfuseplotmark{o}}\end{pgfpicture}}

\newbox\bMark
\setbox\bMark\hbox{\begin{pgfpicture}\textcolor{red}{\pgfuseplotmark{star}}\end{pgfpicture}}

\newrefformat{fig}{Fig.~\ref{#1}}
\newrefformat{tab}{Table~\ref{#1}}
\newrefformat{sec}{Section~\ref{#1}}
\newrefformat{app}{Appendix~\ref{#1}}
\newrefformat{alg}{Algorithm~\ref{#1}}
\newrefformat{property}{Property~\ref{#1}}
\newrefformat{theorem}{Theorem~\ref{#1}}
\newrefformat{lemma}{Lemma~\ref{#1}}
\newrefformat{corollary}{Corollary~\ref{#1}}
\newrefformat{proposition}{Proposition~\ref{#1}}
\newrefformat{def}{Definition~\ref{#1}}
\newrefformat{eq}{equation~(\ref{#1})}

\definecolor{steel}{rgb}{0, 0.2, 0.9} 

\SetKwInput{kwDeclare}{Declare}

\pgfplotsset{compat=newest}

\pgfplotsset{compat=1.11,
    /pgfplots/ybar legend/.style={
    /pgfplots/legend image code/.code={%
       \draw[##1,/tikz/.cd,yshift=-0.25em]
        (0cm,0cm) rectangle (3pt,0.8em);},
   },
}

\SetCommentSty{mycommfont}

\DeclareMathOperator*{\argmin}{argmin}

\makeatletter
\tcbset{
    myhbox/.style 2 args={%
        enhanced, 
        breakable,
        colback=white,
        colframe=steel!50!black,
        attach boxed title to top left={yshift*=-\tcboxedtitleheight}, 
        title={#2},
        boxed title size=title,
        boxed title style={%
            sharp corners, 
            rounded corners=northwest, 
            colback=tcbcolframe, 
            boxrule=0pt,
        },
        underlay boxed title={%
            \path[fill=tcbcolframe] (title.south west)--(title.south east) 
                to[out=0, in=180] ([xshift=5mm]title.east)--
                (title.center-|frame.east)
                [rounded corners=\kvtcb@arc] |- 
                (frame.north) -| cycle; 
        },
        #1
    }
}   
\makeatother

\def\signed #1{{\leavevmode\unskip\nobreak\hfil\penalty50\hskip2em
  \hbox{}\nobreak\hfil(#1)%
  \parfillskip=0pt \finalhyphendemerits=0 \endgraf}}

\newsavebox\mybox

  \newcommand{\squart}[4]{\begin{adjustbox}{max width=.1\textwidth}\begin{picture}(100,5)
    {\color{black}\put(0,5){\line(1,0){100}}\color{black}\put(0,5){\line(0,1){10}}\put(50,5){\line(0,1){10}}\put(100,5){\line(0,1){10}}\put(25,5){\line(0,1){5}}\put(75,5){\line(0,1){5}}\put(-2,-8){\LARGE$0$}\put(42,-8){\LARGE$0.5$}\put(96,-8){\LARGE$1$}}\end{picture}\end{adjustbox}}

  \newcommand{\quart}[4]{\begin{adjustbox}{max width=.1\textwidth}\begin{picture}(100,5)
    {\color{black}\put(#1,5){\line(1,0){#2}}\color{black}\put(#1,2){\line(0,1){6}}\color{black}\put(\fpeval{#1+#2},2){\line(0,1){6}}\color{steel}\put(#3,5){\circle*{7}}\color{black}\put(#3,5){\circle{7}}}\end{picture}\end{adjustbox}}
    
      \newcommand{\bquart}[4]{\begin{adjustbox}{max width=.1\textwidth}\begin{picture}(100,5)
    {\color{black}\put(#1,5){\line(1,0){#2}}\color{black}\put(#1,2){\line(0,1){6}}\color{black}\put(\fpeval{#1+#2},2){\line(0,1){6}}\color{red}\put(#3,5){\circle*{7}}\color{black}\put(#3,5){\circle{7}}}\end{picture}\end{adjustbox}}

      \newcommand{\quartexp}[4]{\begin{adjustbox}{max width=.1\textwidth}\begin{picture}(20,5)
    {\color{black}\put(#1,3){\line(1,0){#2}}\color{black}\put(#1,0){\line(0,1){6}}\color{black}\put(\fpeval{#1+#2},0){\line(0,1){6}}\color{steel}\put(#3,3){\circle*{4}}\color{black}\put(#3,3){\circle{4}}}\end{picture}\end{adjustbox}}
    
          \newcommand{\bquartexp}[4]{\begin{adjustbox}{max width=.1\textwidth}\begin{picture}(20,5)
    {\color{black}\put(#1,3){\line(1,0){#2}}\color{black}\put(#1,0){\line(0,1){6}}\color{black}\put(\fpeval{#1+#2},0){\line(0,1){6}}\color{red}\put(#3,3){\circle*{4}}\color{black}\put(#3,3){\circle{4}}}\end{picture}\end{adjustbox}}

\begin{document}
%
\title{MMO: Meta Multi-Objectivization for Software Configuration Tuning}
%
%
%
%

\author{Pengzhou~Chen,
        Tao~Chen,~\IEEEmembership{Member,~IEEE,}     
        and~Miqing~Li,~\IEEEmembership{Senior Member,~IEEE}
\IEEEcompsocitemizethanks{
\IEEEcompsocthanksitem Tao Chen is the corresponding author (t.chen@bham.ac.uk).\protect
\IEEEcompsocthanksitem Pengzhou Chen is with the School
of Computer Science and Engineering, University of Electronic Science and Technology of China, China.\protect
\IEEEcompsocthanksitem Tao Chen and Miqing Li are with the School
of Computer Science, University of Birmingham, UK.\protect
}
}

%
%

\markboth{Journal of \LaTeX\ Class Files,~Vol.~14, No.~8, August~2024}%
{Shell \MakeLowercase{\textit{et al.}}: Bare Demo of IEEEtran.cls for Computer Society Journals}
%



\IEEEtitleabstractindextext{%
\begin{abstract}



Software configuration tuning is essential for optimizing a given performance objective (e.g., minimizing latency).
Yet, due to the software's intrinsically complex configuration landscape and expensive measurement,
there has been a rather mild success, 
particularly in preventing the search from being trapped in local optima. 
To address this issue, 
in this paper 
we take a different perspective. Instead of focusing on improving the optimizer, 
we work on the level of optimization model and propose a meta multi-objectivization (MMO) model that considers an auxiliary performance objective (e.g., throughput in addition to latency).
What makes this model distinct is that we do not optimize the auxiliary performance objective, 
but rather use it to make similarly-performing while different configurations less comparable 
(i.e. Pareto nondominated to each other), 
thus preventing the search from being trapped in local optima. Importantly, by designing a new normalization method, we show how to effectively use the MMO model without worrying about its weight---the only yet highly sensitive parameter that can affect its effectiveness. 
Experiments on 22 cases from 11 real-world software systems/environments confirm that our MMO model with the new normalization performs better than its state-of-the-art single-objective counterparts on 82\% cases while achieving up to $2.09 \times$ speedup. For 68\% of the cases, the new normalization also enables the MMO model to outperform the instance when using it with the normalization from our prior FSE work under pre-tuned best weights, saving a great amount of resources which would be otherwise necessary to find a good weight. We also demonstrate that the MMO model with the new normalization can consolidate recent model-based tuning tools on 68\% of the cases with up to $1.22 \times$ speedup in general. 




\end{abstract}

}

\maketitle

\IEEEdisplaynontitleabstractindextext

%
\IEEEpeerreviewmaketitle

\IEEEraisesectionheading{\section{Introduction}\label{sec:introduction}}

\IEEEPARstart{M}{any} software systems are highly configurable, such that the configuration options can be flexibly adjusted for performance, including database systems, machine learning systems, and cloud systems, to name a few. For example, \textsc{Apache Storm}, a stream processing system, can be tuned by changing some key configuration options such as \texttt{splitters}. However, a daunting number of configuration options will inevitably introduce a high risk of inappropriate or even poor software configurations set by software engineers. It has been reported that 59\% of the software performance issues worldwide are related to ill-suited configuration rather than code~\cite{DBLP:conf/esem/HanY16}. In 2017-2018, configuration-related performance issues costed at least 400,000 USD per hour for 50\% of the software companies\footnote{ \href{https://www.evolven.com/blog/downtime-outages-and-failures-understanding-their-true-costs.html}{\textcolor{blue}{https://www.evolven.com/blog/downtime-outages-and-failures-understanding-their-true-costs.html}}.}. 

Indeed, adjusting the configurations will affect the outcomes of different performance attributes, such as latency, throughput, and CPU load~\cite{DBLP:conf/sigsoft/ShahbazianKBM20,Chen2018FEMOSAA,nair2018finding,DBLP:journals/pieee/ChenBY20,DBLP:conf/wosp/0001BWY18,DBLP:journals/csur/ChenBY18}. 
However, 
there are many cases wherein only the optimization of a single performance attribute is of interest, whose minimization/maximization serves as a sole performance objective in consideration. For example, in the finance sector, 
a millisecond decrease in the trade delay may boost a high-speed firm's earnings by about 100 million USD per year~\cite{tian2015latency}. 
Another example is related to the 
machine learning systems deployed by large organizations 
(e.g., GPT-4~\cite{gpt4}), 
or those in the health care domain~\cite{DBLP:conf/bcb/AhmadET18}, 
where the concern is mainly on the accuracy, 
while caring little about the overhead/resource incurred for training. 
This has been well-echoed from the literature on software configuration tuning, in the majority of which only a single performance attribute is considered at a time~\cite{DBLP:journals/jmlr/BergstraB12,DBLP:conf/sigmetrics/YeK03,DBLP:conf/sigsoft/OhBMS17,DBLP:conf/www/XiLRXZ04,DBLP:conf/hpdc/LiZMTZBF14,DBLP:conf/sc/BehzadLHBPAKS13,DBLP:conf/icse/LiX0WT20,DBLP:conf/kbse/LiXCT20}.

Despite only a single performance attribute being of concern,
such an optimization scenario is not easy to deal with for any optimizer that tunes the software configuration. 
This is because 

\begin{enumerate}

\item  The configurable systems involve a daunting number of configuration options with complex interactions, rendering a black-box to the software engineers~\cite{DBLP:conf/sigsoft/XuJFZPT15,DBLP:conf/icse/Chen19b,DBLP:journals/tse/ChenB17,DBLP:journals/tsc/ChenB17}.

\item The measurement of each configuration through running the software system is often expensive~\cite{DBLP:conf/mascots/JamshidiC16}, hence exhaustively exploring every configuration is unrealistic.

\item There is generally a high degree of sparsity in the configurable software systems~\cite{nair2018finding,DBLP:conf/wcre/Chen22,DBLP:conf/seams/Chen22}, 
i.e., similar configurations can also have radically different performance.

\end{enumerate}

\begin{figure}[t!]
	\centering
	\includegraphics[width=0.9\columnwidth]{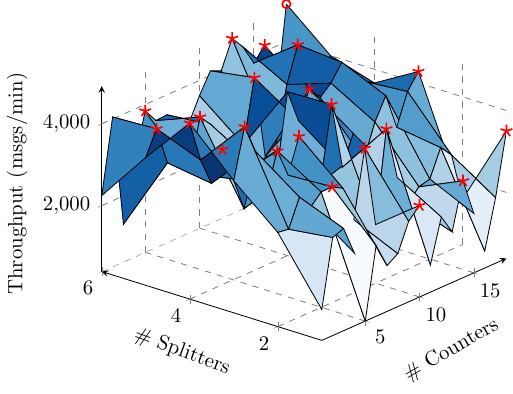}
	\caption{A projected landscape of the performance objective \textit{Throughput} 
		with respect to configuration options \texttt{Splitters} and \texttt{Counters} 
		for \textsc{Storm} under the \textsc{WordCount} benchmark.	
		\copy\aMark~is the global optimum and \copy\bMark~denotes the local optima of throughput that an optimizer needs to escape from.	
}
	\label{fig:example}
\end{figure}

The last characteristic poses a particular challenge to the automatic software configuration tuning in finding the optimal configuration (performance), because 
firstly different configurations  
may achieve locally good, but globally inadequate performance (e.g., local optima); 
and secondly, the landscape of a (local) optimum's neighborhood can be steep and rugged---if the tuning is trapped in a local optimum, 
it may be hard to escape from it as their neighboring configurations often perform significantly worse than it.
As an example, 
Figure~\ref{fig:example} shows the projected configuration landscape for \textsc{Apache Storm} (2 out of 6 configuration options), 
where it can be clearly seen that even with this simplified version, 
the landscape is rather rugged and contains steep ``local optimum traps'',
resulting in significant difficulty in the tuning.

In light of the above challenges,
a number of optimizers from the Search-Based Software Engineering (SBSE) paradigm have been presented, 
such as random search~\cite{DBLP:journals/jmlr/BergstraB12,DBLP:conf/sigmetrics/YeK03,DBLP:conf/sigsoft/OhBMS17}, hill climbing~\cite{DBLP:conf/www/XiLRXZ04,DBLP:conf/hpdc/LiZMTZBF14}, genetic algorithm~\cite{DBLP:conf/sc/BehzadLHBPAKS13,DBLP:conf/sigsoft/ShahbazianKBM20,DBLP:conf/icac/RamirezKCM09,DBLP:conf/ssbse/SinhaCC20}, and simulated annealing~\cite{garvin2009improved,guo2010evaluating}. 
To seek the global optimum (best performance of the concerned performance attribute) while avoiding being trapped in local optima, 
such methods focus on the ``internal'' components of the optimizer.
They work on designing novel search operators (i.e., the way to change the configuration structure, for example, increasing the
neighbourhood size of randomly mutated configurations~\cite{DBLP:conf/sigsoft/OhBMS17}),
or developing various search strategies 
(i.e., the way to balance exploration and exploitation, for example, 
restarting the search in hill climbing~\cite{DBLP:conf/www/XiLRXZ04}). 
However, 
a common limitation of such single-objective optimizers is 
that the goal to find the global optimum is ``less oriented'' 
as there is no clear ``incentive'' to encourage them to traverse the wide search space 
and locating as many local optima as possible, 
thus finding the best one in a resource-efficient manner.  

To better mitigate the local optima, in this paper and our prior FSE'21 work~\cite{ChenMMO21} (we call it FSE work thereafter), 
we tackle this software configuration tuning problem from a different perspective. 
In contrast to the effort made by the existing works on the development of the optimizer,
we work on the optimization model, i.e., the ``external'' part of an optimizer.
This is achieved by proposing a meta multi-objectivization (MMO) model for this single-objective problem, 
to help the search avoid being trapped in local optima and progressively explore the entire objective space. 

In a nutshell, MMO seeks to optimize two \textbf{meta-objectives}, each of which has two components.
The first component of both meta-objectives is the target performance objective (e.g., latency), 
thereby only those configurations that perform well on the target objective being in favor. 
The second component, 
which is related to the other given auxiliary performance objective (e.g., throughput), 
is a completely conflicting term for the two meta-objectives. 
The reason for this design is that 
we hope to keep the target performance objective as a primary term in the model to preserve the tendency towards its optimality, 
but at the same time, 
we want the configurations with different values on the auxiliary performance objective to be incomparable.
We are not interested in minimizing/maximizing the auxiliary performance objective 
since we do not know which value of it can lead to the best result on the target performance objective,
but we wish to keep a good amount of configurations with diverse values of the auxiliary performance objective in the search, 
thus not being trapped in local optima (we will elaborate on this in Section~\ref{sec:method}). The contributions from both this work and the FSE work are:

\begin{itemize}
    \item Unlike existing work for the software configuration tuning which puts effort on 
    the ``internal part'' of the optimization (i.e., improving the search operators of various optimizers), 
    we work on the ``external part''---multi-objectivizing this single-objective optimization scenario.
    
    \item We present a meta multi-objectivization model, MMO, as opposed to the existing multi-objectivization model 
    considered in other SBSE scenarios which directly optimizes the target and auxiliary objectives simultaneously (referred to as plain multi-objectivization or PMO). 
    We show, analytically and experimentally, 
    why MMO is more suitable than PMO for software configuration tuning.
\end{itemize}


However, 
MMO requires a weight parameter to aggregate the target objective component and the auxiliary objective component. 
It is a critical parameter to balance searching for a good target performance objective value and maintaining diverse auxiliary performance objective values,
requiring fine-tuning from the software engineers for every configurable software/environment,
as done in our FSE work~\cite{ChenMMO21}.
This, if done inappropriately, could lead to poor outcomes, 
as we will show in Section~\ref{sec:why-new}. 
Yet, since the measurement of configurations is often expensive, 
finding the best weight in a case-by-case manner is not always realistic, 
which is a major threat to the applicability of the MMO model.


Therefore,
in this paper, 
we also tackle this unwelcome issue.
We show why the weight can be a highly sensitive parameter in the MMO model and propose a way to make the model weight-free without compromising the result. 
This is achieved by presenting a new normalization method, which is simple, but works very well---it leads to results that are even better than those of the FSE work under its best-tuned weight~\cite{ChenMMO21} for the majority of the problems. 
To sum up, the unique contributions of this paper are:



\begin{itemize}

\item A sound and formal analysis of the principle behind MMO, derived from the perspective of geometric transformation in the performance objective space, that explains its intention and what role the parameter $w$ and the normalization play therein. This then enables us to formally reflect on the limitation posed by the MMO model design proposed in the FSE work.


\item Drawing on insights from the analysis, 
we design a new normalization method as part of the MMO model,
capturing the bounds of both performance objectives adaptively. 
This allows us to keep the strengths and characteristics of the MMO model while removing the weight (i.e., setting $w=1$ for all cases). 



\item An extensive evaluation that expands to 11 systems/environments that are of very different domains. 
Since a system comes with two performance objectives, 
each of these is used in turn as the target performance objective, 
leading to 22 cases. 
Under these cases, we compare MMO model using the new normalization with the PMO model and four single-objective counterparts, 
as well as with the MMO model using the normalization from the FSE work.

\item An investigation on how our MMO model with the new normalization can consolidate \textsc{Flash}~\cite{nair2018finding} and \textsc{BOCA}~\cite{DBLP:conf/icse/0003XC021}, 
which are state-of-the-art model-based tuning methods for software configuration tuning.

\end{itemize}

Our experiment results are encouraging: we show that the MMO model with the new normalization achieves better results over the best single-objective counterpart and PMO (on 18 and 20 out of 22 cases wherein 14 and 15 of them are considerably better, respectively), while being much more resource-efficient overall (with up to $2.09 \times$ speedup over the single-objective optimizers and use significantly less resource than that of the PMO). In contrast to using the MMO model with the normalization from FSE work under its best weight, the MMO model with the new normalization shows better results on 15 cases (7 of which are significant) and competitive resource efficiency. 
Notably, 
this is achieved without the need of setting the weight, which can be undesirable as in 13 out of 22 cases, it requires at least 50\% of the search budget as the extra resource to identify the best weight. The MMO model with the new normalization can also consolidate the model-based tuning methods like \textsc{Flash} and \textsc{BOCA}: with minimal code change, both can be improved for 15 out of 22 cases (with 12 or 13 cases of statistically significant improvement) while having a $1.22 \times$ and $1.06 \times$ speedup in general, respectively.

To promote the open-science practice, a GitHub repository that contains all source code and data in this work can be accessed at: \href{https://github.com/ideas-labo/mmo}{\texttt{\textcolor{blue}{https://github.com/ideas-labo/mmo}}}.

The rest of this paper is organized as follows. 
Section~\ref{sec:prob} introduces some background information. 
Section~\ref{sec:method} elaborates on the design of the MMO and PMO model, as well as why and how we design the new normalization. 
Section~\ref{sec:exp} presents our experiment methodology, 
followed by a detailed discussion of the results in Section~\ref{sec:result}. Section~\ref{sec:practice} delineates how to apply MMO in practical software engineering scenarios. The threats to validity are discussed in Section~\ref{sec:discussion}. Sections~\ref{sec:related} and~\ref{sec:con} analyze the related work and conclude the paper, respectively.

\section{Preliminaries}
\label{sec:prob}

In this section, we describe the necessary background information and context for this work.

\subsection{Software Configuration Tuning Problem}


A configurable software system often comes with a set of critical configuration options such that the $i$th option is denoted as $x_i$, 
which can be either a binary or integer variable,
where $n$ is the total number of options.
The search space, 
$\mathbfcal{X}$, 
is the Cartesian product of the possible values for all the $x_i$. 
Formally, 
when only a single performance concern is of interest (such as latency, throughput, or accuracy), 
the goal of software configuration tuning is to achieve\footnote{Without loss of generality, we assume the performance objective to be minimized.}:
\begin{align}
	\argmin~f(\vect{x}),~~\vect{x} \in \mathbfcal{X}
	\label{Eq:SOP}
\end{align}
where $\vect{x} = (x_1, x_2, ..., x_n)$.
This is a classic \textit{single-objective optimization model} and the measurement of $f$ is entirely case-dependent according to the target software and the corresponding performance attribute; 
thus we make no assumption about its characteristics.

\subsection{Multi-objectivization}

Multi-objectivization is the method of transforming a single-objective optimization problem 
into a multi-objective one, 
in order to make the search easier to find the global optimum.
It can be realized by adding a new objective (or several objectives) to the original objective 
or replacing the original objective with a set of objectives. 
The motivation is that 
since in a complex problem landscape, 
the search may get trapped in local optima when considering the original objective 
(due to the total order relation between solutions with respect to that objective), 
considering multiple objectives may make similarly-performed solutions incomparable 
(i.e., Pareto nondominated to each other),
thus helping the search jump out of local optima~\cite{Knowles2001}.

Two solutions being Pareto nondominated means   
that one is better than the other on some objective and worse on some other objective. 
Formally,
for two solutions $\vect{x}$ and $\vect{y}$, 
we call $\vect{x}$ and $\vect{y}$ nondominated to each other 
if $\vect{x} \nprec \mathbf{y} \wedge \vect{y} \nprec \vect{x}$,
where $\nprec$ is the negation of ``to Pareto dominate'' ($\prec$), 
the superiority relation between solutions for multi-objective optimization.
That is, 
considering a minimization problem with $m$ objectives,
$\vect{x}$ is said to \textit{(Pareto) dominate} $\vect{y}$ 
(denoted as $\vect{x}\prec \vect{y}$) 
if $f_i(\vect{x}) \leq f_i(\vect{y})$ for $1 \leq i \leq m$ and 
there exists at least one objective $j$ on which $f_j(\vect{x}) < f_j(\vect{y})$.
Pareto dominance is a partial order relation, 
and thus there typically exist multiple optimal solutions in multi-objective optimization.
For a solution set $\vect{X}$,
a solution $\vect{x} \in \vect{X}$ is called \textit{Pareto optimal} to $\vect{X}$
if there is no solution $\in \vect{X}$ that dominates $\vect{x}$. 
When $\vect{X}$ is the collection of all feasible solutions for a multi-objective problem,
$\vect{x}$ becomes an optimal solution to the problem, 
and the set of all Pareto optimal solutions of the problem is called its \textit{Pareto optimal set}. 

Multi-objectivization is not uncommon in the modern optimization realm, 
particularly to the evolutionary computation community~\cite{Knowles2001,Cai2006,Ishibuchi2007,Song2014,Steinhoff2020}.
To tackle various challenging single-objective optimization problems, 
researchers put much effort in introducing/designing additional objectives,
e.g., creating sub-problems (sub-objectives) of the original objective~\cite{Knowles2001}, 
converting the constraints into an additional objective~\cite{Cai2006},
constructing similar adjustable objectives~\cite{Ishibuchi2007},
considering one of the decision variables~\cite{Song2014},
or even adding a man-made less relevant objective function~\cite{Steinhoff2020}. 




 
\section{Multi-objectivization for Software Configuration Tuning}
\label{sec:method}

In this section, we present the designs of the multi-objectivization models and how they are derived from the key properties in software configuration tuning.

\subsection{Properties in Configuration Tuning}

We observed that, in general, software configuration tuning bears the following properties.

\underline{\textbf{Property 1:}} As shown in Figure~\ref{fig:example} and what has already been reported~\cite{nair2018finding,DBLP:conf/mascots/JamshidiC16,DBLP:journals/tosem/ChenL23a,DBLP:journals/tosem/ChenL23}, 
the configuration landscape of different performance objectives for most configurable software systems is rather rugged with numerous local optima at varying slopes. 
Therefore the tuning, 
once the search is trapped at a local optimum, 
would be difficult to progress. 
This is because all the surrounding configurations of a local optimum are significantly inferior to it,
and the search focus would have no much drive to move away from that local optimum 
(if only the concerned performance attribute is used to guide the search). 
As a result, 
a good optimization model has some additional ``tricks'' to avoid comparing configurations solely based on a single performance attribute. 

\underline{\textbf{Property 2:}} A single measurement of configuration is often expensive. For example, Valov \textit{et al.}~\cite{DBLP:conf/wosp/ValovPGFC17} reported that sampling all values of 11 configuration options for \textsc{x264} needs 1,536 hours. This means that the resource (search budget) in software configuration tuning is highly valuable, hence utilizing them efficiently is critical.

\begin{figure}[t!]
	\centering
	\includegraphics[width=\columnwidth]{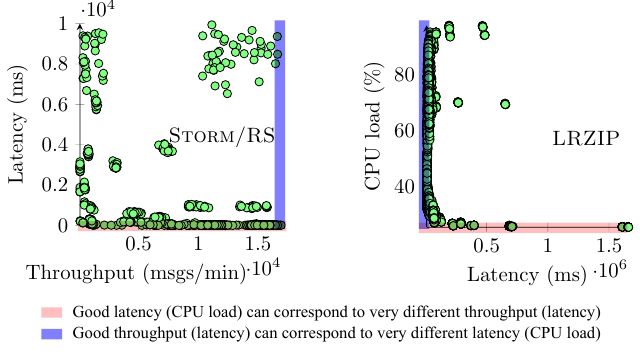}
	\caption{Measured configurations for system \textsc{Storm/RS} and \textsc{LRZIP}. The points that \textbf{Property 3} refers to are highlighted: very good or bad results on one performance objective can both correspond to similarly good values on the other.}
	\label{fig:prop3}
\end{figure}

\underline{\textbf{Property 3:}} The correlation between different performance attributes is often uncertain, as different configurations may have different effects on distinct attributes.
We observed that the configurations may achieve extremely good or bad performance on one while having similarly good results on the other, as illustrated in Figure~\ref{fig:prop3}. 
Taking the system \textsc{Storm} with \textsc{RollingSort} benchmark (denoted \textsc{Storm/RS}) from Figure~\ref{fig:prop3} (left) as an example, suppose that in a multi-threaded and multi-core environment with 100 successful messages, if a configuration $\vect{A}$ enables each of these messages to be processed at 30ms, then the latency and throughput are ${{100 \times 30} \over {100}} = 30$ms and ${100 \over 30} = 3.33$ msgs/ms, respectively. 
In contrast, another configuration $\vect{B}$ may restrict the parallelism (e.g., lower \texttt{spout\_num}), hence there could be 50 messages processed at 20ms each\footnote{The relief of peak CPU load could allow the process of each message faster.} while the other 50 are handled at 40ms each (including 20ms queuing time due to reduced parallelism). Here, the latency remains at ${{50 \times 20 + 50 \times 40} \over {100}} = 30$ms but the throughput is changed to ${100 \over 40} = 2.5$ msgs/ms, which is a 25\% drop. Therefore, we should neither presume a strict conflict nor a harmonic correlation between the performance attributes.

Clearly, 
a good optimization model for software configuration tuning needs to take the above properties into account.

\subsection{Plain Multi-Objectivization (PMO) Model}

A straightforward idea to perform multi-objectivization is to add an auxiliary objective to optimize, 
along with optimizing the target performance objective.
This is what has been commonly used in SBSE scenarios (e.g.,~\cite{derakhshanfar2020good,DBLP:journals/ese/MkaouerKCHD17,DBLP:conf/ssbse/BuzdalovaBP13,DBLP:conf/issta/AbdessalemPNBS20}).
That PMO model can be formulated as:
\begin{align}
\begin{split}
\text{minimize}
\begin{cases}
f_a(\vect{x})\\
f_t(\vect{x})\\
\end{cases}
\end{split}
\label{Eq:pmo_model}
\end{align}
where $f_t(\vect{x})$ denotes the target performance objective (i.e., the concerned one) 
and $f_a(\vect{x})$ denotes the auxiliary performance objective\footnote{Without loss of generality, 
	 we use the minimization form of the performance objectives; 
	the maximization ones can be trivially converted, e.g., by multiplying $-1$.}.

Putting it in the context of software configuration tuning, 
the PMO model may cover \textbf{Property~1}, 
because the natural Pareto relation with respect to the two objectives ensures 
that the target performance objective is no longer a sole indicator to guide the search.
However, 
it does not fit \textbf{Property~2} as PMO additionally optimizes the auxiliary performance objective. 
As such, 
configurations that perform well on the auxiliary performance objective but poorly on the target performance objective
are still regarded as optimal in PMO, 
despite being meaningless to the original problem. 
This can result in a significant waste of resources. 
In addition,
PMO does not consider \textbf{Property~3} as it often assumes conflicting correlation between the two objectives~\cite{DBLP:conf/ssbse/MkaouerKBC14,derakhshanfar2020good}, 
which is hard to assure in software configuration tuning.

\begin{figure*}[t!]

	\centering
    \begin{subfigure}[h]{0.32\textwidth}
		\includegraphics[width=\textwidth]{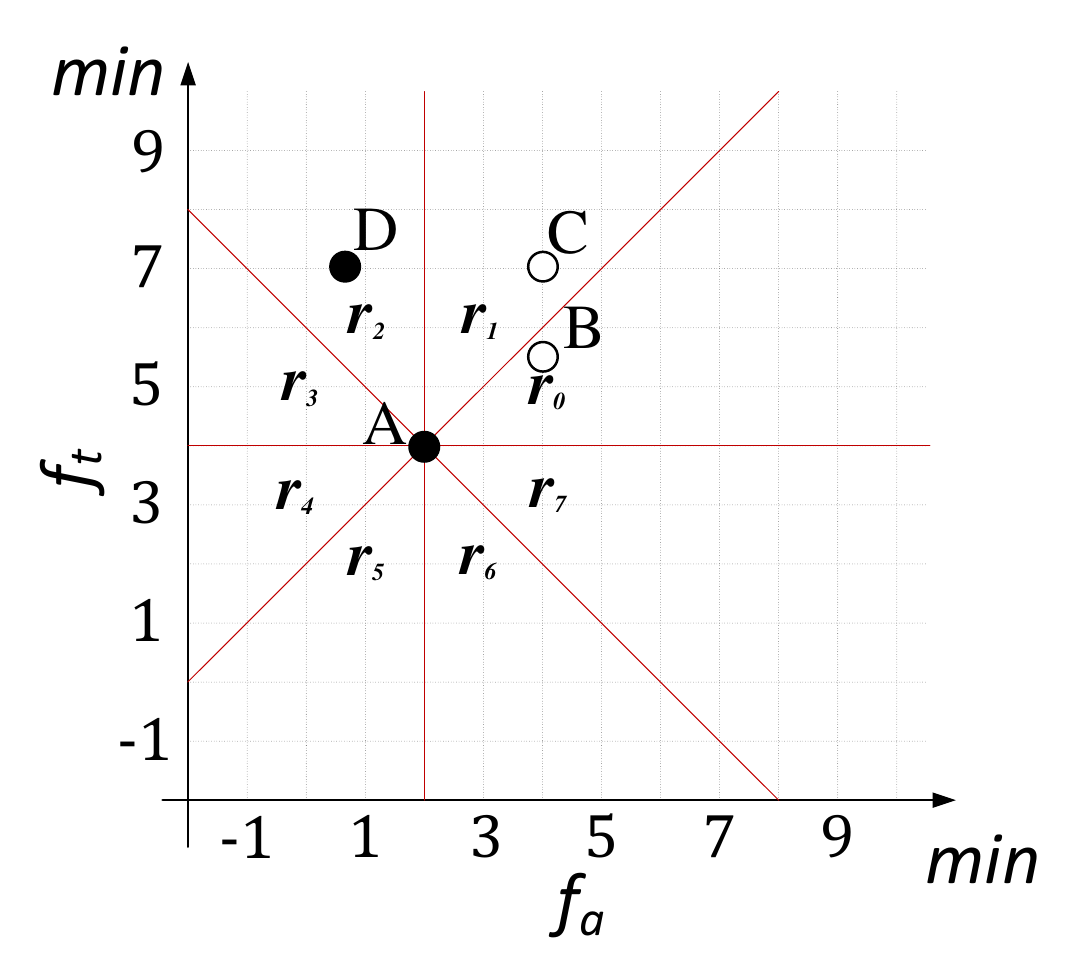}
		\subcaption{{Original space}}
	\end{subfigure}
	~\hfill
	\begin{subfigure}[h]{0.32\textwidth}
		\includegraphics[width=\textwidth]{figures/new_pic/w=1_old.pdf}
	\subcaption{{Scaling at $w=1$ in MMO}}
	\end{subfigure}
		~\hfill
	\begin{subfigure}[h]{0.32\textwidth}
		\includegraphics[width=\textwidth]{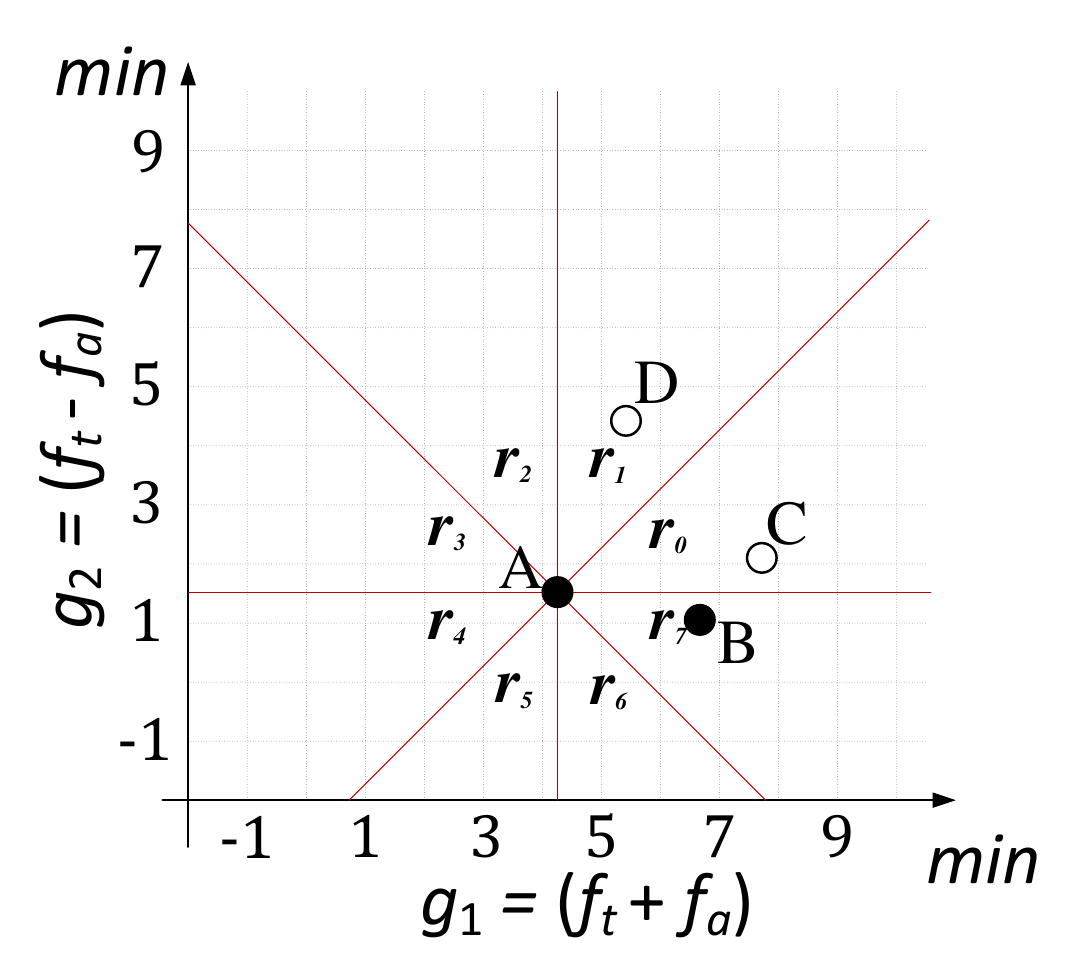}
	\subcaption{{Rotation/dilation at $w=1$ in MMO}}
	\end{subfigure}

		\caption{An illustration of the rotation effect in MMO. $\vect{A}$, $\vect{B}$, $\vect{C}$, and $\vect{D}$ are four configurations with the auxiliary and target performance objective values as $(2, 4)$, $(4, 5.5)$, $(4, 7)$, and $(0.75, 7)$, respectively. The auxiliary and target performance objectives are to be minimized if they are both of concern. $\circ$ denotes the nondominated configurations while $\bullet$ means those that are dominated by at least one other within the corresponding current space.}
	\label{fig:rotation}
\end{figure*}

\subsection{Meta Multi-Objectivization (MMO) Model}

Unlike PMO, 
our meta multi-objectivization (MMO) model creates two meta-objectives based on the performance attributes. 
The aim is to drive the search towards the optimum of the target performance objective and at the same time, not to be trapped in local optima. 
In particular,
we want to achieve two goals:

\begin{itemize}
    \item[---] \textbf{Goal 1:} optimizing the target performance objective still plays a primary role, 
    thus no resource waste on, for example, optimizing the auxiliary one (this fits in \textbf{Property 2});

    \item[---] \textbf{Goal 2:} but those with different values of the auxiliary performance objective are more likely to be incomparable (i.e., Pareto nondominated), 
    hence the search would not be trapped in local optima (this relates to \textbf{Properties 1} and \textbf{3}).
\end{itemize}



Formally, 
the MMO model with two meta-objectives $g_1(\vect{x})$ and $g_2(\vect{x})$ is constructed as\footnote{In our FSE work~\cite{ChenMMO21}, 
	we found that different forms of the auxiliary performance objectives (e.g., linear and quadratic) 
	do not lead to significantly different results, 
	hence in this work, we use the linear form, which is the simplest version of the MMO model.}:
\begin{align}
\begin{split}
\text{minimize}
\begin{cases}
g_1(\vect{x}) = f_t(\vect{x}) + w f_a(\vect{x})\\
g_2(\vect{x}) = f_t(\vect{x}) - w f_a(\vect{x})\\
\end{cases}
\end{split}
\label{Eq:model}
\end{align}
whereby
each of the two meta-objectives shares the same target performance objective $f_t(\vect{x})$, 
but differs (effectively being opposite) 
regarding the auxiliary performance objective $f_a(\vect{x})$.
The auxiliary objective can be a readily available one and whose result is of no interest 
(e.g., throughput or CPU load, in addition to latency). 
The weight $w$ is a critical parameter that balances the target and auxiliary performance objectives. 

\subsubsection{Formal Analysis of MMO}
\label{sec:theory}

Compared with the original space of $f_t$ and $f_a$, \textsc{MMO} essentially does two main geometric operations to transform the original space of the two performance objectives into a meta-objective space: (1) it scales (stretches or shrinks) the configurations along the $f_a$ axis by a factor of $w$; (2) it rotates the scaled configurations by $45^{\circ}$ clockwise and then dilates them on both $f_t$ and $f_a$ by a factor of $\sqrt{2}$. Geometrically, \textsc{MMO} in Equation~\ref{Eq:model} can be decomposed via the following transformation metrics in linear algebra:
\begin{equation}
\begin{aligned}
&\begin{bmatrix}
g_1(\vect{x})\\
g_2(\vect{x})
\end{bmatrix}
=
\overbrace{
\sqrt{2}
\begin{bmatrix}
\cos{{{\pi}\over{4}}} & \sin{{{\pi}\over{4}}}\\
-\sin{{{\pi}\over{4}}} & \cos{{{\pi}\over{4}}}
\end{bmatrix}
}^{\text{rotation/dilation matrix}}
\overbrace{
\begin{bmatrix}
w & 0\\
0 & 1
\end{bmatrix}
}^{\text{scaling matrix}}
\overbrace{
\begin{bmatrix}
f_a(\vect{x}) \\
f_t(\vect{x}) 
\end{bmatrix}
}^{\text{original space}}\\ \\
&=
\overbrace{
\begin{bmatrix}
f_t(\vect{x}) + wf_a(\vect{x})\\
f_t(\vect{x}) - wf_a(\vect{x})
\end{bmatrix}
}^{\text{MMO space}}
\label{Eq:MMO-detail}
\end{aligned}
\end{equation}
whereby $\sin{{{\pi}\over{4}}}={\sqrt{2}\over2}$ and $\cos{{{\pi}\over{4}}}={\sqrt{2}\over2}$, hence the rotation angle is ${{\pi}\over{4}}$ (i.e., $45^{\circ}$ clockwise) and both axes are dilated by a factor of $\sqrt{2}$ thereafter to create a rotation matrix of $1$ and $-1$.


\begin{figure*}[t!]

	\centering
    \begin{subfigure}[h]{0.32\textwidth}
		\includegraphics[width=\textwidth]{figures/new_pic/w=1_old.pdf}
	\subcaption{{Original space}}
	\end{subfigure}
		~\hfill
	\begin{subfigure}[h]{0.32\textwidth}
		\includegraphics[width=\textwidth]{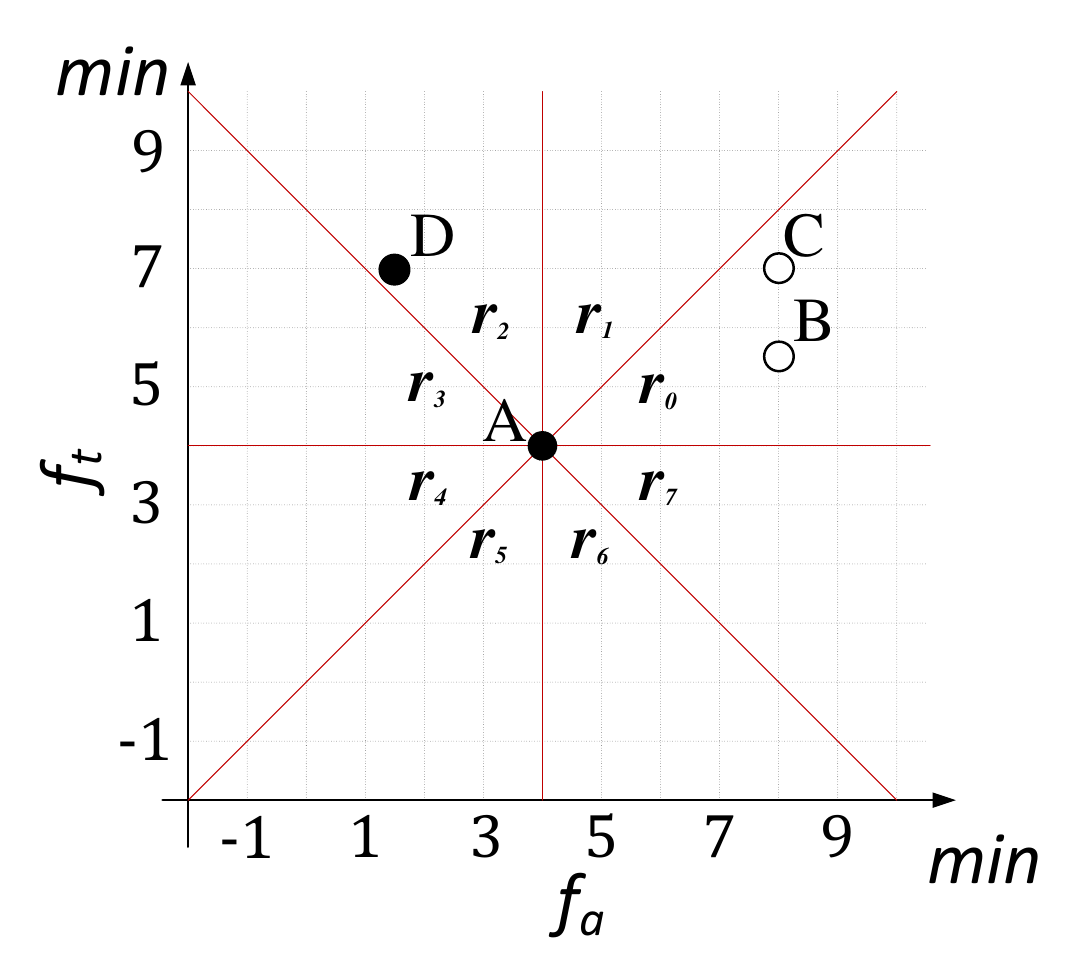}
	\subcaption{{Scaling at $w=2$ in MMO}}
	\end{subfigure}
		~\hfill
	\begin{subfigure}[h]{0.32\textwidth}
		\includegraphics[width=\textwidth]{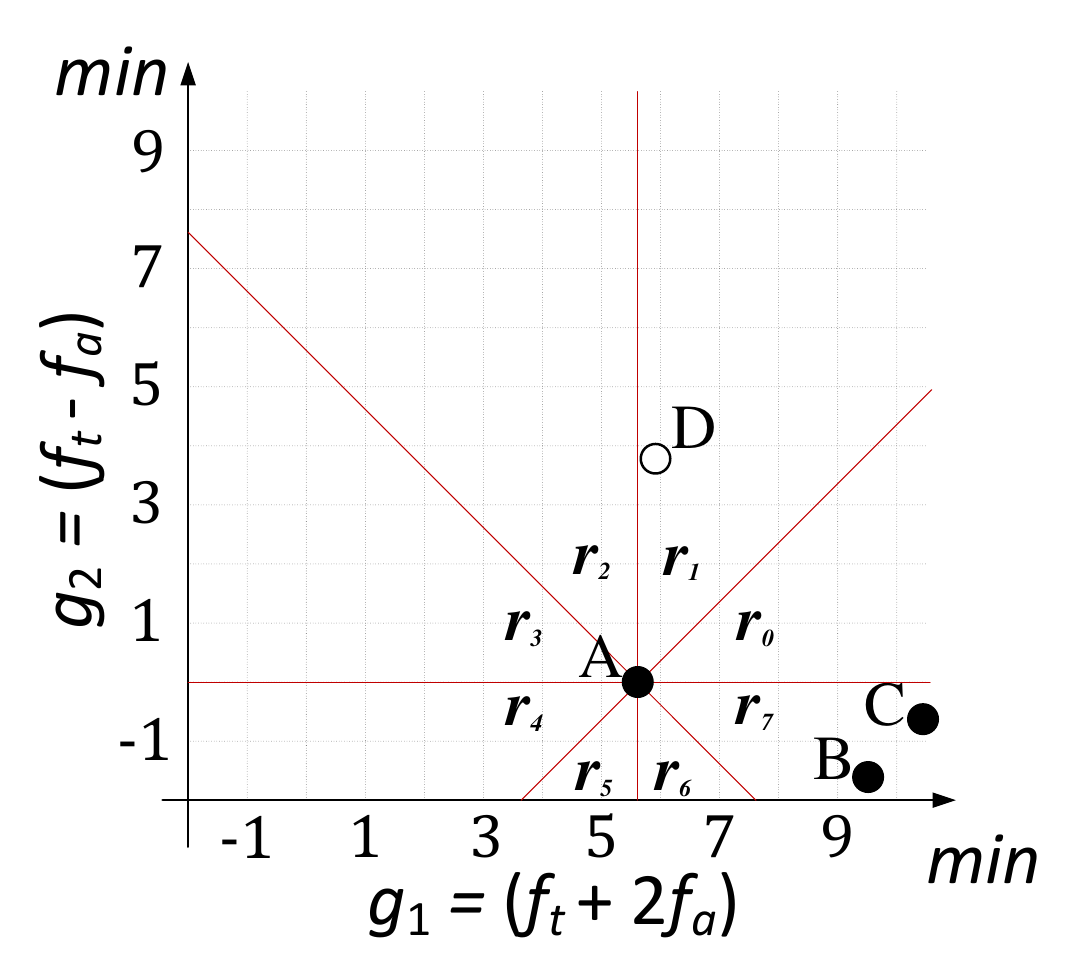}
	\subcaption{{Rotation/dilation at $w=2$ in MMO}}
	\end{subfigure}

\centering
    \begin{subfigure}[h]{0.32\textwidth}
		\includegraphics[width=\textwidth]{figures/new_pic/w=1_old.pdf}
			\subcaption{{Original space}}
	\end{subfigure}
	~\hfill
	\begin{subfigure}[h]{0.32\textwidth}
		\includegraphics[width=\textwidth]{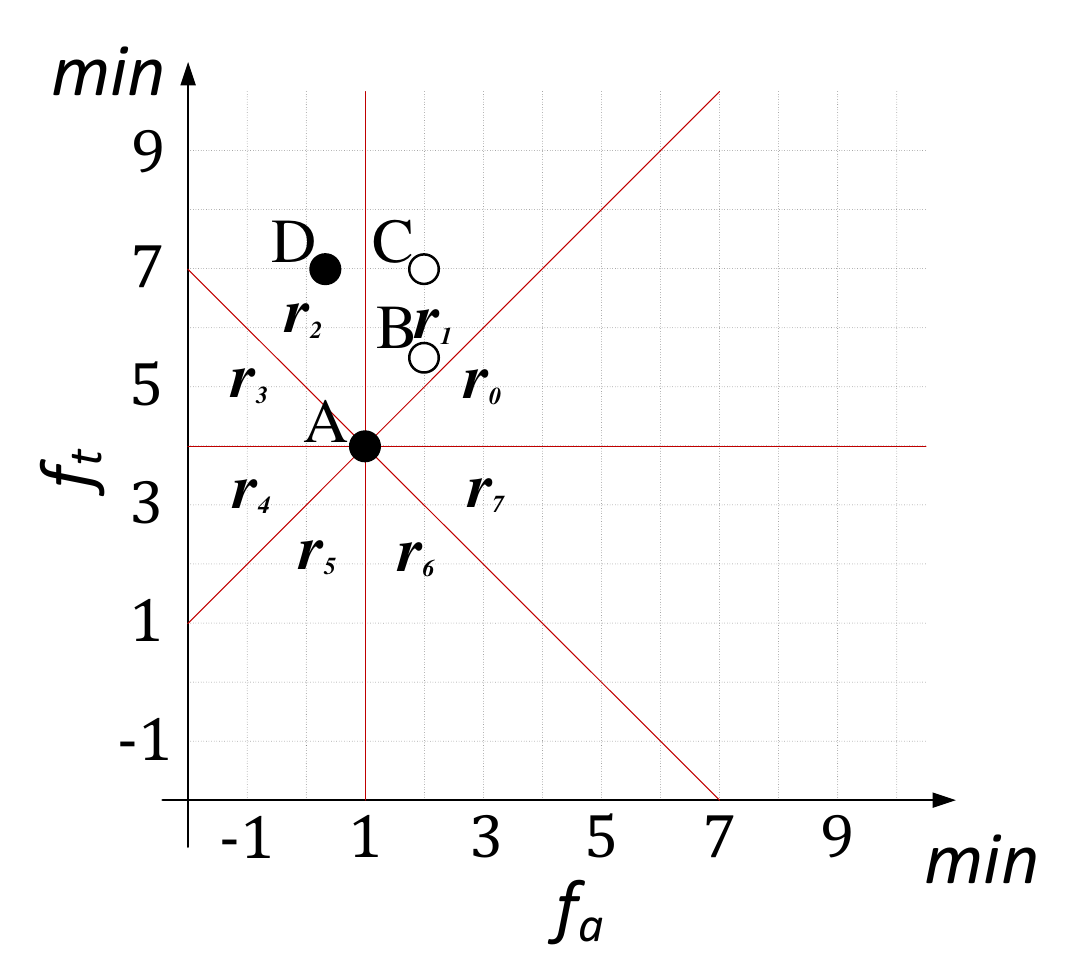}
		\subcaption{{Scaling at $w=0.5$ in MMO}}
	\end{subfigure}
		~\hfill
	\begin{subfigure}[h]{0.32\textwidth}
		\includegraphics[width=\textwidth]{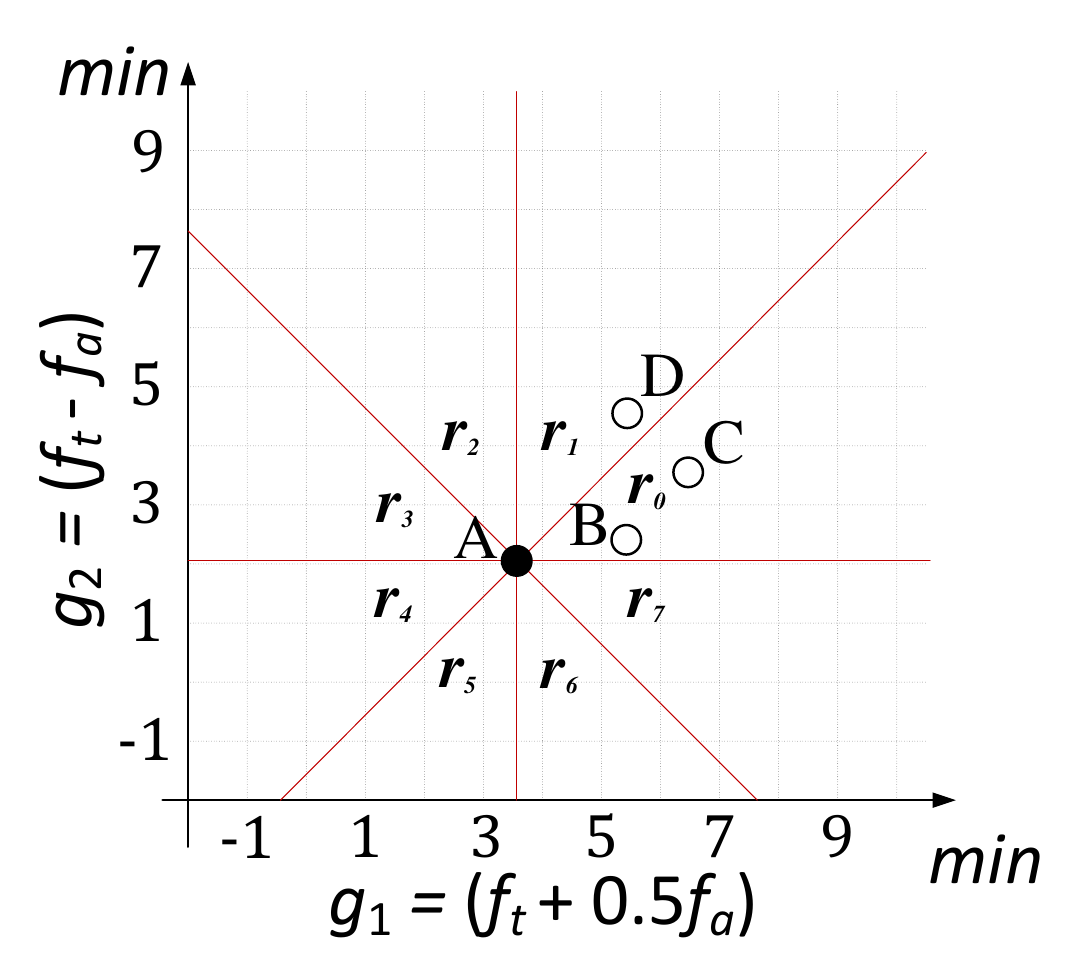}
	\subcaption{{Rotation/dilation at $w=0.5$ in MMO}}
	\end{subfigure}

		\caption{An illustration of the impact of $w$ in MMO. The formats are the same as in Figure~\ref{fig:rotation}. (a), (b), and (c) delineate the effect of an increased weight, i.e., $w=2$; (d), (e), and (f) explain the effect of a decreased weight, i.e., $w=0.5$.}
	\label{fig:w's influence}
\end{figure*}

To better understand how the transformation works in MMO, suppose that there are four configurations $\vect{A}$, $\vect{B}$, $\vect{C}$, and $\vect{D}$, where $\vect{A}$ is one of the nondominated configurations, as shown in Figure~\ref{fig:rotation}a. The areas around $\vect{A}$ can be divided into eight regions every $45^{\circ}$, starting counterclockwise from the region where $\vect{B}$ is located and they are marked as $r_0,r_1,...,r_7$, respectively. Note that since $f_a$ and $f_t$ are to be minimized in the original space of PMO (the same for $g_1$ and $g_2$ in the MMO space), the configurations that are dominated by $\vect{A}$ will be those in its first quadrant, i.e., in regions $r_0$ and $r_1$. Therefore, we can precisely define the dominance relations between configurations and $\vect{A}$ with respect to the regions of $\vect{A}$ via the following:
\begin{itemize}
    \item Configurations in $r_0$ and $r_1$ of $\vect{A}$ (including the boundaries) will be dominated by $\vect{A}$.
    \item Likewise, configurations in $r_4$ and $r_5$ of $\vect{A}$ (including the boundaries) will dominate $\vect{A}$ (not applicable if $\vect{A}$ is Pareto optimal).
    \item Finally, configurations in $r_2$, $r_3$, $r_6$, and $r_7$ will be nondominated to $\vect{A}$, including the boundaries except those adjacent to $r_0$, $r_1$, $r_4$, and $r_5$. 
\end{itemize}

Following the rotation in MMO, the configurations in the regions with respect to $\vect{A}$ will also be rotated with $\vect{A}$, and hence they would be in different regions compared with where they were before, which might cause shifts in their relative dominance relationship to $\vect{A}$---the key that makes MMO works effectively in tuning software configuration. Assuming that $w=1$, i.e., no scaling (Figure~\ref{fig:rotation}b) and hence we can focus on discussing the impact of rotation in MMO, from Figure~\ref{fig:rotation}c it is not difficult to see that, compared to $\vect{A}$, all the configurations will be moved clockwise to their adjacent regions after the rotation of $45^{\circ}$. Using the case of $\vect{A}$ as an example again, configuration $\vect{D}$ in $r_2$ will be moved to $r_1$; $\vect{C}$ in $r_1$ will be moved to $r_0$; $\vect{B}$ in $r_0$ will be moved to $r_7$. As such, we can generalize the following rule: 
\begin{tcbitemize}[%
    raster columns=1, 
    raster rows=1
    ]
  \tcbitem[myhbox={}{MMO Rule}] A configuration in a region $r_i$ will be in a new region $r_j$ following the rotation in MMO, and they satisfy the condition below:
\begin{align}
j=(i+7) \bmod 8
\label{Eq:divide_region}
\end{align}
\end{tcbitemize}

\noindent Finally, $f_a$ and $f_t$ of the configurations are dilated by $\sqrt{2}$ times for normalizing the coefficients of the rotation matrix to 1, and hence such dilatation has no effect on the dominance relationships between configurations. 

From the above, it is intuitive to understand how the MMO can change the dominance relationships in the space: by moving the configurations between the regions with respect to a particular configuration, it is possible to make them change from dominated (comparable) to nondominated (incomparable), e.g., from $r_0$ to $r_7$; or vice versa, e.g., from $r_2$ to $r_1$, thereby altering the number of incomparable configurations (in the sense of Pareto dominance) during the tuning. Such an amendment of the Pareto dominance relationships determines the effectiveness of MMO, as either too many or too few incomparable configurations throughout the tuning would be harmful since the former causes a loss of search direction while the latter increases the possibility of being trapped at local optima.


\begin{figure*}[tbp]
	\begin{center}
 	\setlength{\tabcolsep}{0.2mm}
		\footnotesize
		\begin{tabular}{@{}cc@{}}
			\vspace{-1pt}
			\includegraphics[scale=0.3]{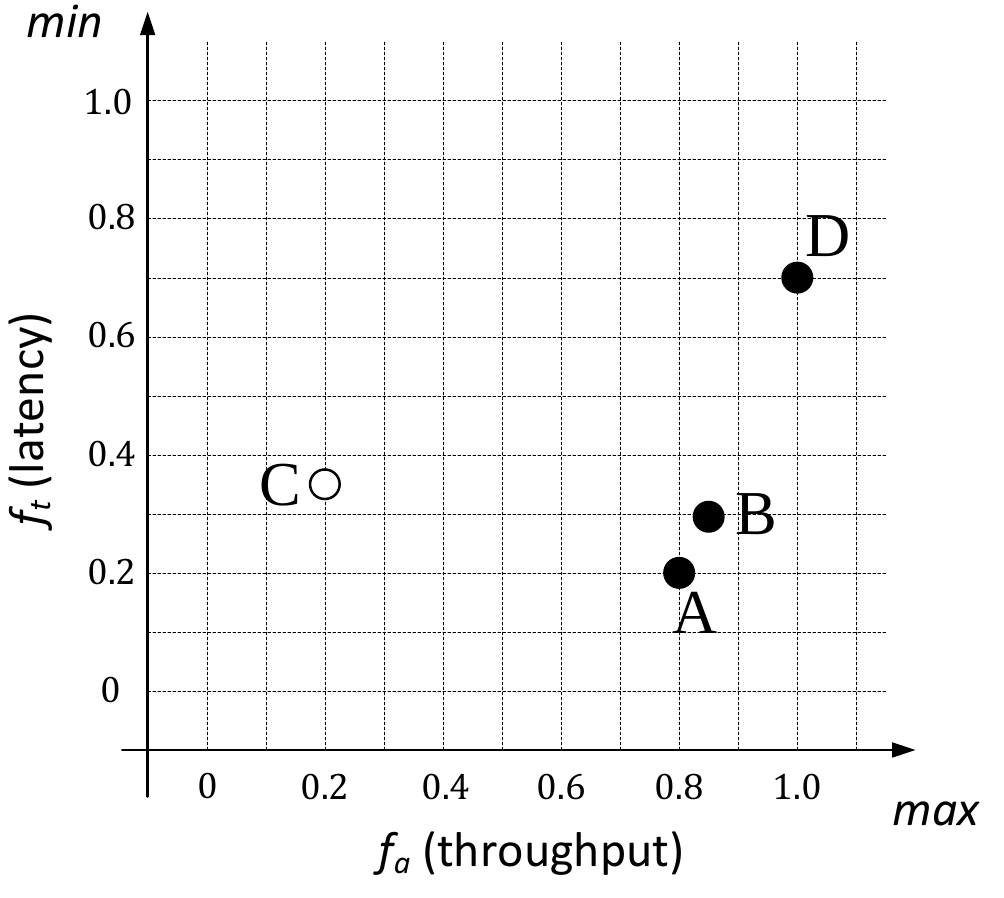}&
			\includegraphics[scale=0.3]{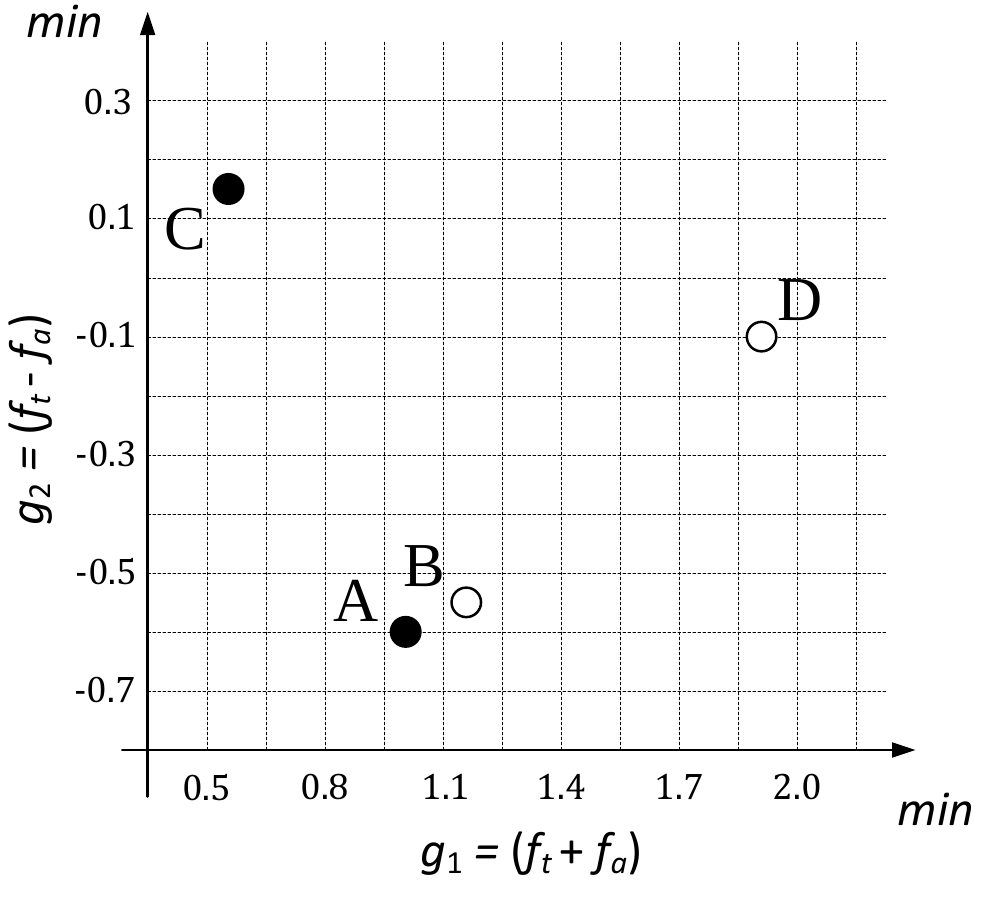}\\
			(a) The original target-auxiliary space (i.e. the PMO model)~~~&~~~ (b) The meta-objective space (i.e. the MMO model) \\
		\end{tabular}
	\end{center}
	\vspace{0pt}
	\caption{{An illustration of comparison between (a) the PMO model and (b) the MMO model on \textsc{Storm},
			where the target performance objective is latency (to minimize) and 
			the auxiliary performance objective is throughput (to maximize). 
			Both of them are normalized and the weight is 1.0 in the MMO model.
			Let us say $\vect{A}$, $\vect{B}$, $\vect{C}$ and $\vect{D}$ be a set of four configurations to be considered.
			Out of them, one needs to select two 
			(e.g., in order to put some better configurations into the next-generation population in a multi-objective optimizer).  
			The solid circle means the configuration being Pareto optimal to the set,
			and the hollow one is the dominated configuration. 			 }  
	}
	\label{Fig:model}
\end{figure*}

Once it is clear how rotation affects the configurations in the MMO space, we can now explain the role of $w$ in the MMO model. Since the value of weight $w$ determines the factor of scaling, it is not hard to imagine that increasing $w$ stretches all configurations horizontally on $f_a$; conversely, decreasing $w$ shrinks the configurations horizontally over $f_a$. Figures~\ref{fig:w's influence}a,~\ref{fig:w's influence}b, and~\ref{fig:w's influence}c give a concrete example using the same configurations as before. When changing from $w=1$ to $w=2$, the $f_a$ for all configurations are stretched by $2$ times, causing configuration $\vect{C}$ to move from region $r_1$ to $r_0$ before the rotation (Figure~\ref{fig:w's influence}b). This means that the relative positions of configurations $\vect{A}$, $\vect{B}$, $\vect{C}$, and $\vect{D}$ will rotate from Figure~\ref{fig:w's influence}b to Figure~\ref{fig:w's influence}c. As such, $\vect{C}$, which should be dominated by $\vect{A}$ when rotating under $w=1$, now will become nondominated to $\vect{A}$ after the rotation. In contrast, if we change $w=1$ to $w=0.5$ (Figures~\ref{fig:w's influence}d,~\ref{fig:w's influence}e, and~\ref{fig:w's influence}f), the configurations will be shrunk on $f_a$ by $0.5$ times, where configuration $\vect{B}$ will be moved from regions $r_0$ to $r_1$ before rotation (Figure~\ref{fig:w's influence}e). As a result, the relative positions of configurations $\vect{A}$, $\vect{B}$, $\vect{C}$, and $\vect{D}$ will rotate from Figure~\ref{fig:w's influence}e to Figure~\ref{fig:w's influence}f. In this case, $\vect{B}$, which should be nondominated to $\vect{A}$ when rotating under $w=1$, now will be dominated by $\vect{A}$ after the rotation.

The above indicates a simple rule to understand the role of $w$ in MMO: a larger $w$ suggests a bigger stretch on $f_a$, making more of the configurations become incomparable following the rotation, which encourages the \textbf{\textit{exploration}} in the search space to find more diverse configurations. 
In the extreme case where $w=\infty$,
the differences between configurations on $f_a$ become infinitely large, making the two meta-objectives linearly conflicted,
hence all configurations are incomparable if they differ on $f_a$ and render the tuning with no guidance. 
On the other hand, A smaller $w$ means a bigger shrink on $f_a$, thus more configurations become comparable after the rotation, putting more emphasis on optimizing $f_t$, i.e., \textbf{\textit{exploitation}}. 
In the extreme case when $w=0$,
$f_a$ is completely ruled out, leaving the two meta-objectives identical, and as such all configurations are comparable provided that they differ on $f_t$, thereby making the tuning more difficult to jump out of the local optima. In general, neither too large nor too small $w$ is ideal; yet, how large (or small) is considered as too large (or too small) is really case-dependent.

While here we use a nondominated configuration $\vect{A}$ as the example, it is worth noting that the analysis discussed thereof is applicable to any configurations. As such, the dominance relationships between any pair of configurations can be potentially changed by the scaling of $w$ and rotation introduced in MMO. This as a whole would affect the behaviors and focus (exploration vs. exploitation) of the configuration tuning process regardless of the underlying multi-objective optimizer.


\subsubsection{The Characteristics of MMO}
\label{sec:Characteristics of MMO}

To better understand the characteristics of the MMO model derived from our analysis and how the aforementioned two goals can be achieved, 
Figure~\ref{Fig:model} gives an example of \textsc{Storm} on how it distinguishes between different configurations, 
in comparison with the PMO model,
where we assume that latency is the target performance objective $f_t$ and throughput is the auxiliary performance objective $f_a$. 
Suppose that there is a set of four configurations $\vect{A}$, $\vect{B}$, $\vect{C}$ and $\vect{D}$.
Let us say if we want to select two from them based on their fitness 
(e.g., in order to put some better configurations into the next-generation population in a multi-objective optimizer, such as NSGA-II).
For the PMO model (Figure~\ref{Fig:model}a) that minimizes latency and maximizes throughput,
the configuration $\vect{D}$, which performs extremely poor on latency, will certainly be selected by any multi-objective optimizer,
since it is Pareto optimal and also less crowded than the other Pareto optimal configuration $\vect{A}$ and $\vect{B}$.
In contrast,
for the MMO model (Figure~\ref{Fig:model}b) which minimizes the two meta objectives,
the two configurations that will be selected are $\vect{A}$ and $\vect{C}$
(since they are the only two Pareto optimal ones).

It is worth noting that for the single-objective optimization model (which only considers latency), 
the two chosen configurations will be $\vect{A}$ and $\vect{B}$.
However,
since $\vect{C}$ and $\vect{A}$ behave much more differently than $\vect{B}$ and $\vect{A}$ on the throughput, 
it is more likely that they are located in distant regions in the configuration landscape; 
thus preserving $\vect{C}$ rather than $\vect{B}$ (when $\vect{A}$ is preserved) 
is generally more likely to help the search to escape from the local optimum.

In the following, 
we provide several remarks to help further grasp the characteristics of the MMO model.

\begin{figure*}[t!]
	\centering
    \begin{subfigure}[h]{0.32\textwidth}
		\includegraphics[width=\textwidth]{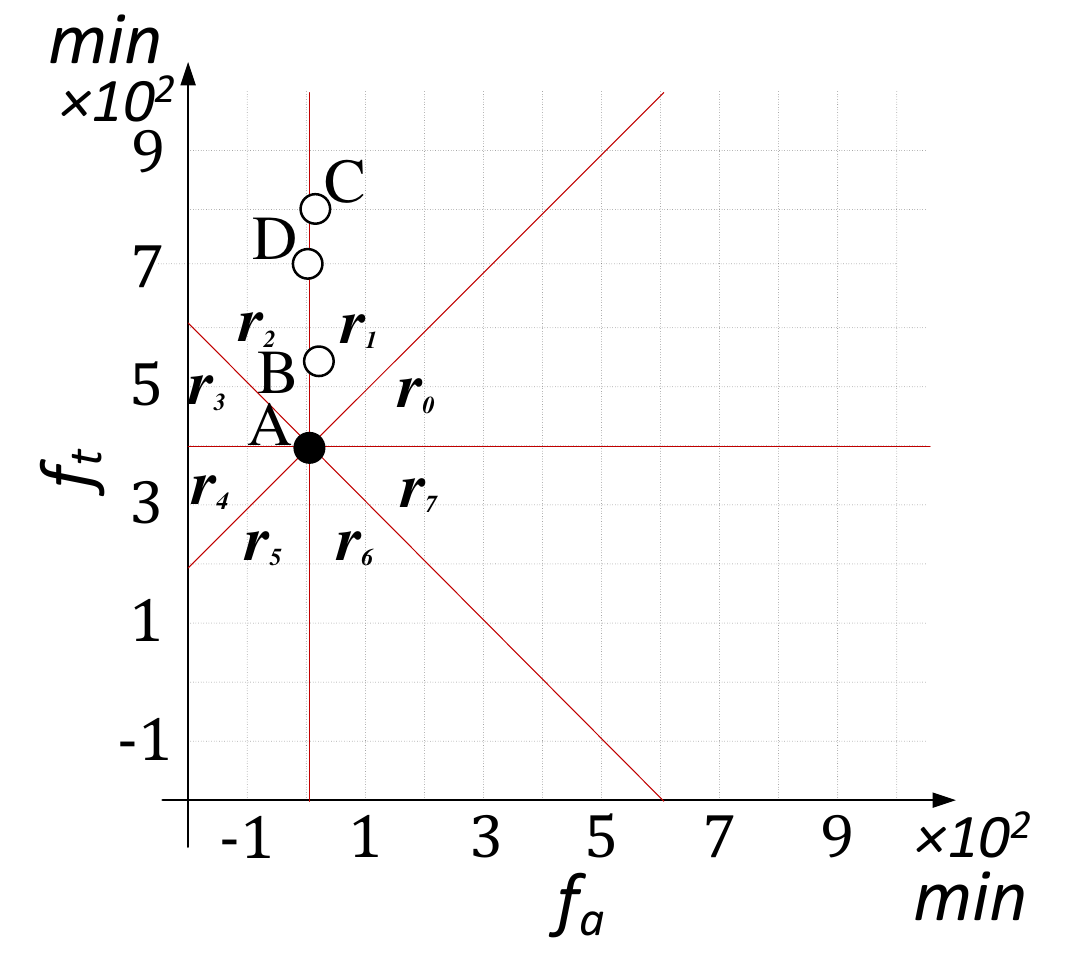}
		\subcaption{{Original space where $f_a$ has a much smaller scale than that of $f_t$}}
	\end{subfigure}
	~\hfill
	\begin{subfigure}[h]{0.32\textwidth}
		\includegraphics[width=\textwidth]{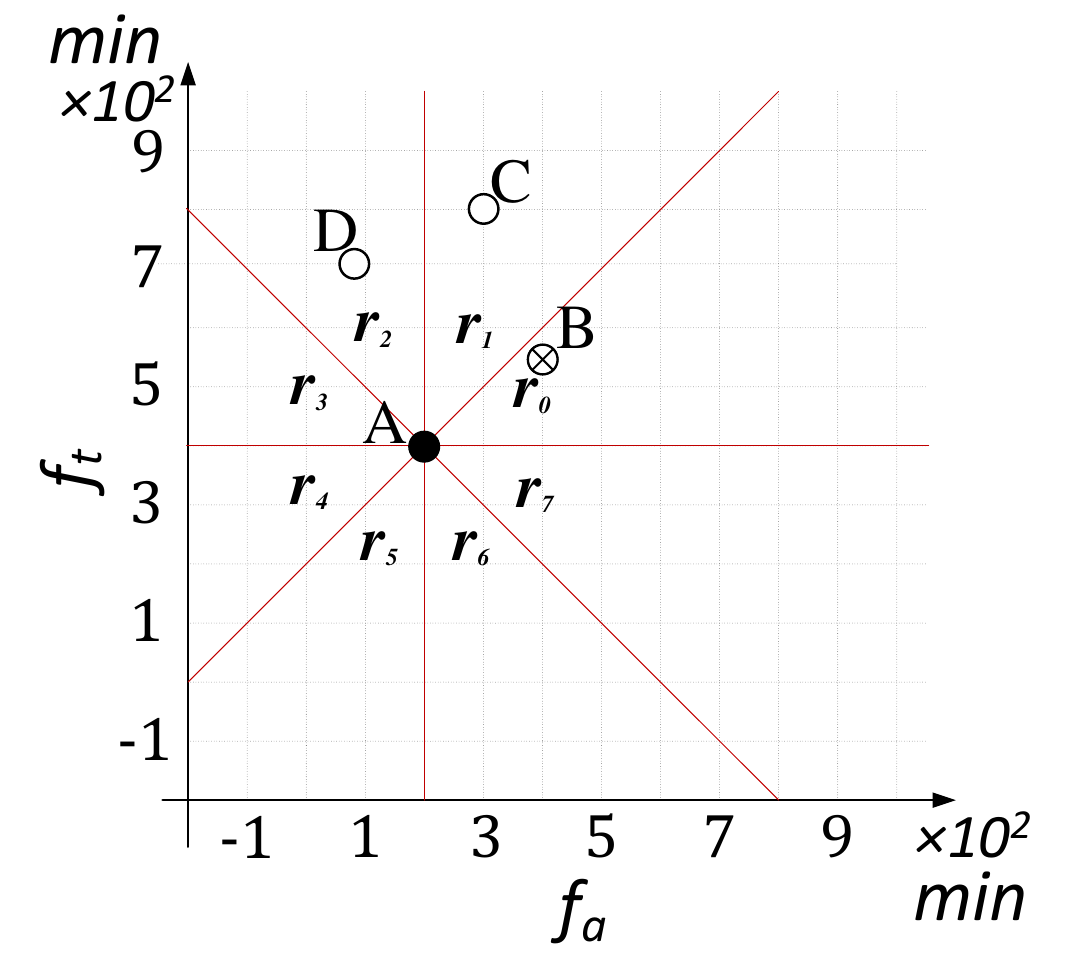}
	\subcaption{{If no normalization is used; scaling at $w=100$ in MMO}}
	\end{subfigure}
		~\hfill
	\begin{subfigure}[h]{0.32\textwidth}
		\includegraphics[width=\textwidth]{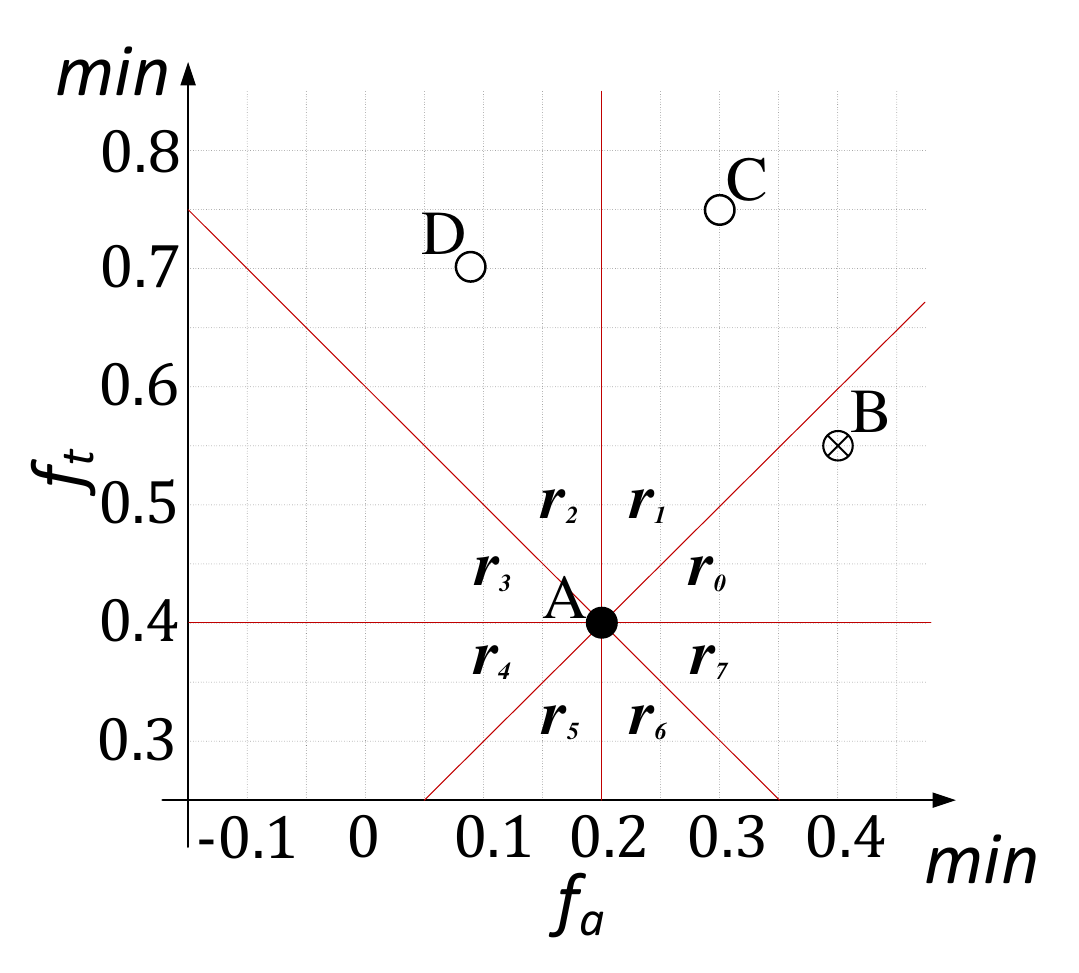}
  \subcaption{{Anticipated effect of normalization; scaling at $w=1$ in MMO}}
	\end{subfigure}
		\caption{An illustration of the impact of the scale discrepancy between performance objectives and the anticipated normalization effect in MMO. $\vect{A}$, $\vect{B}$, $\vect{C}$, and $\vect{D}$ are four configurations with the auxiliary and target performance objective values as $(2\times10^2, 4)$, $(4\times10^2, 5.5)$, $(3\times10^2, 8)$, respectively. \textbf{$\otimes$ and $\Circle$ denote the configurations that will still be/become nondominated and dominated, respectively, after the rotation in MMO;} $\bullet$ is the reference nondominated configuration.}
	\label{fig:Our targets}
\end{figure*}


\vspace{5pt}
\noindent \textbf{Remark 1.} The global optimum of the original single-objective problem 
(i.e., the configuration with the best target performance objective) 
is Pareto optimal (e.g., the configuration $\vect{A}$ in the example of Figure~\ref{Fig:model}).
This can be derived immediately by contradiction from Equation~(\ref{Eq:model}) or the analysis of rotation from Section~\ref{sec:theory}.


\vspace{5pt}
\noindent \textbf{Remark 2.}
A similar but more general observation is that 	
a configuration will never be dominated by another that has a worse target performance objective.
That is, 
if configuration $\vect{x_1}$ has a better target performance objective than $\vect{x_2}$ 
(i.e., $f_t(\vect{x_1}) < f_t(\vect{x_2})$), 
then whatever their auxiliary performance objective values are,
$\vect{x_2}$ will not be better than $\vect{x_1}$ on both $g_1$ and $g_2$;
in the best case for $\vect{x_2}$, 
they are nondominated to each other 
(e.g., the configuration $\vect{B}$ versus $\vect{C}$ in Figure~\ref{Fig:model}). Indeed, according to the analysis from Section~\ref{sec:theory}, there is no way for $\vect{x_1}$ to be moved to the region $r_0$ or $r_1$ of $\vect{x_2}$ as $\vect{x_1}$ can only be in the $r_4$ to $r_7$ of $\vect{x_2}$ before the rotation.

\vspace{5pt}
\noindent \textbf{Remark 3.}
The above two remarks apply to the target performance objective, 
but not to the auxiliary performance objective. 
This is a key difference from the PMO model, 
where both objectives hold these remarks.
An example of the consequence is the configuration $\vect{D}$ of Figure~\ref{Fig:model}, 
which is meaningless to the original problem, 
but treated as being optimal in PMO and not in MMO.

\vspace{5pt}
\noindent \textbf{Remark 4.}
Our MMO model does not bias to a higher or lower value on the auxiliary performance objective,
in contrast to PMO. 
Indeed, 
as in \textbf{Property 3}, 
we do not know for certain what value of the auxiliary performance objective corresponds to the
best target performance objective.

\vspace{5pt}
\noindent \textbf{Remark 5.}
Configurations with dissimilar auxiliary performance objective values tend to be incomparable
	(i.e., nondominated to each other) 
	even if one is
	fairly inferior to the other on the target performance objective.
	For example,
	the configuration $\vect{C}$ in Figure~\ref{Fig:model},
	which has worse latency than $\vect{A}$,
	is not dominated by $\vect{A}$ as their throughput are rather different.
	In contrast, 
	the configuration $\vect{B}$,
	which even has better latency than $\vect{C}$,
	is dominated by $\vect{A}$, 
	as they are similar on throughput. 
	This enables the model to keep exploring diverse promising configurations during the search, 
	thereby a higher chance to find the global optimum.


\vspace{5pt}
\noindent \textbf{Remark 6.}
If two configurations have the same value of auxiliary performance objective $f_a$,
then they are always subject to the dominance relation (i.e., either dominating or being dominated). This is because, if configuration $\vect{x_1}$ has the same $f_a$ value as $\vect{x_2}$, then $\vect{x_1}$ can only be on the boundary between $r_1$ and $r_2$ (or $r_5$ and $r_6$) for $\vect{x_2}$, meaning that after the rotation, it can only be on the boundary between $r_0$ and $r_1$ (or $r_4$ and $r_5$) for $\vect{x_2}$, in which case their relationships are always dominated (comparable) regardless the $f_t$ value.

\vspace{5pt}
\noindent \textbf{Remark 7.}
If two configurations have the same value of the target performance objective $f_t$, 
then they are always nondominated to each other in the MMO model.
This is because, if configuration $\vect{x_1}$ has the same $f_t$ value as $\vect{x_2}$, then $\vect{x_1}$ can only be on the boundary between $r_0$ and $r_7$ (or $r_3$ and $r_4$) for $\vect{x_2}$, meaning that after the rotation, it can only be on the boundary between $r_7$ and $r_6$ (or $r_2$ and $r_3$) for $\vect{x_2}$, in which case their relationships are always nondominated (incomparable) regardless the $f_a$ value and its scaling.


\vspace{5pt}
From \textbf{Remarks 1--5}, 
we can see that the MMO model is capable of focusing on optimizing the target performance objective (\textbf{Goal 1})
while mitigating the search from being trapped in local optima (\textbf{Goal 2}). 
In the following sections, 
on the basis of \textbf{Remarks 6} and \textbf{7}, together with the analysis from Section~\ref{sec:theory},
we will explain why and how the weight parameter $w$ in the MMO model can be removed by changing the normalization method for the model.


\subsection{Normalization for MMO Model in the FSE work~\cite{ChenMMO21}}
\label{sec:norm-theory}

The above analysis assumes an ideal scenario, i.e., the target and auxiliary performance objectives are of similar scale. This is, however, unrealistic for practically tuning configuration. For example, measuring the difference between latency results often reaches the magnitude of $5$ order while CPU load merely differs at the scale of a few percentages. The consequence is that the appropriate $w$ value, which enables a good balance between exploitation and exploitation for MMO, can be either very large or very small depending on the case, leading to high difficulty in setting the $w$.

Figure~\ref{fig:Our targets} gives an example. As can be seen, from Figure~\ref{fig:Our targets}a, $f_t$ has roughly $100\times$ greater scale than that of $f_a$, and thereby from the coordinates, the configurations will be shrunk along $f_a$ to the boundary between $r_1$ and $r_2$ for $\vect{A}$. This means that, after the rotation in MMO, $\vect{A}$ will dominate all other configurations and make them comparable, which is harmful. To mitigate such, one would need to give a rather large $w$ value, i.e., $w=100$, that enables a more reasonable incomparability among the configurations, i.e., some are nondominated while some others are dominated by $\vect{A}$ (Figure~\ref{fig:Our targets}b). Yet, since we do not normally have a precise understanding as to what extent the scales between different performance objectives differ beforehand, one would need to examine a wide range of possible $w$ values.



To ease the above, in the FSE work~\cite{ChenMMO21},
we obtain more commensurable $f_t(\vect{x})$ and $f_a(\vect{x})$ via the following normalization (we call it FSE normalization thereafter): 
\begin{equation}
f(\vect{x})  = {{f^{o}(\vect{x}) - f^o_{lower} \over {f^o_{upper} - f^o_{lower}}}}
\label{Eq:norm}
\end{equation}
where $f^{o}(\vect{x})$ denotes the original value of the configuration $\vect{x}$ on the performance objective $f$, 
and $f^o_{lower}$ and $f^o_{upper}$ are the global lower and upper bounds on that performance objective for the software, 
respectively. That is, the true scale of the performance objective is used as the bounds.

In practical software configuration tuning, however, $f^o_{lower}$ and $f^o_{upper}$ are likely to be unknown a priori. 
Therefore in our FSE work, 
these bounds are updated by using the maximum and minimum values discovered so far during the tuning to approximate the true scales. 
Note that using the true scales of the objectives (if known) or their close approximations for normalization is a widely used method in SBSE~\cite{10.1145/3392031,DBLP:conf/sigsoft/ShahbazianKBM20,DBLP:conf/ecows/Strunk10,DBLP:journals/asc/CremeneSPD16,DBLP:conf/icsoc/AlrifaiRDN08}.

Essentially, normalization plays a similar role to $w$ in that they both scale the relative positions of the configurations (but the $w$ primarily works on $f_a$). As such, with the FSE normalization, our hope was that its resulting scaling could reduce the range of the ideal $w$ values, hence relieving the effort of adjusting it. In the best scenario, we anticipate that the configurations on $f_t$ and $f_a$ can be naturally scaled to ideal positions even when $w=1$ (i.e., no scaling), thereby there is a good mix of incomparable and comparable configurations after the rotation,
leading to more balanced exploitation and exploration, e.g., in Figure~\ref{fig:Our targets}c.


\begin{figure}[t!]
	\centering

	\begin{subfigure}[h]{0.45\columnwidth}
		\includegraphics[width=\columnwidth]{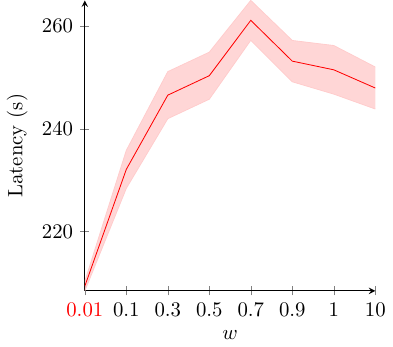}
		\subcaption{\textsc{Storm}}
	\end{subfigure}
		 ~\hfill
	\begin{subfigure}[h]{0.45\columnwidth}
	\includegraphics[width=\columnwidth]{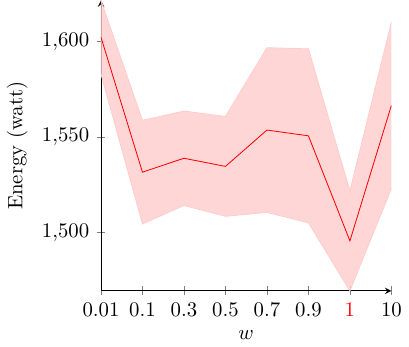}
			\subcaption{\textsc{x264}}
	\end{subfigure}
		\caption{The performance of eight $w$ values for the original MMO model with the FSE normalization on two exampled systems (the best $w$ is highlighted). With 50 repeated runs under 600 and 2500 measurement budgets respectively, \textsc{Storm} consumes a total of $600\times8\times50=2.4\times10^5$ measurements while \textsc{x264} needs $2500\times8\times50=10^6$ measurements. Suppose that each measurement takes one second, it would need around 2.7 and 11.6 days merely to identify the best $w$ setting.}
	\label{fig:weigh-perf-exp}
\end{figure}

\subsection{What was Wrong?}
\label{sec:why-new}


Indeed, we have shown that the FSE normalization method can be effective in narrowing down the ideal range of $w$ to achieve superior results~\cite{ChenMMO21}, 
but with one ineffective outcome: 
our analysis thereafter reveals that the weight $w$ in the MMO model remains a highly sensitive parameter, even within a narrower range,
and finding the right setting for a system still requires much effort of trial and error.
In~\cite{ChenMMO21} and this work (Section~\ref{sec:result}), 
we examined a set of the weight settings for MMO model (i.e., $0.01, 0.1, 0.3, 0.5, 0.7, 0.9, 1.0, 10$)\footnote{We chose these values because they are originally used in the FSE work and it is found that $w$ values outside $[0.01,10]$ only degrade the results.}.
A key finding is that the weight achieving the best performance differs drastically on different configurable software systems: as shown in Figure~\ref{fig:weigh-perf-exp}, some systems work better with a tiny weight value, e.g., $w=0.01$ for the latency of \textsc{Storm} under the \textsc{WordCount} benchmark, while some others do best with a much bigger value, e.g., $w=1$ for the energy usage of system \textsc{x264}. In what follows, we will explain what caused this issue that deviates from our original expectation by means of both theoretical analysis and empirical evidence.

\begin{figure*}[t!]
	\centering
     \begin{subfigure}[h]{0.32\textwidth}
		\includegraphics[width=\textwidth]{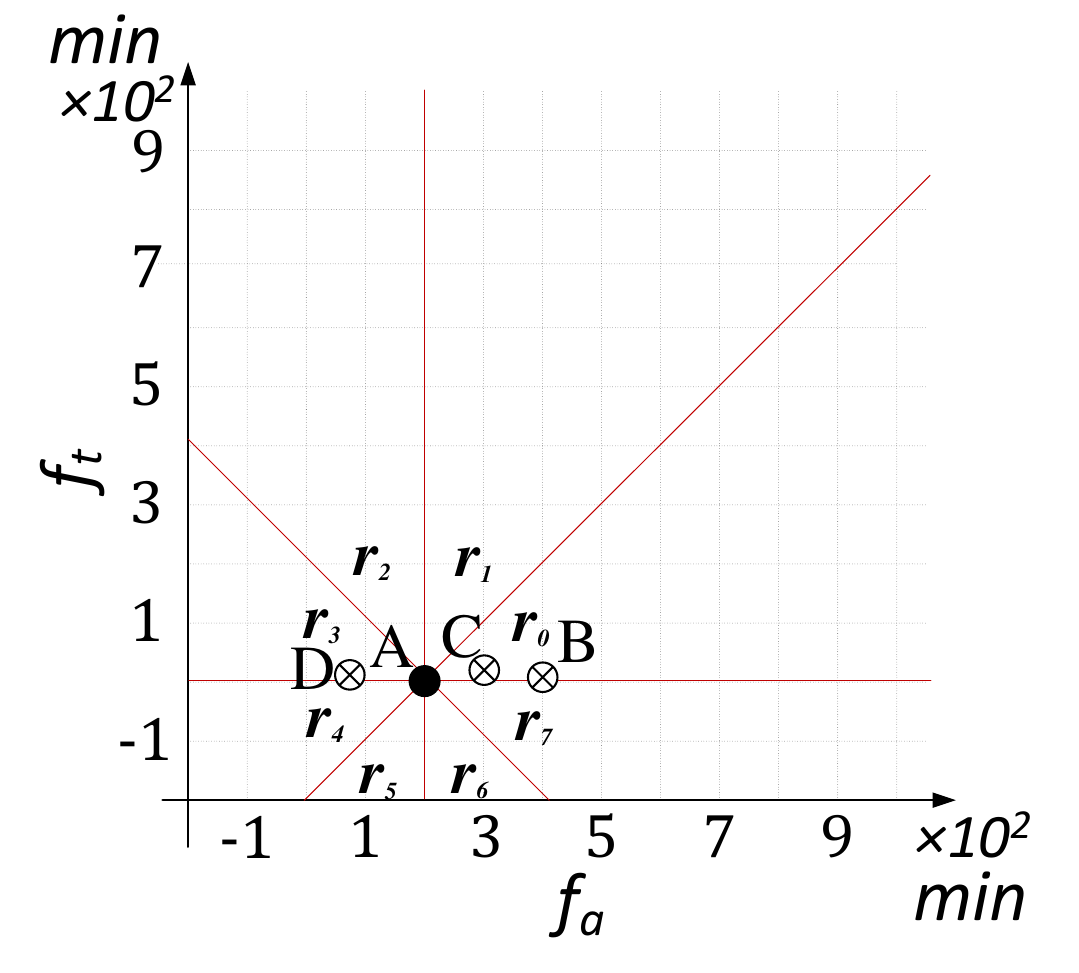}
		\subcaption{{Original space where $f_t$ has a much smaller scale than that of $f_a$}}
	\end{subfigure}
		~\hfill
	\begin{subfigure}[h]{0.32\textwidth}
		\includegraphics[width=\textwidth]{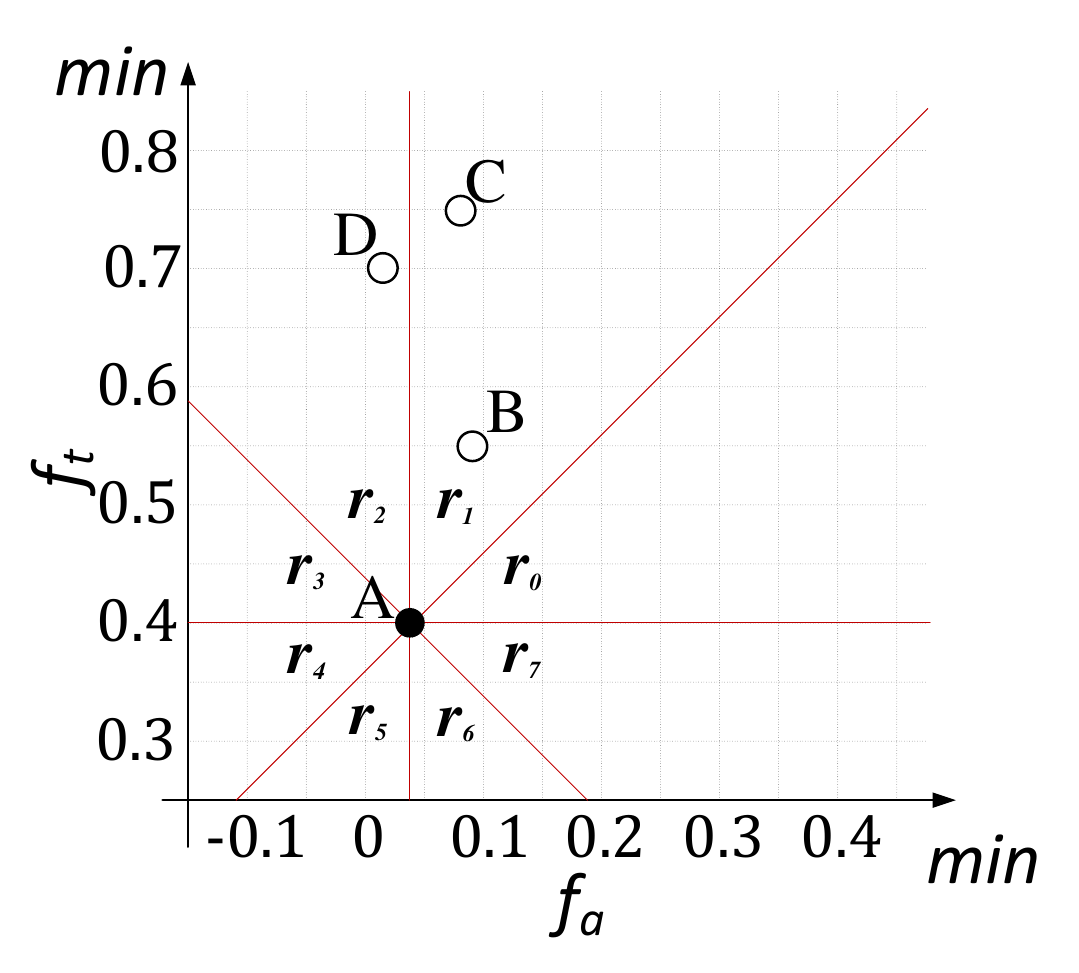}
		\subcaption{{FSE normalization at $f_a \in[0,5000]$ /$f_t \in[0,10]$; scaling at $w=1$ in MMO}}
	\end{subfigure}
		~\hfill
	\begin{subfigure}[h]{0.32\textwidth}
		\includegraphics[width=\textwidth]{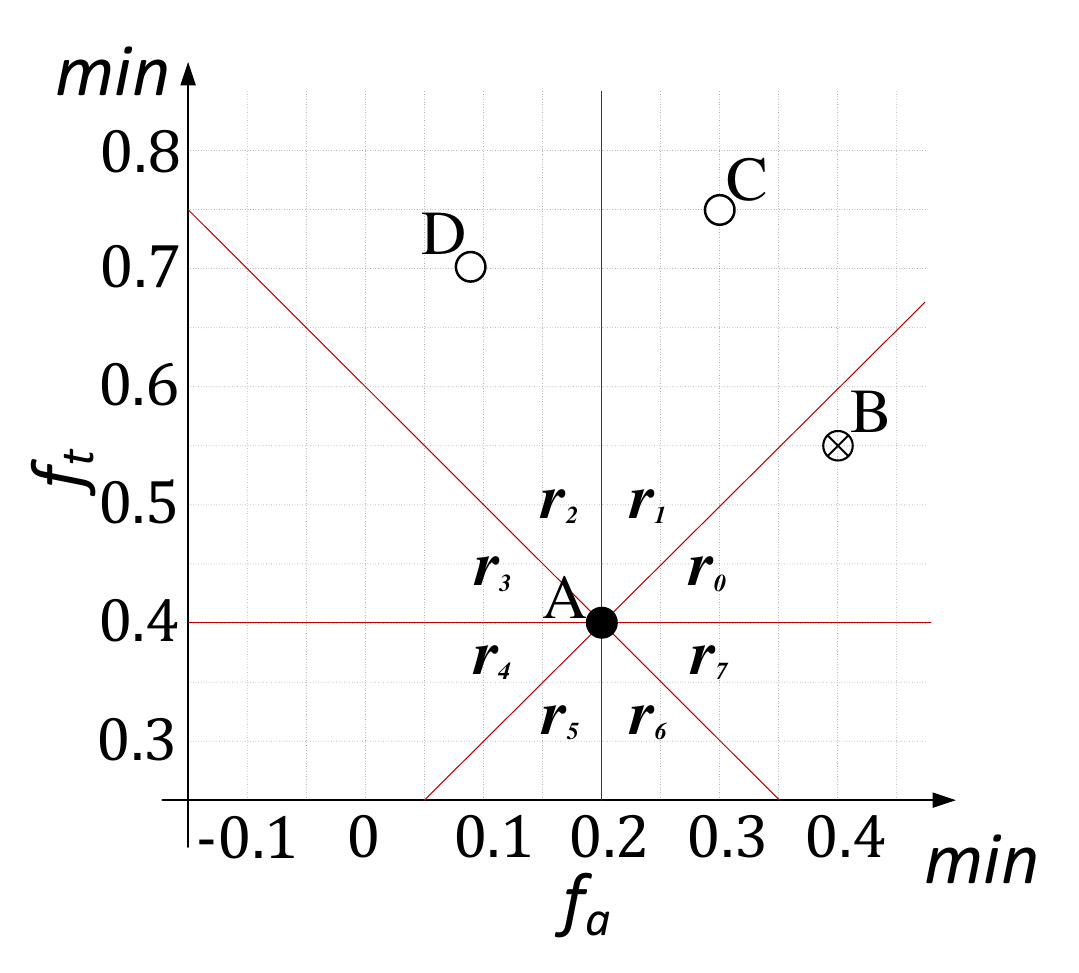}
			\subcaption{{FSE normalization at $f_a \in[0,5000]$ /$f_t \in[0,10]$; scaling at $w=5$ in MMO}}
	\end{subfigure}
	
     \begin{subfigure}[h]{0.32\textwidth}
		\includegraphics[width=\textwidth]{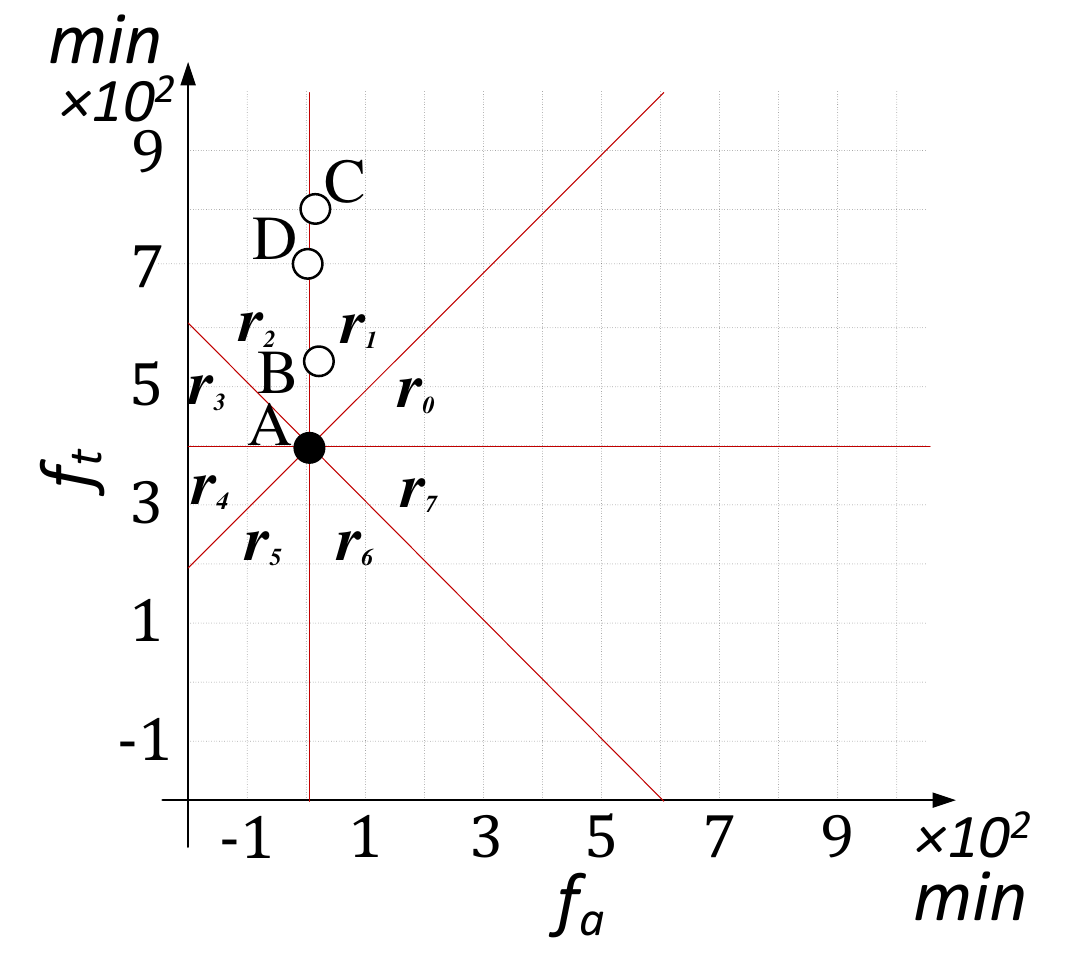}
			\subcaption{{Original space where $f_a$ has a much smaller scale than that of $f_t$}}
	\end{subfigure}
		~\hfill
	\begin{subfigure}[h]{0.32\textwidth}
		\includegraphics[width=\textwidth]{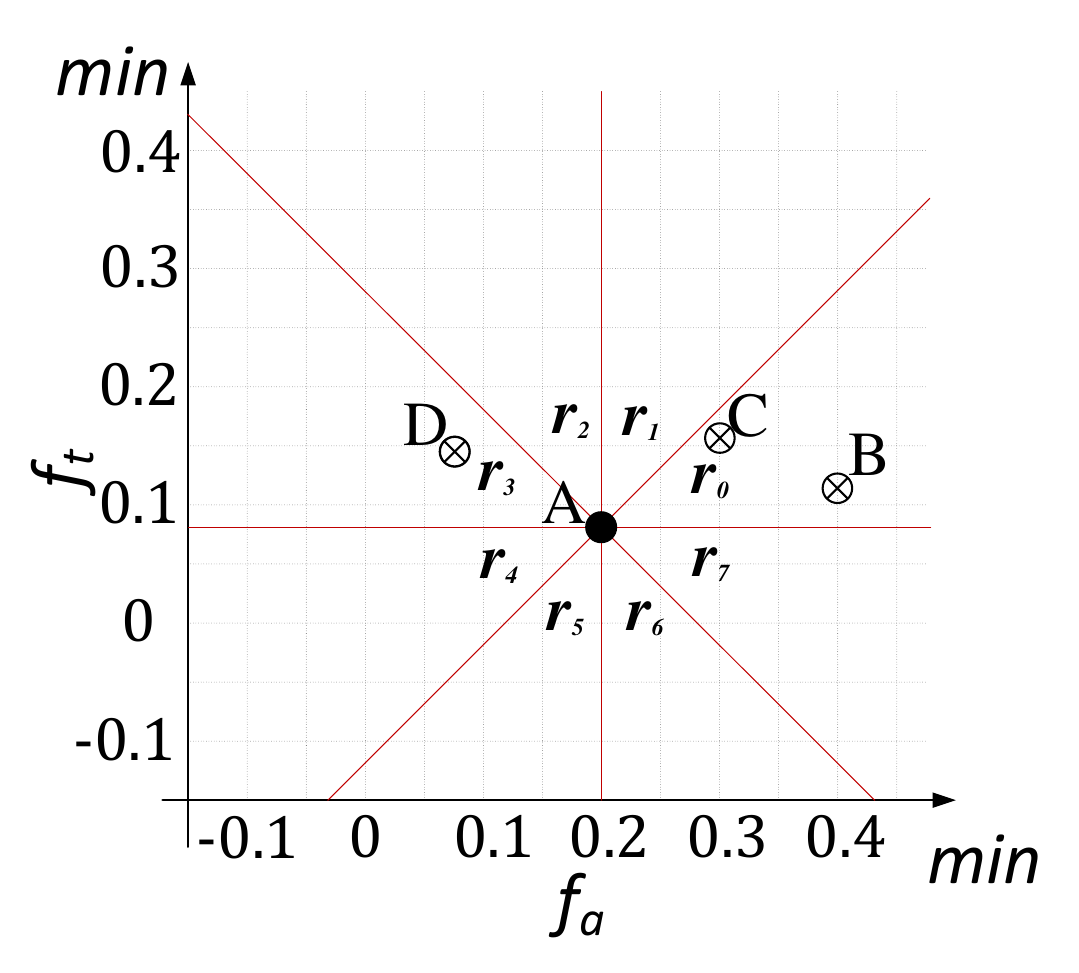}
			\subcaption{{FSE normalization at $f_a \in[0,10]$/$f_t \in[0,5000]$; scaling at $w=1$ in MMO}}
	\end{subfigure}
		~\hfill
	\begin{subfigure}[h]{0.32\textwidth}
		\includegraphics[width=\textwidth]{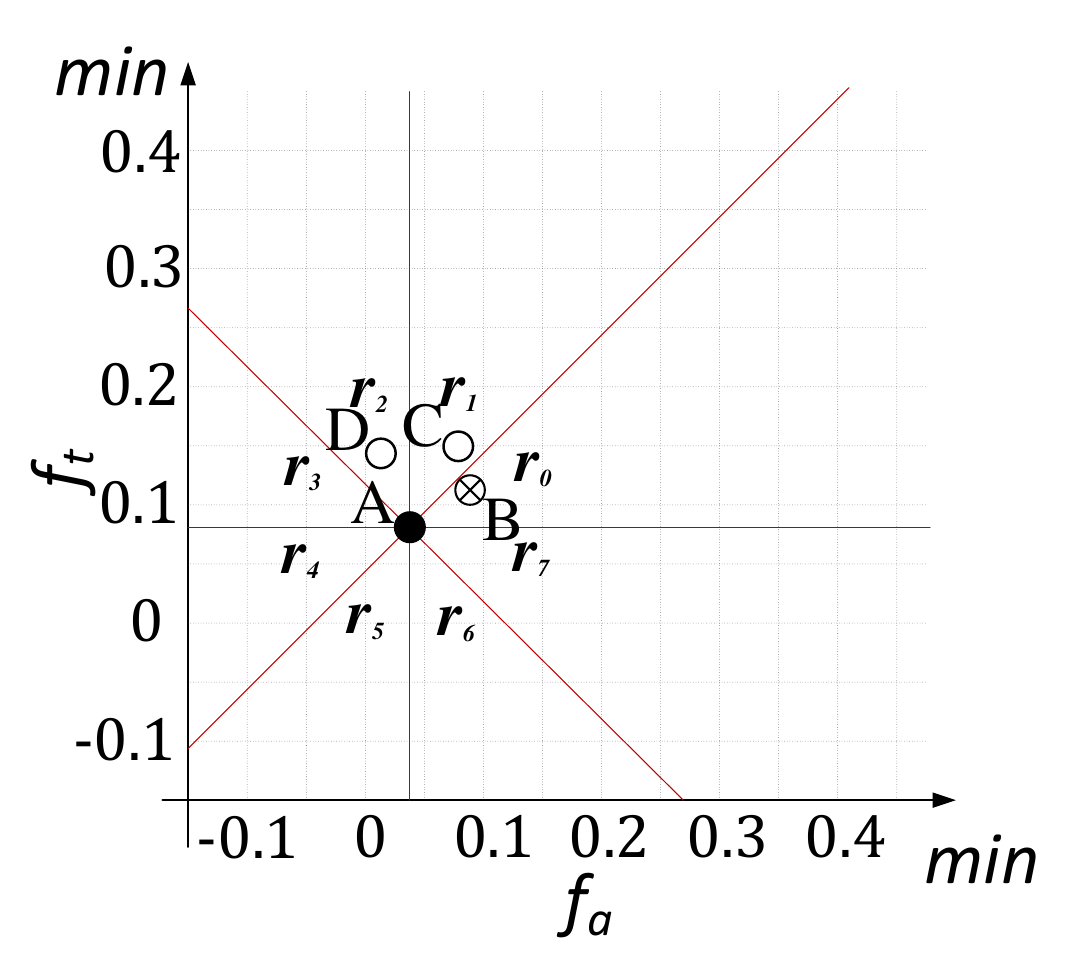}
				\subcaption{{FSE normalization at $f_a \in[0,10]$/$f_t \in[0,5000]$; scaling at $w=0.2$ in MMO}}
	\end{subfigure}
		\caption{Illustration of why the MMO designs in our FSE work is highly sensitive to $w$. $\vect{A}$, $\vect{B}$, $\vect{C}$, and $\vect{D}$ are four configurations with the auxiliary and target
performance objective values as $(2, 4\times10^2)$, $(4, 5.5\times10^2)$, $(3, 8\times10^2)$, and $(0.75, 7\times10^2)$ in (a), respectively. In (d), the values are $(2\times10^2, 4)$, $(4\times10^2, 5.5)$, $(3\times10^2, 8)$, and $(0.75\times10^2, 7)$, respectively. The format is the same as Figure~\ref{fig:Our targets}. (a), (b), and (c) show the case where the $f_a$ is normalized into a tiny range with bounds of $[0,5000]$ as opposed to that of $[0,10]$ for $f_t$; (d), (c), and (e) demonstrate the case where the $f_t$ is normalized into a tiny range with bounds of $[0,5000]$ as opposed to that of $[0,10]$ for $f_a$.}
	\label{fig:w'influence in norm}
\end{figure*}

\subsubsection{An Analysis}

The above occurrence is due to the severe discrepancy between the range of the current search population and the performance objectives' scales in software configuration tuning, which obscures the benefit of the normalization schema we used for the FSE work. To provide a sound analysis thereof, recall the analysis on the effect of $w$ and the rotation from Section~\ref{sec:theory}, using lower/upper bound in the normalization might lead to two cases in the presence of discrepant performance objective values: 

\textbf{Case 1:} If the $f_a$ of the configurations in the population shrinks into a tiny range compared to its true objective scale in the whole search space (while the $f_t$ does not), then the $f_a$ in the population after the normalization (Equation~\ref{Eq:norm}) will be very close (see \textbf{Remark 6}). Figure~\ref{fig:w'influence in norm}a shows an example, from which we see that without normalization, all configurations will be nondominated after rotation, i.e., the scale of $f_t$ (e.g., CPU load) is much smaller than that of $f_a$ (e.g., latency), which is not ideal. However, when the bounds for $f_a$ is $[0,5000]$ while that for $f_t$ is $[0,10]$, with the FSE normalization (Figure~\ref{fig:w'influence in norm}b), all configurations (except $\vect{A}$) will be shrunk towards the boundary between $r_1$ and $r_2$ of $\vect{A}$ (or $r_5$ and $r_6$), meaning that they will become dominated (or dominate to) by $\vect{A}$ after the rotation. This is also devastating to the tuning since too many comparable configurations will over-emphasize exploitation. To mitigate such, we need to set a larger $w$ for stretching $f_a$, i.e., in this case, $w=5$ as shown in Figure~\ref{fig:w'influence in norm}c, thereby the number of incomparable configurations after rotation can be more reasonable to balance the exploration and exploitation.

\begin{figure}[t!]
	\centering
	\begin{subfigure}[h]{0.31\columnwidth}
		\includegraphics[width=\textwidth]{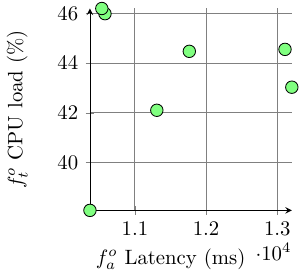}
		\subcaption{\footnotesize Original}
	\end{subfigure}
	~
	\begin{subfigure}[h]{0.31\columnwidth}
		\includegraphics[width=\textwidth]{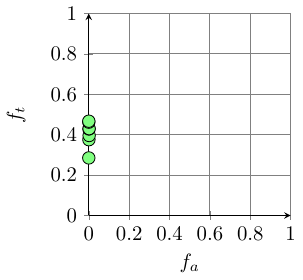}
		\subcaption{\footnotesize Normalized}
	\end{subfigure}
	~
	\begin{subfigure}[h]{0.31\columnwidth}
		\includegraphics[width=\textwidth]{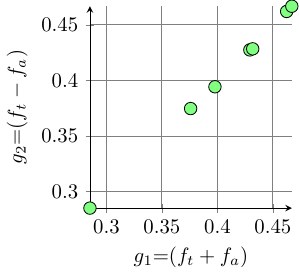}
		\subcaption{\footnotesize In the MMO space}
	\end{subfigure}

	\caption{An intermediate population of configurations for system \textsc{LRZIP} generated by our MMO model with the FSE normalization~\cite{ChenMMO21} on top of NSGA-II. Scales in the original population are adjusted for visibility.}
	\label{fig:lrzip-old}
\end{figure}
\begin{figure}[t!]
	\centering
	\begin{subfigure}[h]{0.31\columnwidth}
		\includegraphics[width=\textwidth]{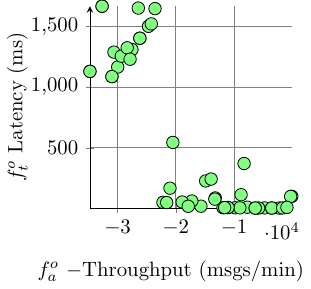}
		\subcaption{\footnotesize Original}
	\end{subfigure}
	~
	\begin{subfigure}[h]{0.31\columnwidth}
		\includegraphics[width=\textwidth]{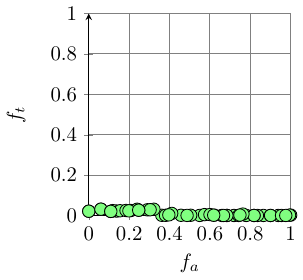}
		\subcaption{\footnotesize Normalized}
	\end{subfigure}
	~
	\begin{subfigure}[h]{0.31\columnwidth}
		\includegraphics[width=\textwidth]{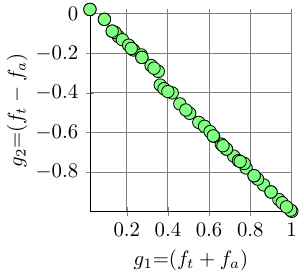}
		\subcaption{\footnotesize In the MMO space}
	\end{subfigure}

	\caption{An intermediate population of configurations for system \textsc{Storm/WC} generated by our MMO model with the FSE normalization~\cite{ChenMMO21} on top of NSGA-II. Scales in the original population are adjusted for visibility.}
	\label{fig:storm-old}
\end{figure}

\textbf{Case 2:} On the other hand, if the $f_t$ in the population evolves into a range
that is tiny compared to its objective scale in the whole search space 
(while the $f_a$ does not),
then the values of the $f_t$ of configurations in the current population after the normalization
(Equation~\ref{Eq:norm}) will be very close (see \textbf{Remark 7}). As can be seen in Figure~\ref{fig:w'influence in norm}d, suppose that before normalization the scale of $f_a$ (e.g., CPU load) is much smaller than that of $f_t$ (e.g., latency), then this will cause $\vect{A}$ to dominate all other configurations when rotating, which is harmful. With the FSE normalization under the bounds for is $[0,5000]$ for $f_t$ and $[0,10]$ for $f_a$, as in Figure~\ref{fig:w'influence in norm}e, all configurations (except $\vect{A}$) will be shrunk towards the boundary between $r_3$ and $r_4$ (or $r_0$ and $r_7$) of $\vect{A}$, causing more configurations to become nondominated to $\vect{A}$ after the rotation, hence creating many incomparable configurations that focus too much on exploration that would also harm the guidance of tuning. Likewise, to relieve such a case, we need to set a smaller $w$ for shrinking $f_a$, e.g., $w=0.2$ as shown in Figure~\ref{fig:w'influence in norm}f, thereby the number of incomparable configurations after rotation is more appropriate, enabling the tuning to favor towards exploitation that reaches a balance.

As a result, from the above, it is clear that although the FSE normalization helps to reduce the ideal ranges of $w$ values ($0.2$ and $5$ instead of the range on, e.g., $[0.001,1000]$), it can still negatively influence the appropriate number of incomparable configurations following the rotation in MMO, because a very large upper bound for one performance objective might be reached and being used throughout the tuning, even if such an extreme configuration has been ruled out later on. Therefore, we still need to non-trivially adjust $w$ to mitigate such a side-effect from the FSE normalization, which collectively influences the effects of rotation. This is the key reason that the MMO designs in our FSE work remain highly sensitive to its only parameter $w$.

\subsubsection{Some Empirical Evidences}

\begin{figure*}[t!]
	\centering

	\begin{subfigure}[h]{0.24\textwidth}
		\includegraphics[width=\textwidth]{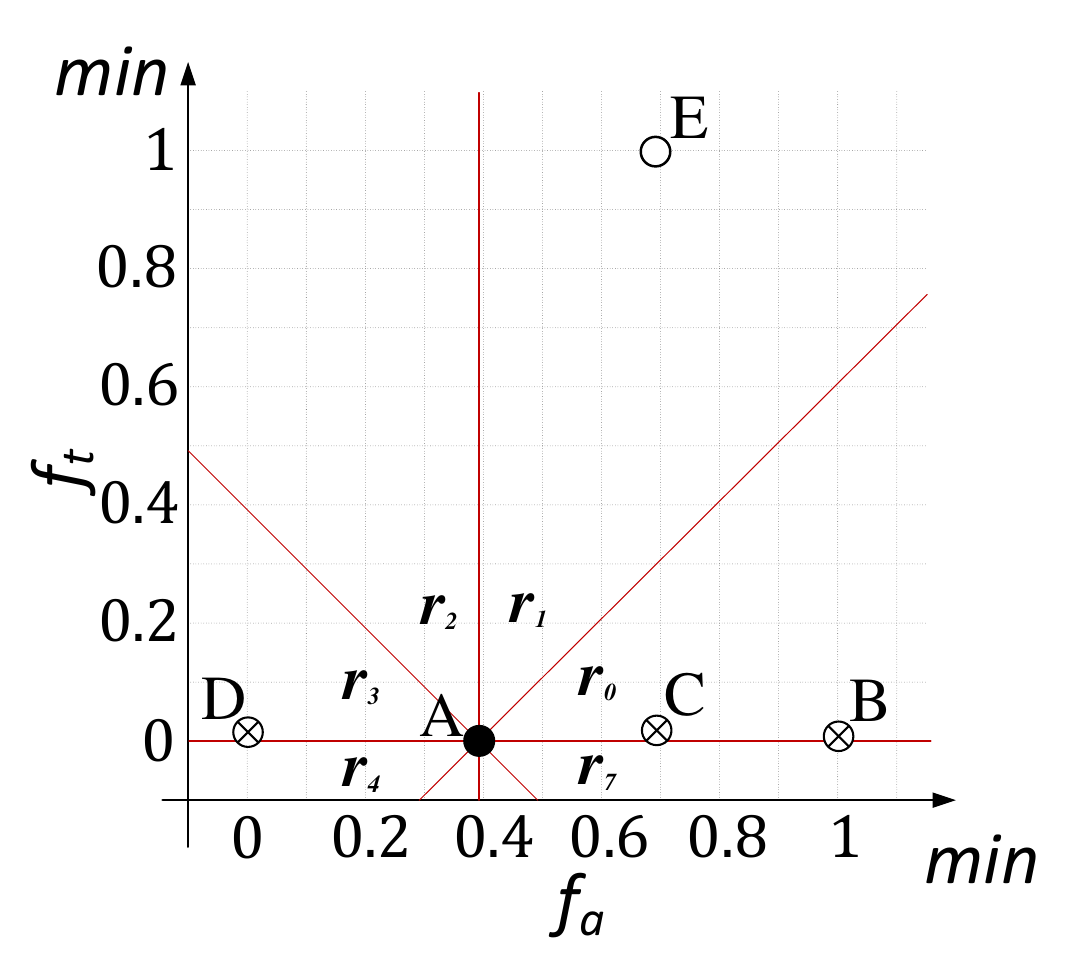}
		\subcaption{{FSE normalization in MMO at an iteration where $\vect{E}$ is newly discovered}}
	\end{subfigure}
		 ~\hfill
	\begin{subfigure}[h]{0.24\textwidth}
		\includegraphics[width=\textwidth]{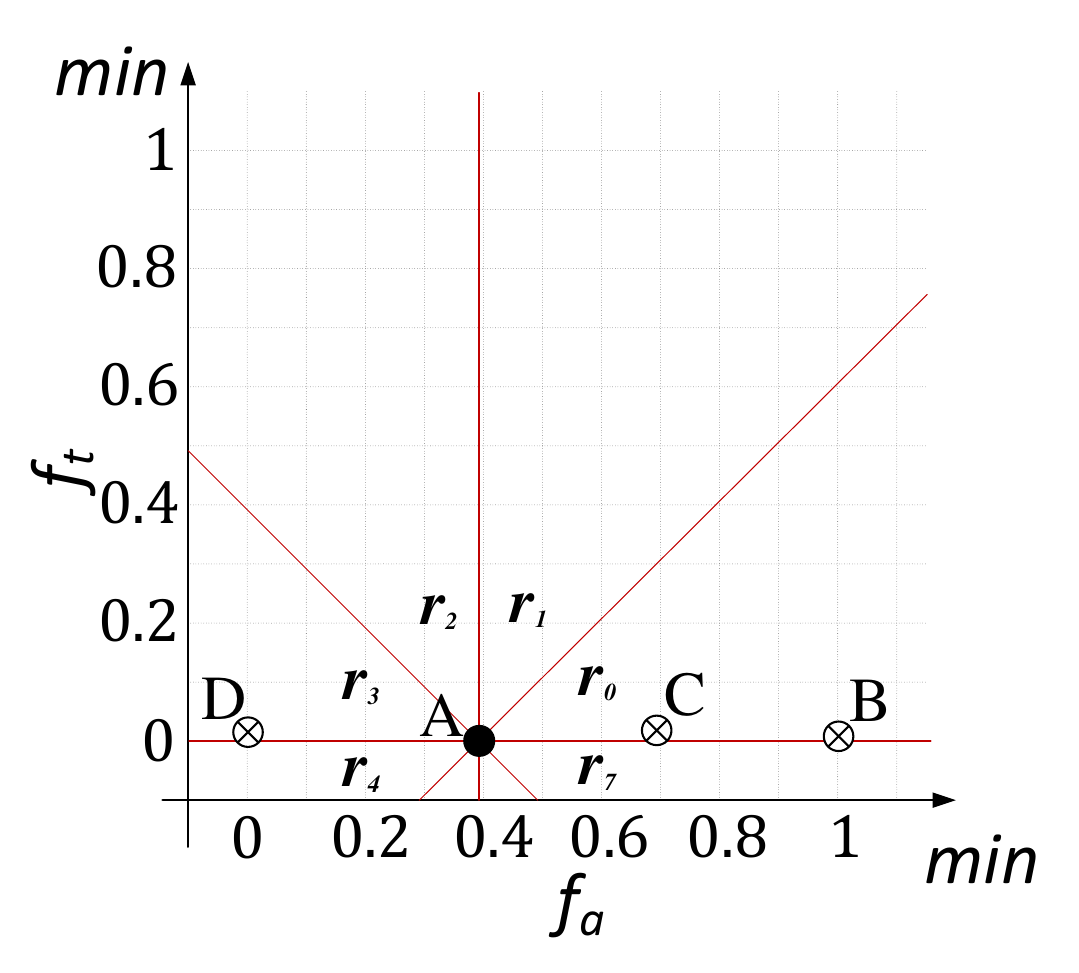}
			\subcaption{{FSE normalization in MMO after some iterations when $\vect{E}$ is eliminated}}
	\end{subfigure}
	 ~\hfill
	\begin{subfigure}[h]{0.24\textwidth}
		\includegraphics[width=\textwidth]{figures/new_pic_1/2_norm_1.pdf}
			\subcaption{{New normalization in MMO at an iteration where $\vect{E}$ is newly discovered}}
	\end{subfigure}
	 ~\hfill
	\begin{subfigure}[h]{0.24\textwidth}
		\includegraphics[width=\textwidth]{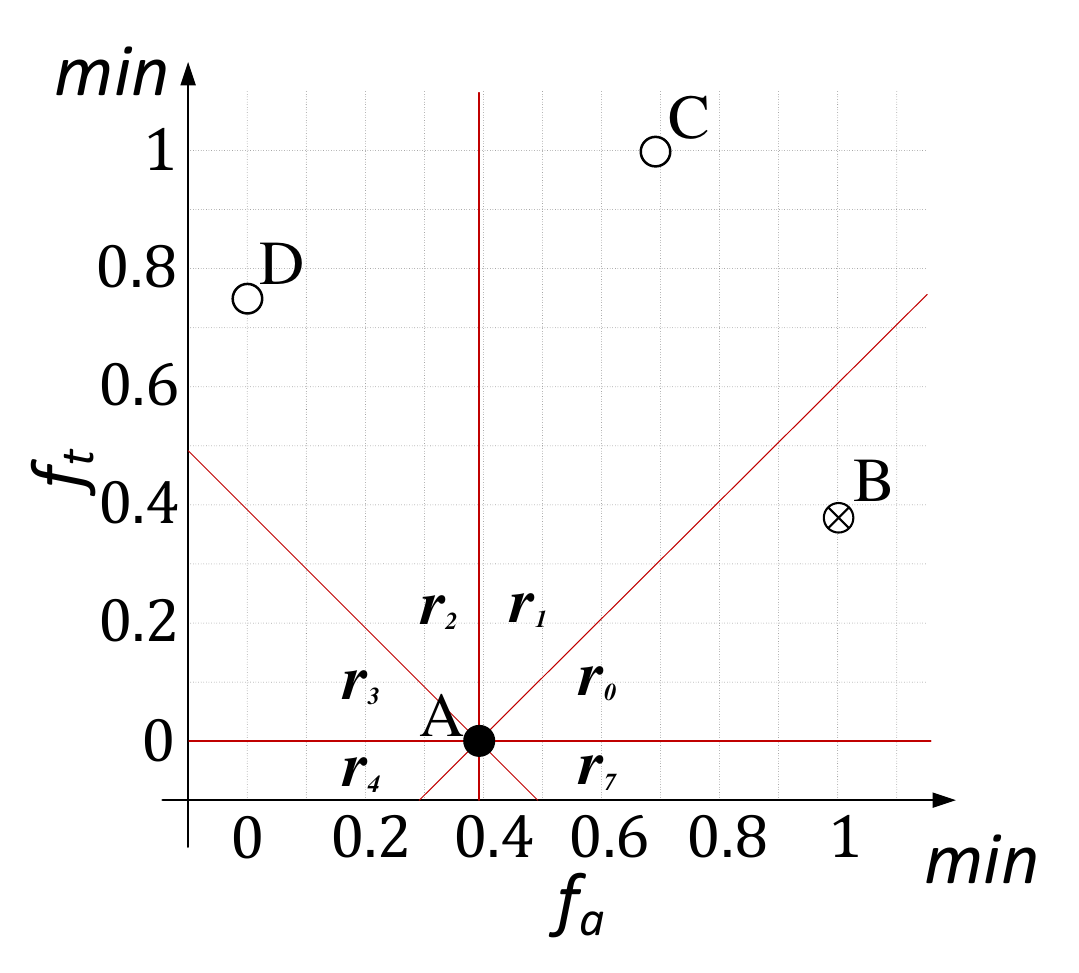}
				\subcaption{{New normalization in MMO after some iterations when $\vect{E}$ is eliminated}}
	\end{subfigure}
		\caption{A simple illustration of why the new normalization is more useful. $\vect{A}$, $\vect{B}$, $\vect{C}$, $\vect{D}$ and $\vect{E}$ are five configurations with the auxiliary and target performance objective values as $(2, 4)$, $(4, 5.5)$, $(3, 8)$, $(0.75, 7)$, and $(3, 5\times10^2)$, respectively. The format is the same as Figure~\ref{fig:Our targets}. (a) and (b) show the case of MMO under the FSE normalization; (c) and (d) demonstrate the case of MMO under the new normalization. All cases have no scaling in MMO, i.e., $w=1$.}
	\label{fig:The contrast between norms}
\end{figure*}

To demonstrate the devastating impact of FSE normalization on the MMO model, Figure~\ref{fig:lrzip-old} illustrates an intermediate population of MMO model during the tuning for system \textsc{LRZIP}. As can be seen from Figure~\ref{fig:lrzip-old}b, 
since the range of auxiliary performance objective in the population becomes ``very small'' (around $[1.03\times10^4ms, 1.3\times10^4ms]$), 
compared to the objective scale ($[10^4ms, 1.7\times10^6ms]$,
after the normalization the auxiliary performance objective's values become tiny, 
condensing in the range of $[0, 1.7\times10^{-3}]$ only. 
This, as we discussed during the analysis, can lead to all the configurations in the population being either dominating or dominated by each other 
within the transformed space of our MMO model (Figure~\ref{fig:lrzip-old}c), which effectively means that the problem degenerates to the original single-objective problem, 
where the configurations are discriminative virtually based on their target performance objective 
(thus easily being trapped in local optima).

Figure~\ref{fig:storm-old} gives yet another example, where we visualize an intermediate population of MMO model (with the FSE normalization) 
during the tuning for system \textsc{Storm/WC}.   
As can be seen in Figure~\ref{fig:storm-old}b, 
since the range of the target performance objective (latency) in the population becomes ``very small'' (around $[95ms, 1800ms]$), 
compared to the objective scale ($[3ms, 55209ms]$),
after the normalization the values of the target performance objective become tiny, 
condensing in the range of $[0, 0.03]$ only. 
The auxiliary performance objective (throughput), 
in contrast, 
are more evenly spread over the range of $[0,1]$ after the normalization.
As with our analysis, this can lead to all the configurations in the population being nondominated to each other 
within the transformed MMO space (Figure~\ref{fig:storm-old}c).
Unfortunately, 
all configurations in the population being nondominated is detrimental to the search 
since there is no selection pressure (i.e., discriminative power); 
everyone is incomparable even the one with the best target performance objective.

\subsection{A New Normalization}

The above analysis and observations suggest that the FSE normalization based on the (approximate) true scales of the performance objectives may not be suitable for the MMO model.
Fortunately, 
this can be fixed by considering the current population as the basis of bounds in the normalization, which is the key extension in this work. 
That is, we replace Equation~\ref{Eq:norm} with the following:

\begin{equation}
f(\vect{x})  = {{f^{o}(\vect{x}) - f^o_{min}} \over {f^o_{max} - f^o_{min}}}
\label{Eq:norm_new}
\end{equation}
where $f^{o}(\vect{x})$ denotes the original value of the configuration $\vect{x}$ on the performance objective $f$, 
and $f^o_{min}$ and $f^o_{max}$ are the maximum and minimum values of the current population on $f$, 
respectively. As such, instead of using the global bounds throughout the search, the local bounds 
(in the population of configurations of every generation along with the evolution) are used in the normalization.

To better explain how the new normalization differs from the FSE normalization in the MMO space, Figure~\ref{fig:The contrast between norms} shows an example. Here, from Figure~\ref{fig:The contrast between norms}a and~\ref{fig:The contrast between norms}c, suppose that during the tuning a configuration $\vect{E}$ with a very large value of $f_t$ is discovered, then this will cause both normalizations to shrink the configurations along $f_t$, leaving a negative impact as most configurations will become nondominated (incomparable) after rotation. However, it is possible that $\vect{E}$ will be subsequently ruled out due to it being the only configuration dominated by $\vect{A}$ in the MMO space. Yet, with the FSE normalization (Figure~\ref{fig:The contrast between norms}b), the bounds remain unchanged hence all configurations are still nondominated, i.e., the side effect left by $\vect{E}$ will remain present. In contrast, with the new normalization in this work (Figure~\ref{fig:The contrast between norms}d), the bounds are updated locally within the population of preserved configurations and hence they can be scaled more reasonably, leading to a better mix of comparable and incomparable configurations after $\vect{E}$ is eliminated. This will strike a good balance between imposing the selection pressure toward the best target performance objective and preserving the diversity of the auxiliary performance objective.

\begin{figure}[t!]
	\centering
	\begin{subfigure}[h]{0.31\columnwidth}
		\includegraphics[width=\textwidth]{figures/details/lrzip-original.pdf}
		\subcaption{\footnotesize Original}
	\end{subfigure}
	~
	\begin{subfigure}[h]{0.31\columnwidth}
		\includegraphics[width=\textwidth]{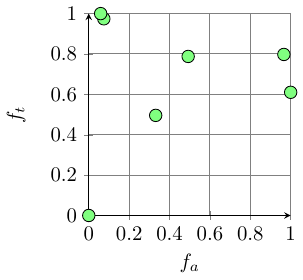}
		\subcaption{\footnotesize Normalized}
	\end{subfigure}
	~
	\begin{subfigure}[h]{0.31\columnwidth}
		\includegraphics[width=\textwidth]{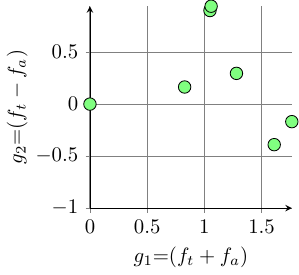}
		\subcaption{\footnotesize In the MMO space}
	\end{subfigure}

		\caption{The same population of Figure~\ref{fig:lrzip-old} under the MMO model with the new normalization method in this work. Scales in the original population are adjusted for visibility.}
	\label{fig:lrzip-new}
\end{figure}
\begin{figure}[t!]
	\centering
	\begin{subfigure}[h]{0.31\columnwidth}
		\includegraphics[width=\textwidth]{figures/details/storm-original.pdf}
		\subcaption{\footnotesize Original}
	\end{subfigure}
	~
	\begin{subfigure}[h]{0.31\columnwidth}
		\includegraphics[width=\textwidth]{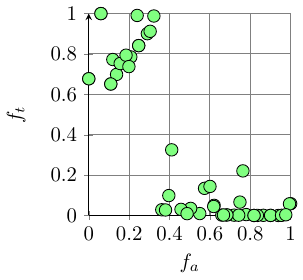}
		\subcaption{\footnotesize Normalized}
	\end{subfigure}
	~
	\begin{subfigure}[h]{0.31\columnwidth}
		\includegraphics[width=\textwidth]{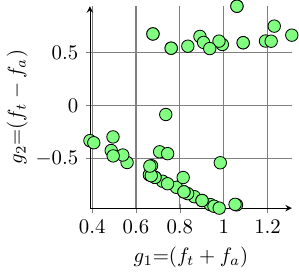}
		\subcaption{\footnotesize In the MMO space}
	\end{subfigure}

			\caption{The same population of Figure~\ref{fig:storm-old} under the MMO model with the new normalization method in this work. Scales in the original population are adjusted for visibility.}
	\label{fig:storm-new}
\end{figure}

Figures~\ref{fig:lrzip-new} and~\ref{fig:storm-new} give the results of the examples from Figures~\ref{fig:lrzip-old} and~\ref{fig:storm-old} after the new normalization is implemented, 
respectively.
As can be seen,
the configurations in the population after the normalization do not concentrate into one value on either objective (Figures~\ref{fig:storm-new}b and \ref{fig:lrzip-new}b), 
and in our MMO space there now exist both dominated and nondominated configurations in the population (Figures~\ref{fig:storm-new}c and \ref{fig:lrzip-new}c). 
In this case,
there is less need to adjust the $w$ in the MMO model to mitigate the side-effect of normalization, which balances the number of incomparable configurations after rotation, 
since the two performance objectives after the normalization are always commensurable. As such, we can generally remove the weight, i.e., setting $w=1$ for all cases.





%

\subsection{Integrating with an Optimizer}

Since MMO model is an optimization model, it can fit with different population-based multi-objective optimizers such as NSGA-II. A pseudo-code for using the MMO model with the normalization on top of NSGA-II has been demonstrated in Algorithm~\ref{alg:mmo-new}. As can be seen, there are two amendments required (the \textcolor{red!80!black}{red crossed} statements are the code for the FSE work and the \textcolor{green!50!black}{green} ones are the code changed in this work): 
\begin{enumerate}
\item Keeping track of the bounds on $f_t(\vect{x})$ and $f_a(\vect{x})$ for normalizing both the target and auxiliary performance objectives (lines 6--7 and 23--24). 
The definitions of those bounds differ depending on the normalization methods, i.e., with lines 23--24 instead of lines 21--22, the bounds are locally restricted to the current population or otherwise, they would be the global bounds so far.

\item Performing the normal Pareto search procedure in NSGA-II within the transformed meta-objective space ($g_1(\vect{x})$ and $g_2(\vect{x})$) of MMO model without a weight, 
instead of the original target-auxiliary space ($f_t(\vect{x})$ and $f_a(\vect{x})$), as shown at lines 10, 15, 28, and 30.
\end{enumerate}

Indeed, we do not need to make a significant amount of refactoring on MMO at the code level, but as we will show later on (Section~\ref{sec:result}), such a simple change can lead to dramatic improvements in its effectiveness while saving the overhead of adjusting the weight. It is worth noting that, proposing a simple method that leads to large improvements is not easy, as this requires in-depth understanding and reasoning about the principles/causes behind the observations, as what we have shown in our theoretical and empirical analysis, which requires a large amount of effort.

\section{Experimental Evaluation}
\label{sec:exp}

\begin{algorithm}[t!]
	\DontPrintSemicolon
	\footnotesize
	
	\caption{\textsc{MMOonNSGA-II}}
	\label{alg:mmo-new}
	\KwIn{Configuration space $\mathcal{V}$; the system $\mathcal{F}$; \textcolor{red!80!black}{\sout{weight $w$}}}
	
	\KwOut{${s}_{best}$ the best configuration on $f_t(\vect{x})$}
	\kwDeclare{bound vectors $\mathbf{\overline{z}_{max}}$ and $\mathbf{\overline{z}_{min}}$}
	Randomly initialize a population of $n$ configurations $\mathcal{P}$\\

	$\mathbf{\overline{z}_{max}}=\emptyset$; $\mathbf{\overline{z}_{min}}=\emptyset$\\
	\tcc{measuring $f_t$ and $f_a$ of the configurations in $\mathcal{P}$ on the system}
	\textsc{measure($\mathcal{P},\mathcal{F}$)}\\
	
	\tcc{initializing the bounds for normalization}
	{{$\mathbf{\overline{z}_{max}}=$ \textsc{updateUpperBounds($\mathcal{P}$)}}}\\
	{{$\mathbf{\overline{z}_{min}}=$ \textsc{updateLowerrBounds($\mathcal{P}$)}}}\\

	\tcc{updating $g_1$ and $g_2$}
	\textcolor{red!80!black}{\sout{\textsc{computeMMOModel($w, \mathcal{P},\mathbf{\overline{z}_{max}},\mathbf{\overline{z}_{min}}$)}}}\\
	\textcolor{green!50!black}{\textsc{computeMMOModel($\mathcal{P},\mathbf{\overline{z}_{max}},\mathbf{\overline{z}_{min}}$)}}\\
	
	
	\While{The search budget is not exhausted}
	{  
		
		$\mathcal{P'}=\emptyset$\\
		
		\While{$\mathcal{P'}<n$}
		{ 
			\tcc{selecting parents based on $g_1$ and $g_2$}
			$\{{s}_x,{s}_y\}\leftarrow$\textsc{mating($\mathcal{P}$)}
			
			$\{{o}_x,{o}_y\}\leftarrow$\textsc{doCrossoverAndMutation($\mathcal{V}, {s}_x,{s}_y$)}\\
			\tcc{measuring $f_t$ and $f_a$ for configurations ${o}_x$ and ${o}_y$ on the system (if unique)}
			\textsc{measure(${o}_x, {o}_y,\mathcal{F}$)}\\
			$\mathcal{P'}\leftarrow\mathcal{P'}\bigcup\{{o}_x,{o}_y\}$\\
		}
		\tcc{the bounds are reset based on the current population at each generation regardless of the previous bound values}
		
		
		\textcolor{red!80!black}{\sout{$\mathbf{\overline{z}_{max}}=$ \textsc{updateUpperBounds($\mathcal{P},\mathbf{\overline{z}_{max}}$)}}}\\
		\textcolor{red!80!black}{\sout{$\mathbf{\overline{z}_{min}}=$ \textsc{updateLowerrBounds($\mathcal{P},\mathbf{\overline{z}_{min}}$)}}}\\

		\textcolor{green!50!black}{{$\mathbf{\overline{z}_{max}}=$ \textsc{updateUpperBounds($\mathcal{P}$)}}}\\
		\textcolor{green!50!black}{{$\mathbf{\overline{z}_{min}}=$ \textsc{updateLowerrBounds($\mathcal{P}$)}}}\\
		
		$\mathcal{U'}\leftarrow$$\mathcal{P}\bigcup\mathcal{P'}$\\
		
		\tcc{updating $g_1$ and $g_2$}
		\textcolor{red!80!black}{\sout{\textsc{computeMMOModel($w, \mathcal{U'},\mathbf{\overline{z}_{max}},\mathbf{\overline{z}_{min}}$)}}}\\
		\textcolor{green!50!black}{\textsc{computeMMOModel($\mathcal{U'},\mathbf{\overline{z}_{max}},\mathbf{\overline{z}_{min}}$)}}\\
		\tcc{sorting based on $g_1$ and $g_2$}
		$\mathcal{U}\leftarrow$\textsc{nondominatedSorting($\mathcal{U'}$)}\\
		$\mathcal{P}\leftarrow$top $n$ configurations from $\mathcal{U}$
	}
	
	\Return ${s}_{best}\leftarrow$\textsc{bestConfiguration($\mathcal{P}$)}
	
\end{algorithm}

In this section, we articulate the experimental methodology for evaluating our MMO model with the new normalization. To better distinguish this work and the FSE work~\cite{ChenMMO21}, we use the following terminology:

\begin{itemize}
    \item \textbf{MMO-FSE:} This refers to our MMO model with the normalization method from the FSE work.
    \item \textbf{MMO:} This refers to our MMO model with the new normalization proposed in this work.
\end{itemize}

All optimization models and optimizers are implemented in Java, using jMetal~\cite{DBLP:journals/aes/DurilloN11} and Opt4J~\cite{DBLP:conf/gecco/LukasiewyczGRT11}.

\begin{table*}[t!]
\caption{Configurable software systems studied.}
\label{tb:sys}
\centering
\footnotesize
\begin{tabular}{lllllll}\toprule

\textbf{Software System}&\textbf{Domain}&\multicolumn{2}{c}{\textbf{Performance Objectives}}&\textbf{$\#$ Options}&\textbf{Search Space}&\textbf{Used By}\\

\midrule

\textsc{MariaDB}&SQL database&\textsc{O1:} latency& \textsc{O2:} CPU load&10&864&\cite{DBLP:journals/corr/abs-2106-02716}\\

\rowcolor{steel!10}\textsc{Storm/WC}&stream processing&\textsc{O1:} throughput& \textsc{O2:} latency&6&2,880&\cite{nair2018finding,DBLP:conf/mascots/JamshidiC16,DBLP:journals/corr/abs-2106-02716}\\

\textsc{VP9}&video encoding&\textsc{O1:} latency& \textsc{O2:} CPU load&12&3,008&\cite{DBLP:journals/corr/abs-2106-02716}\\

\rowcolor{steel!10}\textsc{Storm/RS}&stream processing&\textsc{O1:} throughput& \textsc{O2:} latency&6&3,839&\cite{nair2018finding,DBLP:conf/mascots/JamshidiC16,DBLP:journals/corr/abs-2106-02716}\\

\textsc{LRZIP}&file compression&\textsc{O1:} latency& \textsc{O2:} CPU load&12&5,184&\cite{DBLP:journals/corr/abs-2106-02716}\\

\rowcolor{steel!10}\textsc{MongoDB}&no-SQL database&\textsc{O1:} latency& \textsc{O2:} CPU load&15&6,840&\cite{DBLP:journals/corr/abs-2106-02716}\\

\textsc{Keras-DNN/SA}&deep learning&\textsc{O1:} AUC& \textsc{O2:} inference time&12&16,384&\cite{DBLP:conf/mascots/MendesCRG20,DBLP:conf/sigsoft/JamshidiVKS18}\\

\rowcolor{steel!10}\textsc{Keras-DNN/Adiac}&deep learning&\textsc{O1:} AUC& \textsc{O2:} inference time&12&24,576&\cite{DBLP:conf/mascots/MendesCRG20,DBLP:conf/sigsoft/JamshidiVKS18}\\

\textsc{x264}&video encoding&\textsc{O1:} PSNR& \textsc{O2:} energy usage&17&53,662&\cite{nair2018finding,DBLP:conf/icse/SiegmundKKABRS12,DBLP:journals/corr/abs-2106-02716}\\

\rowcolor{steel!10}\textsc{LLVM}&compiler&\textsc{O1:} latency& \textsc{O2:} CPU load&16&65,436&\cite{nair2018finding,DBLP:journals/corr/abs-2106-02716}\\

\textsc{Trimesh}&triangle mesh&\textsc{O1:} $\#$ iteration& \textsc{O2:} latency&13&239,260&\cite{nair2018finding,DBLP:conf/icse/SiegmundKKABRS12}\\


\bottomrule
\end{tabular}
\end{table*}

\subsection{Research Questions}
\label{sec:rq}


Our experiment answers a few research questions (RQs):

\begin{itemize}
    \item[---] \textbf{RQ1:} How effective is the MMO?
\end{itemize}

As the most fundamental question, we ask \textbf{RQ1} to verify whether our MMO can better help to mitigate the issue of local optima, i.e., by providing better results than the MMO-FSE, PMO, and state-of-the-art single-objective counterparts. However, even if the MMO can lead to promising results by mitigating local optima, it would be less useful if it requires a significantly large amount of resources to do so. Under the same settings as \textbf{RQ1}, our second research question is, therefore: 

\begin{itemize}
    \item[---] \textbf{RQ2:} How resource-efficient is the MMO?
\end{itemize}

\noindent In \textbf{RQ2}, we are interested in examining whether the MMO can utilize the resource (the number of measurements) efficiently when reaching a certain level of performance. 

One of the key novelties for MMO, compared with MMO-FSE, is weight-free. Yet, this would be meaningless if the MMO-FSE achieves similarly promising results over different weights on the systems studied; or if the effort for finding the best weight is trivial. Hence, our next \textbf{RQ} is:

\begin{itemize}
    \item[---] \textbf{RQ3:} How meaningful is the weight-free design in MMO?
\end{itemize}

\noindent \textbf{RQ3} seeks to understand two aspects: (1) how does MMO perform when compared with MMO-FSE under different weights; and (2) How much extra resource is required to tune the MMO-FSE for finding a promising weight value.

To reduce unnecessary noise, we investigate \textbf{RQ1-3} by directly measuring the systems, which belongs to the measurement-based tuning methods for software configuration tuning~\cite{DBLP:conf/sigsoft/OhBMS17,DBLP:conf/cloud/ZhuLGBMLSY17,DBLP:conf/www/XiLRXZ04}. However, 
there exist studies leveraging on the model-based tuning methods where a surrogate is built to serve as a cheap evaluator 
to predict the performance of a configuration, 
under the assumption of the single-objective model. 
Since the key difference between measurement-based and model-based tuning methods lies in whether a surrogate is used to guide the search, the MMO, which itself is an optimization model, can be considered complementary to the model-based alternative. Therefore, our final research question is concerned with:

\begin{itemize}
    \item[---] \textbf{RQ4:} Can MMO consolidate the existing model-based tuning method?
\end{itemize}

To that end, we extend \textsc{Flash}~\cite{nair2018finding} and \textsc{BOCA}~\cite{DBLP:conf/icse/0003XC021}---two recent tools from the Software Engineering community for configuration tuning---with our MMO and examine whether its performance can be improved.



\subsection{Software Systems}

To improve the generality of this work, we chose systems from existing studies according to the following criteria:

\begin{itemize}
      \item To ensure complexity, we exclude simple systems, i.e., those with less than 10 configuration options and all of them are binary.
      
      \item The system should involve at least two performance objectives.
      
    \item To expedite the experiments, the system should contain readily available data of the measurements on all the valid configurations.
    
    \item To improve the diversity of the subject, for the system under different benchmarks, we use the one with the largest search space and the one with the highest deviation on the performance, providing that the above points are satisfied.

\end{itemize}

As shown in Table~\ref{tb:sys}, we experiment on 11 real-world software systems and environments that have been commonly used in prior work~\cite{nair2018finding,DBLP:conf/mascots/JamshidiC16,DBLP:conf/mascots/MendesCRG20,DBLP:conf/sigsoft/JamshidiVKS18,DBLP:journals/corr/abs-2106-02716}. They come from diverse domains, e.g., SQL database, video encoding, and stream processing, while having different performance attributes, scale, and search space of valid configurations. Each software system has two performance objectives, which are chosen from prior work~\cite{nair2018finding,DBLP:conf/mascots/JamshidiC16,DBLP:conf/mascots/MendesCRG20,DBLP:conf/sigsoft/JamshidiVKS18,DBLP:journals/corr/abs-2106-02716}. In all experiments, we use each of their two performance attributes as the target performance objective in turn while the other serves as the auxiliary performance objective, leading to 22 cases in total. We apply the same configuration options and their ranges as studied previously since those have been shown to be the key ones for the software systems under the related environment. 



Noteworthily, it can be rather expensive even for a single measurement under those systems, e.g., it may take up to 341 seconds to measure a configuration on \textsc{MongoDB}. To ensure realism and expedite the experiments, we use the datasets of those systems collected by existing work, in which each measurement is extracted from 3-5 repeats~\cite{DBLP:conf/mascots/JamshidiC16,DBLP:journals/corr/abs-2106-02716}.




\subsection{Settings for RQ1, RQ2, and RQ3}
\label{sec:settings1}

\subsubsection{Optimizers}
\label{sec:settings1-opt}

For the single-objective optimization model, we examine four state-of-the-art optimizers that are widely used in software configuration tuning, all of which deal with local optima in different ways:

\begin{itemize}
    \item Random Search (RS) with a high neighbourhood radius to escape from the local optima, as used in~\cite{DBLP:journals/jmlr/BergstraB12,DBLP:conf/sigmetrics/YeK03,DBLP:conf/sigsoft/OhBMS17}.
    \item Stochastic Hill Climbing with restart (SHC-r), 
    which is exploited by~\cite{DBLP:conf/www/XiLRXZ04,DBLP:conf/hpdc/LiZMTZBF14}, aiming to avoid local optima by using different starting points.
    \item Single-Objective Genetic Algorithm (SOGA) from~\cite{DBLP:conf/sc/BehzadLHBPAKS13,DBLP:conf/sigsoft/ShahbazianKBM20,DBLP:conf/icac/RamirezKCM09,DBLP:conf/ssbse/SinhaCC20} that seeks to escape local optima by using variation operators.
    \item Simulated Annealing (SA) that tackles local optima by stochastically accepting inferior configurations as used in~\cite{garvin2009improved,guo2010evaluating}.  
\end{itemize}





While the MMO does not tie to any specific multi-objective optimizer, 
we use NSGA-II for the MMO, MMO-FSE, and PMO in this work, 
because (1) it has been predominately used for software configuration tuning in prior work when multiple performance attributes are of interest~\cite{Chen2018FEMOSAA,DBLP:conf/wosp/SinghBSH16,DBLP:journals/infsof/ChenLY19,DBLP:conf/icpads/KumarBCLB18,DBLP:journals/jss/SobhyMBCK20}; (2) it shares many similarities with the SOGA that we compare in this work.
However, it is worth noting that MMO may not be able to work with some multi-objective optimizers designed for SBSE problems 
where the objectives are not treated equally, 
such as~\cite{Panichella2015,Hierons2016,Hierons2020}.



\subsubsection{Weight Values for MMO-FSE}

In our experiments, we evaluate a set of weight values, i.e., $w \in \{0.01,0.1,0.3,0.5,0.7,0.9,1.0,10\}$, for the MMO-FSE. Those are merely pragmatic settings, but we found that weight beyond this range only degraded the performance, please kindly refer to Section~\ref{sec:norm-theory} for a theoretical explanation. Further, a similar setting is also what has been commonly followed for SBSE work in general~\cite{DBLP:journals/tosem/ChenL23}. In this way, we aim to examine whether the MMO-FSE can perform as well as MMO under the best weight chosen from a set of diverse weight values (or indeed worse on all of them).

\begin{table}[t!]
\caption{Measurement search budgets and population sizes.}
\vspace{-0.4cm}
\label{tb:settings}
\setlength{\tabcolsep}{0.6mm}
\centering
\begin{tabular}{ccc||ccc}
&&&&&\\
\toprule
\textbf{Software}&\textbf{Size}&\textbf{Budget}&\textbf{Software}&\textbf{Size}&\textbf{Budget}\\
\midrule

\textsc{MariaDB}&20&400&\textsc{Storm/WC}&50&600\\
\rowcolor{steel!10}\textsc{VP9}&30&700&\textsc{Storm/RS}&50&900\\
\textsc{LRZIP}&20&400&\textsc{MongoDB}&20&500\\
\rowcolor{steel!10}\textsc{Keras-DNN/SA}&20&400&\textsc{Keras-DNN/Adiac}&20&400\\
\textsc{x264}&50&2,500&\textsc{LLVM}&20&600\\
\rowcolor{steel!10}\textsc{Trimesh}&20&1,000&&&\\


\bottomrule
\end{tabular}
\end{table}





\subsubsection{Search Budget}

In this work, we use the number of measurements to quantify the search budget and resource consumed, as it is language-/platform-independent and does not suffer from the interference caused by the background processes of the operating system.

Since one of our goals is to examine how badly a model/optimizer can suffer from trapping at undesired local optima when tuning software configuration, it is important to study the result under reasonable convergence, i.e., increasing the search budget is unlikely to change the outcomes. To that end, for every optimizer/model on each system (and its performance objectives), we examine different search budgets from $\{100,200,...,X\}$ where $X$ refers to the smaller one between $3,000$ and the size of the search space. The purpose is to set a search budget as the smallest number of measurements for all optimizers/models, such that they all have less than 10\% changes of configuration in the population (or no better configuration found when no population is involved) within the last 10\% of the successive measurement count\footnote{For SOGA and NSGA-II, the population size is initially fixed to 10, which is the smallest size that we will examine subsequently.}. The settings are similar to those used by existing work which were found in a similar way, e.g., Gerasimou \textit{et al.} ~\cite{DBLP:journals/ase/GerasimouCT18} set no better configuration found for the last 20\% of the iteration count as a sign of convergence; Krall \textit{et al.}~\cite{DBLP:journals/tse/KrallMD15} use 15\% as an indicator of convergence. Note that we additionally monitor the percentage of the population change rather than purely whether a better configuration is found, since Harman~\cite{DBLP:conf/icse/Harman07} suggests that a good sign of little realistic chance of further improvements on population-based optimizers is that the population has become homogeneous. To ensure the realism of the setting, we make sure that the actual time taken for exhausting the search budget does not exceed 48 hours for a run overall. The identified search budgets are then used as the termination criterion in our experiments, as shown in Table~\ref{tb:settings}. It is worth noting that the search budget identified remains much smaller than the corresponding search space. For example, it only allows for measuring 0.42\% of the configurations for \textsc{Trimesh}.

Since each measurement has considered the noise~\cite{DBLP:conf/mascots/JamshidiC16,DBLP:journals/corr/abs-2106-02716} and only the profiling of systems is expensive in practice, in each run, we cached the measurements of distinct configurations, which can be reused directly when the same configuration appears again during the tuning. As such, only the distinct configurations would consume the budget.

To account for the stochastic nature of the optimizers, we repeat all experiments 50 runs under the search budget.

\subsubsection{Other Parameters}

For the other key parameters of the optimizers, we apply the binary tournament for mating selection, together with the boundary mutation and uniformed crossover in SOGA and NSGA-II, as used in prior work~\cite{Chen2018FEMOSAA,DBLP:conf/sigsoft/ShahbazianKBM20,DBLP:journals/infsof/ChenLY19}. The mutation and crossover rates are set to 0.1 and 0.9, respectively, as commonly set in software configuration tuning~\cite{Chen2018FEMOSAA}. 

What we could not decide easily is the population size for SOGA and NSGA-II. Therefore, for each software system, we additionally examine a set of population sizes, i.e., $\{10,20,...,100\}$, under the search budget identified previously. Similarly, we set the largest population size that can still ensure there are less than 10\% changes of configurations at the last 10\% of the measurement count. The results are shown in Table~\ref{tb:settings}. In this way, we seek to reach a good balance between convergence (smaller population change) and diversity (larger population size) under a budget.

\subsection{Settings for RQ4}
\label{sec:settings2}

\subsubsection{Optimizers}

For model-based tuning methods, we consider \textsc{Flash}\footnote{Note that when only a single performance objective matters, \textsc{Flash} uses a single-objective model like the RS, SHC-r, SOGA, and SA studied in this work. Hence, we use the single-objective version of \textsc{Flash}.} (TSE'20)~\cite{nair2018finding} and \textsc{BOCA} (ICSE'21)~\cite{DBLP:conf/icse/0003XC021} in this work, because

\begin{itemize}
    \item they are recent efforts from the software engineering community to tune software configuration. 
    
    \item they have been specifically tailored to cater for the key properties of the tuning problem, e.g., high sparsity and expensive measurements.
    
    \item their authors have shown that they outperform other more general configuration tuning approaches, e.g., \textsc{BOCA} is better than TPE~\cite{DBLP:conf/icse/0003XC021} and \textsc{Flash} is superior to $\epsilon$-PAL~\cite{nair2018finding}, as well as better than some older methods for software configuration tuning, e.g., the one by Jamshidi and Casale~\cite{nair2018finding}.
    
    \item both have been tested on some of the systems studied in this work, e.g., \textsc{x264} and \textsc{LLVM}.
\end{itemize}

\begin{algorithm}[t]
    \DontPrintSemicolon
    \footnotesize

    \caption{\textsc{Flash}}
    \label{alg:flash}
    \KwIn{Configuration space $\mathcal{V}$; the system $\mathcal{F}$}

    \KwOut{${s}_{best}$ the best configuration on $f_t(\vect{x})$}
        \kwDeclare{vector of surrogates $\mathcal{M}$ (one for each performance objective)}
    Randomly initialize a size of $k$ configurations $\mathcal{P}$\\
  \textsc{measure($\mathcal{P},\mathcal{F}$)}\\
     \tcc{removing measured configurations}
    $\mathcal{V}\leftarrow$$\mathcal{V} - \mathcal{P}$\\
    \While{The search budget is not exhausted}
    {  
    $\mathcal{M}=$ \textsc{trainCARTs($\mathcal{P}$)}\\
       \tcc{searching an estimated-best configuration for measurement}
     ${o}=$ \textsc{findBestConfiguration($\mathcal{V},\mathcal{M}$)}\\
             \textsc{measure(${o},\mathcal{F}$)}\\
    $\mathcal{V}\leftarrow$$\mathcal{V} - {o}$\\
          $\mathcal{P}\leftarrow$$\mathcal{P} + {o}$\\
          \If{${o}$ is measured to be better than ${s}_{best}$ on $f_t(\vect{x})$} 
          {
          ${s}_{best}={o}$\\
          
          }
    }

        \Return ${s}_{best}$
      
\end{algorithm}


In a nutshell, \textsc{Flash} was derived from the Sequential Model-Based Optimization (SMBO) paradigm, which is a generalization of the Bayesian Optimization (BO)~\cite{DBLP:journals/pieee/ShahriariSWAF16}. As shown in Algorithm~\ref{alg:flash}, the basic idea is to build a surrogate that learns the correlation between configurations and their values of a performance objective (line 6). Such a surrogate is then used to guide the search to decide which promising configuration to measure next via an acquisition function (line 8), after which the surrogate would be updated by using the newly measured configuration. Like other measurement-based tuning methods, the process terminates when the search budget is exhausted. Yet, unlike the classic BO, \textsc{Flash} does two major changes for the problem to tune a single performance objective:

\begin{enumerate}

\item The surrogate is a CART~\cite{DBLP:books/wa/BreimanFOS84} instead of the classic Gaussian Process in BO~\cite{DBLP:journals/pieee/ShahriariSWAF16}.
\item The acquisition function no longer considers uncertainty but solely targets the best-predicted performance value.

\end{enumerate}

Note that \textsc{Flash} originally uses an exhaustive search to find the best-predicted configuration at each sampling iteration (line 8), but this may not be ideal for our study because of two reasons: (1) exhaustively traversing the whole configuration space itself is still a lengthy process, especially on some of the large systems. For example, on each iteration for \textsc{Trimesh}, it can take several minutes on a standard machine to run even for a surrogate. (2) Since the surrogate is not always accurate~\cite{DBLP:conf/cloud/ZhuLGBMLSY17}, the exhaustive search could amplify the side effects caused by the errors in misleading the search. Therefore, we replace the exhaustive search with a random search, which works well and has been recommended as a replacement for SMBO~\cite{DBLP:journals/jmlr/BergstraB12}.

Similar to \textsc{Flash}, \textsc{BOCA} (Algorithm~\ref{alg:boca}) also leverages Bayesian Optimization but it uses Random Forest~\cite{breiman2001random} and Expected Improvement~\cite{DBLP:journals/jgo/JonesSW98} as the surrogate and acquisition function, respectively. Further, \textsc{BOCA} takes the top $K$ most important configuration options into account based on the rank from the Random Forest; it then creates a set of candidate configurations that cover $c$ settings for the unimportant options combined with every setting of the important ones, where $c$ is determined proportionally to the search progress (lines 5-9). The one with the best acquisition value from the set is then measured.

\subsubsection{Search Budget}

To ensure fairness, we set the same search budget as used in the original work of \textsc{Flash}~\cite{nair2018finding}, i.e., 50 measurements. Further, we also use the same initial sample size ($k=30$ in Algorithm~\ref{alg:flash}) to pre-train the surrogate. As for the search process with the surrogate, we allow for 1,000 surrogate evaluations (including redundant ones) which is a typical setting from the other work for optimizing the surrogate when an exhaustive search is undesirable~\cite{DBLP:conf/emo/KnowlesH05,DBLP:conf/gecco/PreussRW10}.


Similar to \textbf{RQ1-3}, each experiment is repeated 50 runs.



\begin{algorithm}[t]
    \DontPrintSemicolon
    \footnotesize

    \caption{\textsc{BOCA}}
    \label{alg:boca}
    \KwIn{Configuration space $\mathcal{V}$; the system $\mathcal{F}$}

    \KwOut{${s}_{best}$ the best configuration on $f_t(\vect{x})$}
        \kwDeclare{vector of surrogates $\mathcal{M}$}
    Randomly initialize a size of $k$ configurations $\mathcal{P}$\\
  \textsc{measure($\mathcal{P},\mathcal{F}$)}\\
    \While{The search budget is not exhausted}
    {  
    $\mathcal{M}=$ \textsc{trainRandomForest($\mathcal{P}$)}\\
    $\mathcal{I}=$\textsc{getAllSettingsOnImportantOptions($\mathcal{M}, K$)}\\
    \For{$i \in \mathcal{I}$}{
    ${c}=$\textsc{decay($j$)}\\
    $\mathcal{I}=$\textsc{getUnimportantSettings($c$)}\\
    $\mathcal{P'}\leftarrow$\textsc{combinedSamples($i,\mathcal{U}$)}
    }
    
       \tcc{searching an estimated-best configuration for measurement}
     ${o}=$ \textsc{findBestConfiguration($\mathcal{P'}$)}\\
             \textsc{measure(${o},\mathcal{F}$)}\\
          $\mathcal{P}\leftarrow$$\mathcal{P} + {o}$\\
          \If{${o}$ is measured to be better than ${s}_{best}$ on $f_t(\vect{x})$} 
          {
          ${s}_{best}={o}$\\
          
          }
    }

        \Return ${s}_{best}$
      
\end{algorithm}

\subsection{Statistical Validation}

We use the following methods for statistical test: 


\begin{itemize}


     \item[---]\textbf{Non-parametric test:} To verify statistical significance, we leverage the Wilcoxon signed-rank test~\cite{Wilcoxon1945IndividualCB} (for paired comparisons between two approaches) and Kruskal-Wallis test~\cite{mckight2010kruskal} (for multiple comparisons). In particular, to understand which pairwise comparisons are significant in the  Kruskal-Wallis test, we use Dunn's test~\cite{upton2014dictionary} as the post-hoc method together with the Holm-Bonferroni correction~\cite{abdi2010holm}, which will significantly reduce the chances of Type-I error. All of the above are widely used non-parametric test for SBSE and has been recommended in software engineering research for their strong statistical power~\cite{ArcuriB11}. The standard $a=0.05$ is set as the significance level over 50 runs. 
     
    
    
    \item[---]\textbf{Effect size:} To ensure the differences are not generated from a trivial effect, we use $\hat{A}_{12}$~\cite{Vargha2000ACA} to verify the effect size of the comparisons on target performance objectives over 50 runs. According to Vargha and Delaney~\cite{Vargha2000ACA}, when comparing our MMO and its counterpart in this work, $\hat{A}_{12}>0.5$ denotes that the MMO is better for more than 50\% of the times (MMO wins); MOO will lose if $\hat{A}_{12}<0.5$ and it is a tie when $\hat{A}_{12}=0.5$. In particular, $0.56\leq \hat{A}_{12}<0.64$ indicates a small effect size while $0.64 \leq \hat{A}_{12} < 0.71$ and $\hat{A}_{12} \geq 0.71$ mean a medium and a large effect size, respectively.

\end{itemize}

As such, we say a result of the comparison is statistically significant only if it has $\hat{A}_{12} \geq 0.56$ (or $\hat{A}_{12} \leq 0.44$) and $p < 0.05$ (after correction if needed). 


\begin{table*}[t!]
\caption{Comparing MMO with the other state-of-the-arts over 50 runs. SO$_{best}$ and MMO-FSE$_{best}$ denote the best single-objective model/optimizer and the MMO-FSE with the best weight, respectively. \quartexp{0}{20}{10}{20} shows the average and standard error (SE) on the target performance objective achieved (all objective values are normalized within $[0,1]$ and the closer to the left, the better). \bquartexp{0}{20}{10}{20} denotes the best average among others. Column ``$\hat{A}_{12}$ ($p$ value)'' shows the $\hat{A}_{12}$ and corrected $p$ value when comparing the corresponding counterpart with MMO. ``T'', ``S'', ``M'', and ``L'' denotes trivial, small, medium, and large effect size, respectively. The \setlength{\fboxsep}{1.5pt}\colorbox{steel!30}{blue cells} denote MMO wins ($\hat{A}_{12} > 0.5$) while \setlength{\fboxsep}{1.5pt}\colorbox{red!10}{red cells} mean it loses ($\hat{A}_{12} < 0.5$); otherwise it is a tie ($\hat{A}_{12} = 0.5$). Statistically significant comparisons, i.e., $\hat{A}_{12} \geq 0.56$ ($\hat{A}_{12} \leq 0.44$) and $p<0.05$ are highlighted in bold.}
    \label{tb:rq1}
    \footnotesize
  \begin{center}
    \begin{adjustbox}{max width = 1\textwidth}

    \begin{tabular}{cc@{}cc@{}cc@{}cc}
        \begin{tabular}{lcc}
            \cellcolor[gray]{1}\textbf{}  & \cellcolor[gray]{1}\textbf{Mean/SE} & \cellcolor[gray]{1}\textbf{$\hat{A}_{12}$ ($p$ value)}  \\
            \hline

SO$_{best}$&  \quart{74.95}{24.36}{87.13}{100} &\cellcolor{steel!30}\textbf{.66 M ($=$.005)}\\ 
MMO-FSE$_{best}$&  \bquart{0.00}{23.94}{11.97}{100} &\cellcolor{red!10}{.43 M ($=$.114)}\\ 
PMO& \quart{77.91}{22.09}{88.96}{100} &\cellcolor{steel!30}\textbf{.65 M ($=$.009)}\\ 
MMO&  \quart{16.89}{24.90}{29.34}{100} &-\\ 
        \end{tabular} & 
        &
       \begin{tabular}{lcccl}
                 \cellcolor[gray]{1}\textbf{}  & \cellcolor[gray]{1}\textbf{Mean/SE} & \cellcolor[gray]{1}\textbf{$\hat{A}_{12}$ ($p$ value)}  \\
            \hline
&  \quart{10.18}{25.55}{22.96}{100} &\cellcolor{steel!30}{.51 T ($=$.814)}\\ 
&  \quart{4.82}{14.81}{12.23}{100} &\cellcolor{steel!30}{.53 T ($=$1.33)}\\ 
&  \quart{69.66}{30.34}{84.83}{100} &\cellcolor{steel!30}\textbf{.71 L ($=$.001)}\\ 
&  \bquart{0.00}{13.92}{6.96}{100} &-\\ 
        \end{tabular} &
        &
       \begin{tabular}{lcccl}
                \cellcolor[gray]{1}\textbf{}  & \cellcolor[gray]{1}\textbf{Mean/SE} & \cellcolor[gray]{1}\textbf{$\hat{A}_{12}$ ($p$ value)}  \\
            \hline
&  \quart{19.53}{16.70}{27.88}{100} &\cellcolor{steel!30}{.55 T ($=$.395)}\\ 
&  \bquart{0.00}{6.18}{3.09}{100} &\cellcolor{red!10}{.46 T ($=$.339)}\\ 
&  \quart{58.75}{41.25}{79.38}{100} &\cellcolor{steel!30}\textbf{.66 M ($<$.001)}\\ 
&  \quart{6.89}{10.25}{12.01}{100} &-\\ 

        \end{tabular} &
        &
       \begin{tabular}{lcccl}
               \cellcolor[gray]{1}\textbf{}  & \cellcolor[gray]{1}\textbf{Mean/SE} & \cellcolor[gray]{1}\textbf{$\hat{A}_{12}$ ($p$ value)}  \\
            \hline
&  \bquart{0.00}{0.00}{0.00}{100} &{.50 T ($=$1.00)}\\ 
&  \quart{83.65}{16.35}{91.82}{100} &\cellcolor{steel!30}\textbf{.99 L ($<$.001)}\\ 
&  \quart{10.30}{5.35}{12.97}{100} &\cellcolor{steel!30}\textbf{.66 M ($=$.017)}\\ 
&  \bquart{0.00}{0.00}{0.00}{100} &-\\  
        \end{tabular}  \\
            (a). \textsc{MariaDB-O1}&& (b). \textsc{MariaDB-O2} && (c). \textsc{Storm/WC-O1}&& (d). \textsc{Storm/WC-O2}

        \\
        \\
           \begin{tabular}{lcccl}
          \cellcolor[gray]{1}\textbf{}  & \cellcolor[gray]{1}\textbf{Mean/SE} & \cellcolor[gray]{1}\textbf{$\hat{A}_{12}$ ($p$ value)}  \\
            \hline
   
SO$_{best}$&   \quart{36.48}{14.60}{43.77}{100} &\cellcolor{steel!30}\textbf{.64 M ($=$.020)}\\ 
MMO-FSE$_{best}$&  \bquart{0.00}{2.22}{1.11}{100} &\cellcolor{red!10}{.44 S ($=$.228)}\\ 
PMO&   \quart{86.44}{13.56}{93.22}{100} &\cellcolor{steel!30}\textbf{.89 L ($<$.001)}\\ 
MMO& \quart{10.05}{10.57}{15.33}{100} &-\\

        \end{tabular} & 
        &
       \begin{tabular}{lcccl}
                     \cellcolor[gray]{1}\textbf{}  & \cellcolor[gray]{1}\textbf{Mean/SE} & \cellcolor[gray]{1}\textbf{$\hat{A}_{12}$ ($p$ value)}  \\
            \hline
   
&  \quart{7.40}{4.64}{9.72}{100} &\cellcolor{steel!30}\textbf{.64 M ($=$.027)}\\ 
&  \bquart{0.00}{3.64}{1.82}{100} &\cellcolor{red!10}{.48 T ($=$.737)}\\ 
&  \quart{70.39}{29.61}{85.20}{100} &\cellcolor{steel!30}\textbf{.87 L ($<$.001)}\\ 
&  \quart{1.79}{4.55}{4.06}{100} &-\\ 
        \end{tabular} &
        &
       \begin{tabular}{lcccl}
            \cellcolor[gray]{1}\textbf{}  & \cellcolor[gray]{1}\textbf{Mean/SE} & \cellcolor[gray]{1}\textbf{$\hat{A}_{12}$ ($p$ value)}  \\
            \hline
&  \quart{10.72}{10.48}{15.96}{100} &\cellcolor{steel!30}{.57 M ($=$.444)}\\ 
&  \quart{3.19}{8.67}{7.52}{100} &\cellcolor{steel!30}{.52 T ($=$.687)}\\ 
&  \quart{86.34}{13.66}{93.17}{100} &\cellcolor{steel!30}\textbf{.90 L ($<$.001)}\\ 
&  \bquart{0.00}{6.87}{3.43}{100} &-\\ 
        \end{tabular} &
        &
       \begin{tabular}{lcccl}
               \cellcolor[gray]{1}\textbf{}  & \cellcolor[gray]{1}\textbf{Mean/SE} & \cellcolor[gray]{1}\textbf{$\hat{A}_{12}$ ($p$ value)}  \\
            \hline
   
&  \bquart{0.00}{0.00}{0.00}{100} &{.50 T ($=$1.00)}\\ 
&  \quart{80.17}{19.83}{90.09}{100} &\cellcolor{steel!30}\textbf{.94 L ($<$.001)}\\ 
&  \quart{1.81}{5.69}{4.66}{100} &\cellcolor{steel!30}{.53 T ($=$1.00)}\\ 
&  \bquart{0.00}{0.00}{0.00}{100} &-\\ 
   
        \end{tabular}  \\ 
                (e). \textsc{VP9-O1}&& (f). \textsc{VP9-O2} && (g). \textsc{Storm/RS-O1}&& (h). \textsc{Storm/RS-O2}
        \\
        
               \\
           \begin{tabular}{lcccl}
             \cellcolor[gray]{1}\textbf{}  & \cellcolor[gray]{1}\textbf{Mean/SE} & \cellcolor[gray]{1}\textbf{$\hat{A}_{12}$ ($p$ value)}  \\
            \hline

SO$_{best}$&  \quart{20.77}{13.56}{27.55}{100} &\cellcolor{steel!30}\textbf{.62 S ($=$.001)}\\
MMO-FSE$_{best}$&   \quart{23.95}{9.99}{28.95}{100} &\cellcolor{steel!30}\textbf{.77 L ($<$.001)}\\ 
PMO&  \quart{56.76}{43.24}{78.38}{100} &\cellcolor{steel!30}\textbf{.89 L ($<$.001)}\\ 
MMO&  \bquart{0.00}{6.51}{3.25}{100} &-\\ 
        \end{tabular} & 
        &
       \begin{tabular}{lcccl}
              \cellcolor[gray]{1}\textbf{}  & \cellcolor[gray]{1}\textbf{Mean/SE} & \cellcolor[gray]{1}\textbf{$\hat{A}_{12}$ ($p$ value)}  \\
            \hline

&  \quart{40.21}{56.42}{68.42}{100} &\cellcolor{steel!30}{.55 T ($=$.236)}\\ 
&  \quart{0.00}{43.10}{21.55}{100} &\cellcolor{steel!30}\textbf{.77 L ($<$.001)}\\ 
&  \quart{42.49}{57.51}{71.24}{100} &\cellcolor{steel!30}\textbf{.81 L ($<$.001)}\\ 
&  \bquart{2.51}{35.35}{20.19}{100} &-\\ 
        \end{tabular} &
        &
       \begin{tabular}{lcccl}
              \cellcolor[gray]{1}\textbf{}  & \cellcolor[gray]{1}\textbf{Mean/SE} & \cellcolor[gray]{1}\textbf{$\hat{A}_{12}$ ($p$ value)}  \\
            \hline
&  \quart{65.79}{34.21}{82.90}{100} &\cellcolor{steel!30}\textbf{.74 L ($<$.001)}\\ 
&  \bquart{3.60}{22.33}{14.76}{100} &\cellcolor{steel!30}{.54 T ($=$.614)}\\ 
&  \quart{33.61}{38.05}{52.64}{100} &\cellcolor{steel!30}{.59 S ($=$.253)}\\ 
&  \quart{0.00}{39.96}{19.98}{100} &-\\    
        \end{tabular} &
        &   
       \begin{tabular}{lcccl}
                \cellcolor[gray]{1}\textbf{}  & \cellcolor[gray]{1}\textbf{Mean/SE} & \cellcolor[gray]{1}\textbf{$\hat{A}_{12}$ ($p$ value)}  \\
            \hline

&  \quart{50.40}{33.46}{67.13}{100} &\cellcolor{steel!30}\textbf{.63 S ($=$.031)}\\ 
&  \quart{57.53}{33.06}{74.06}{100} &\cellcolor{steel!30}\textbf{.64 M ($=$.044)}\\ 
&  \quart{54.03}{45.97}{77.01}{100} &\cellcolor{steel!30}{.60 S ($=$.091)}\\ 
&  \bquart{0.00}{37.83}{18.92}{100} &-\\  
        \end{tabular}  \\
            (i). \textsc{LRZIP-O1}&& (j). \textsc{LRZIP-O2} && (k). \textsc{MongoDB-O1}&& (l). \textsc{MongoDB-O2}
        \\
        
               \\
           \begin{tabular}{lcccl}
               \cellcolor[gray]{1}\textbf{}  & \cellcolor[gray]{1}\textbf{Mean/SE} & \cellcolor[gray]{1}\textbf{$\hat{A}_{12}$ ($p$ value)}  \\
            \hline
  
SO$_{best}$&  \quart{62.77}{37.23}{81.39}{100} &\cellcolor{steel!30}\textbf{.63 S ($=$.023)}\\ 
MMO-FSE$_{best}$&  \quart{21.71}{35.09}{39.25}{100} &\cellcolor{steel!30}{.54 T ($=$.981)}\\ 
PMO&   \bquart{3.59}{7.91}{7.54}{100} &\cellcolor{red!10}{.48 T ($=$.690)}\\ 
MMO&  \quart{0.00}{31.37}{15.69}{100} &-\\ 
        \end{tabular} & 
        &
       \begin{tabular}{lcccl}
             \cellcolor[gray]{1}\textbf{}  & \cellcolor[gray]{1}\textbf{Mean/SE} & \cellcolor[gray]{1}\textbf{$\hat{A}_{12}$ ($p$ value)}  \\
            \hline
&  \quart{9.98}{6.90}{13.43}{100} &\cellcolor{steel!30}\textbf{.62 S ($=$.040)}\\ 
&  \quart{84.87}{15.13}{92.43}{100} &\cellcolor{steel!30}\textbf{.94 L ($<$.001)}\\ 
&  \quart{7.51}{5.67}{10.34}{100} &\cellcolor{steel!30}\textbf{.61 S ($=$.049)}\\ 
&  \bquart{0.00}{6.24}{3.12}{100} &-\\   
        \end{tabular} &
        &
       \begin{tabular}{lcccl}
            \cellcolor[gray]{1}\textbf{}  & \cellcolor[gray]{1}\textbf{Mean/SE} & \cellcolor[gray]{1}\textbf{$\hat{A}_{12}$ ($p$ value)}  \\
            \hline
&  \quart{72.46}{27.54}{86.23}{100} &\cellcolor{steel!30}\textbf{.65 M ($=$.005)}\\ 
&  \quart{40.15}{29.02}{54.66}{100} &\cellcolor{steel!30}{.54 T ($=$.698)}\\ 
&  \bquart{0.00}{8.11}{4.06}{100} &\cellcolor{steel!30}{.55 T ($=$.991)}\\ 
&  \quart{19.35}{28.34}{33.52}{100} &-\\ 
        \end{tabular} &
        &
       \begin{tabular}{lcccl}
             \cellcolor[gray]{1}\textbf{}  & \cellcolor[gray]{1}\textbf{Mean/SE} & \cellcolor[gray]{1}\textbf{$\hat{A}_{12}$ ($p$ value)}  \\
            \hline
&  \quart{2.64}{1.64}{3.46}{100} &\cellcolor{steel!30}\textbf{.64 M ($=$.008)}\\ 
&  \quart{61.75}{38.25}{80.88}{100} &\cellcolor{steel!30}\textbf{.99 L ($<$.001)}\\ 
&  \quart{6.60}{1.64}{7.42}{100} &\cellcolor{steel!30}\textbf{.81 L ($<$.001)}\\ 
&  \bquart{0.00}{0.40}{0.20}{100} &-\\ 

        \end{tabular}  \\
             (n). \textsc{Keras-DNN/SA-O1}&& (m). \textsc{Keras-DNN/SA-O2} && (o). \textsc{Keras-DNN/Adiac-O1}&& (p). \textsc{Keras-DNN/Adiac-O2}
        \\
        
               \\
           \begin{tabular}{lcccl}
               \cellcolor[gray]{1}\textbf{}  & \cellcolor[gray]{1}\textbf{Mean/SE} & \cellcolor[gray]{1}\textbf{$\hat{A}_{12}$ ($p$ value)}  \\
            \hline
  
SO$_{best}$&   \quart{52.76}{47.24}{76.38}{100} &\cellcolor{steel!30}\textbf{.69 M ($=$.001)}\\ 
MMO-FSE$_{best}$&   \bquart{0.00}{63.97}{31.98}{100} &\cellcolor{red!10}{.45 T ($=$.461)}\\ 
PMO& \quart{92.38}{6.04}{95.40}{100} &\cellcolor{steel!30}\textbf{.67 M ($=$.021)}\\ 
MMO& \quart{40.29}{45.15}{62.86}{100} &-\\
        \end{tabular} & 
        &
       \begin{tabular}{lcccl}
              \cellcolor[gray]{1}\textbf{}  & \cellcolor[gray]{1}\textbf{Mean/SE} & \cellcolor[gray]{1}\textbf{$\hat{A}_{12}$ ($p$ value)}  \\
            \hline
&  \quart{51.90}{40.27}{72.04}{100} &\cellcolor{steel!30}\textbf{.61 S ($=$.046)}\\ 
&  \bquart{0.00}{34.14}{17.07}{100} &\cellcolor{steel!30}{.52 T ($=$.688)}\\ 
&  \quart{78.46}{21.54}{89.23}{100} &\cellcolor{steel!30}\textbf{.72 L ($<$.001)}\\ 
&  \quart{27.98}{53.61}{54.79}{100} &-\\ 
        \end{tabular} &
        &
       \begin{tabular}{lcccl}
             \cellcolor[gray]{1}\textbf{}  & \cellcolor[gray]{1}\textbf{Mean/SE} & \cellcolor[gray]{1}\textbf{$\hat{A}_{12}$ ($p$ value)}  \\
            \hline
&  \bquart{0.00}{0.00}{0.00}{100} &{.50 T ($=$1.00)}\\ 
&  \bquart{0.00}{0.00}{0.00}{100} &{.50 T ($=$1.00)}\\ 
&  \quart{2.28}{97.72}{51.14}{100} &\cellcolor{steel!30}{.52 T ($=$.135)}\\ 
&  \bquart{0.00}{0.00}{0.00}{100} &-\\  
        \end{tabular} &
        &
       \begin{tabular}{lcccl}
            \cellcolor[gray]{1}\textbf{}  & \cellcolor[gray]{1}\textbf{Mean/SE} & \cellcolor[gray]{1}\textbf{$\hat{A}_{12}$ ($p$ value)}  \\
            \hline
&  \quart{30.17}{22.86}{41.60}{100} &\cellcolor{steel!30}\textbf{.62 S ($=$.043)}\\ 
&  \quart{18.04}{20.64}{28.36}{100} &\cellcolor{steel!30}{.59 S ($=$.182)}\\ 
&  \quart{83.25}{16.75}{91.63}{100} &\cellcolor{steel!30}\textbf{.78 L ($<$.001)}\\ 
&  \bquart{0.00}{18.72}{9.36}{100} &-\\   
        \end{tabular}  \\
             (q). \textsc{x264-O1}&& (r). \textsc{x264-O2} && (s). \textsc{LLVM-O1}&& (t). \textsc{LLVM-O2}
        \\
        
               \\
           \begin{tabular}{lcccl}
                 \cellcolor[gray]{1}\textbf{}  & \cellcolor[gray]{1}\textbf{Mean/SE} & \cellcolor[gray]{1}\textbf{$\hat{A}_{12}$ ($p$ value)}  \\
            \hline
     
SO$_{best}$&  \bquart{100}{0}{100}{100} &{.50 T ($=$1.00)}\\ 
MMO-FSE$_{best}$& \bquart{100}{0}{100}{100} &{.50 T ($=$1.00)}\\ 
PMO&   \bquart{100}{0}{100}{100} &{.50 T ($=$1.00)}\\ 
MMO&   \bquart{100}{0}{100}{100} &-\\          
        \end{tabular} & 
        &
       \begin{tabular}{lcccl}
                 \cellcolor[gray]{1}\textbf{}  & \cellcolor[gray]{1}\textbf{Mean/SE} & \cellcolor[gray]{1}\textbf{$\hat{A}_{12}$ ($p$ value)}  \\
            \hline
&  \quart{28.02}{17.48}{36.76}{100} &\cellcolor{steel!30}\textbf{.66 M ($=$.017)}\\ 
&  \quart{12.09}{11.78}{17.98}{100} &\cellcolor{steel!30}{.52 T ($=$.744)}\\ 
&  \quart{75.81}{24.19}{87.91}{100} &\cellcolor{steel!30}\textbf{.81 L ($<$.001)}\\ 
&  \bquart{0.00}{16.30}{8.15}{100} &-\\ 
        \end{tabular} &
        
             \begin{minipage}{0.001\textwidth}
         \begin{tabular}{p{5.05cm}ccc}
              &   \textbf{\% Win}& \textbf{\% Lose} & \textbf{\% Tie}   \\
            \hline
MMO vs. SO$_{best}$&82\%&0\%&18\%\\ 
MMO vs. MMO-FSE$_{best}$&68\%&23\%&9\%\\ 
MMO vs. PMO&90\%&5\%&5\%\\ 
&&&\\ 
        \end{tabular}    
     \end{minipage}
        
        &
  
        &
   
         &

       \\
            (u). \textsc{Trimesh-O1}&& (v). \textsc{Trimesh-O2}  & \begin{minipage}{0.001\textwidth}
                        \begin{tabular}{c}
           (w). Overall \% win/loss/tie for MMO versus the others based on $\hat{A}_{12}$
        \end{tabular} 
                 \end{minipage}&
                 &&
        \\
    \end{tabular}
  \end{adjustbox}
   \end{center}
\end{table*}

\section{Results}
\label{sec:result}

In this section, we present and discuss the experiment results. All code and data can be accessed at: \href{https://github.com/ideas-labo/mmo}{\texttt{\textcolor{blue}{https://github.com/ideas-labo/mmo}}}.

\subsection{RQ1: Effectiveness}
\label{sec:rq1}

\subsubsection{Method}

To answer \textbf{RQ1}, we compare MMO with the best state-of-the-art single-objective counterparts (as discussed in Section~\ref{sec:settings1-opt}), as well as the MMO-FSE with a best-tuned weight and PMO, over all the 22 cases of study. Since the best single-objective optimizer (denoted as SO$_{best}$) and the MMO-FSE with the best weight (denoted as MMO-FSE$_{best}$) differ across the systems/objectives, we use the following procedure to select the best representative in each case:

\begin{enumerate}

\item Run all candidates under the full-scale experiment.
\item Rank the results using Scott-Knott test~\cite{scott1974cluster} according to the target performance objective\footnote{Scott-Knott test is a widely used test in SBSE~\cite{xia2018hyperparameter} to distinguish different approaches into clusters based on an indicator (target performance objective in this work), between each of which are guaranteed to have statistically significant differences; the approaches within the same cluster are said to be statistically similar. The clusters are then ranked.}.
\item Select the one with the best rank; if there are multiple candidates under the best rank, the one with the best average (over 50 runs) on the target performance objective would be used.

\end{enumerate}


To ensure statistical significance, the statistical test\footnote{Since there are multiple comparisons, we use Kruskal-Wallis test and the corrected $p$ values (via Holm-Bonferroni correction) of Dunn’s test for all the 3 comparisons between MMO and its counterpart.} and effect size are reported for every pairwise comparison between our MMO and the other counterparts over 50 runs.

\subsubsection{Findings}

From Table~\ref{tb:rq1}, we can see that MMO performs considerably better than the best single-objective counterpart SO$_{best}$ (which can vary depending on the case), winning 18 out of 22 cases within which 14 of them show statistical significance\footnote{All Kruskal-Wallis tests show $p<0.001$ at the global level, hence the details are omitted for simplicity of exposition.} ($\hat{A}_{12} \geq 0.56$ and $p<0.05$); the remaining 4 cases are all tie and there are no cases of losses. The magnitudes of gains are also clear. The improvements over PMO are also clear: MMO wins 20 cases (15 have statistically significant differences) under mostly large magnitude of gains; there are also one tie and one loss. 

When comparing to the MMO-FSE with the best weight (MMO-FSE$_{best}$), 
MMO wins 15 out of the 22 cases with 7 of them showing statistical significance; 
loses on 5 cases with no statistically significant ones, 
together with two ties. 
This means that, 
although MMO-FSE$_{best}$ is competitive, 
MMO can still obtain further improvement in general thanks to the new normalization method. 
This is especially true in some cases, 
such as \textsc{Storm/RS}-O2, 
where the target performance objective values are much more skewed than the auxiliary ones (recall from Figure~\ref{fig:storm-old}). 
Even though MMO-FSE$_{best}$ was pre-tuned with some best weights, 
such finding is not surprising because: 
firstly, despite that the range of good weights can be reduced compared with when no normalization is used, the given set of candidate weights may not be exhaustive. 
Indeed, 
as we will show in Section~\ref{sec:rq3}, 
the weight tuning itself can be profoundly expensive, 
making exhaustive search unrealistic. 
As such, 
the chosen weight may still be far from the truly optimal weight setting. 
Secondly, 
as the population evolves, 
the objective values keep changing, 
particularly on the target performance objective.
A fixed weight typically does not stay ideal during the entire evolution process. 
For example, 
the weight may be a good fit at the beginning of the evolution 
when the population has a relatively large range of the target performance objective values,
but it may become unsuitable 
when the population converges into a tiny region with respect to the target performance objective.



\begin{figure}[t!]
\centering
\includegraphics[width=\columnwidth]{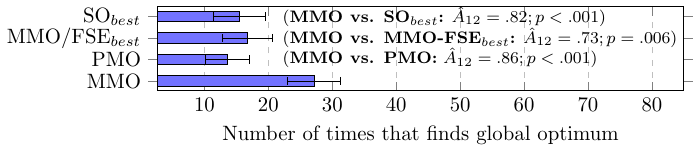}
\caption{The number of times to find the global optimum within 50 runs for all cases. The reported figures are the average and standard error across all cases. For all statistical comparisons with MMO, $\hat{A}_{12}>.50$ means MMO wins. All comparisons show large effect sizes.}
\label{fig:go}
\end{figure}

To examine whether MMO can indeed improve the chances for reaching the global optimum of the target performance objective, in Figure~\ref{fig:go}, we plot the average number of runs that each model/optimizer reaches the global optimum across all cases. We see that, as expected, MMO cannot find the global optimum for all runs under the systems studied. However, in general, it hits the global optimum more regularly than the others with statistical significance and large effect size\footnote{We use the Wilcoxon signed-rank test here since the comparisons cut across the subject systems, i.e., they are paired.}.

\begin{figure}[t!]
\centering
\begin{subfigure}[h]{0.8\columnwidth}
\includegraphics[width=\textwidth]{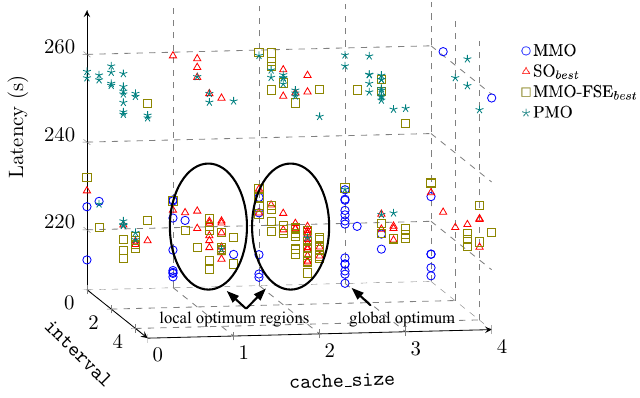}
\subcaption{\footnotesize\textsc{MongoDB-O1}}
\end{subfigure}

\begin{subfigure}[h]{0.8\columnwidth}
\includegraphics[width=\textwidth]{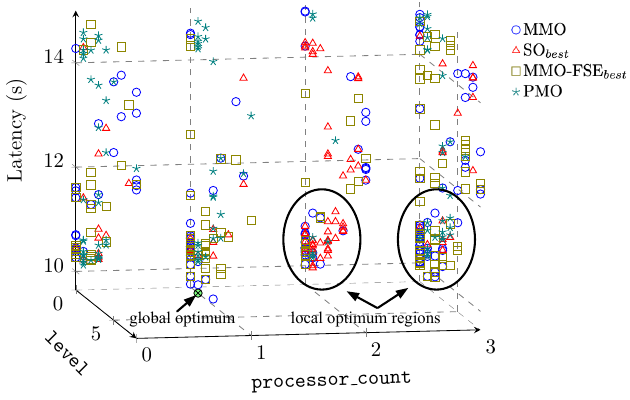}
\subcaption{\footnotesize\textsc{LRZIP-O1}}
\end{subfigure}
\caption{Projected landscapes of the explored configurations for two exampled systems. Each point is a configuration measured in the run, regardless of whether it is preserved or not. (a) represents a case where MMO finds the global optimum while the others do not; (b) showcases the scenario where none of them found the global optimum, but MMO produces results that are much closer than those of the others.}
\label{fig:details}
\end{figure}

To understand why our MMO can outperform the state-of-the-arts, we took a closer look at the configurations explored during the runs. We identified two most common patterns shown in Figure~\ref{fig:details}. As can be seen from Figure~\ref{fig:details}a, the first pattern is where MMO reaches the global optimum while the others do not; the second represents a run where the global optimum has never been found, but MMO produces a result that is much closer to it than that of the others, as shown in Figure~\ref{fig:details}b. It is worth noting that, under both patterns, there exist some large regions of local optima that cause the others to suffer more than MMO. This is evident by Figure~\ref{fig:details} where the highlighted local optima regions are mostly crowded with points explored by the other counterparts. The MMO, in contrast, escapes from these local optima by exploring an even larger area while keeping the tendency towards better target performance objective, which is precisely our \textbf{Goals 1} and \textbf{2} from Section~\ref{sec:method}.

In summary, we can answer \textbf{RQ1} as:

\begin{quotebox}
   \noindent
   \textit{The MMO is effective because we found that}
      \begin{itemize}[leftmargin=0.5cm]
   \item \textit{it provides considerably better results than the SO$_{best}$ (82\%, 18 out of 22 cases) and PMO (90\%, 20 out of 22 cases).}
   \item \textit{it also obtains relatively good improvement over MMO-FSE$_{best}$, thanks to the new normalization.}
         \end{itemize}
\end{quotebox}

\subsection{RQ2: Resource Efficiency}
\label{sec:rq2}
\subsubsection{Method}
To understand the resource efficiency of MMO in \textbf{RQ2}, for each case out of the 22, we use the following procedure:

\begin{enumerate}

\item Identify a baseline, $b$, taken as the smallest number of measurements that the best single-objective counterpart (SO$_{best}$) consumes to achieve its best result of the target performance objective, averaging over 50 runs (says $T$).

\item For each of the others, find the smallest number of measurements, denoted as $m$, at which the average result of the target performance objective (the mean over 50 runs) is equivalent to or better than $T$.

\item The speedup over SO$_{best}$, i.e., $s = {b \over m}$, is reported, according to the metric used by Gao \textit{et al.}~\cite{DBLP:conf/icse/GaoZ0LY21}. 


\end{enumerate}

\begin{figure}[t!]
	\centering
	\begin{subfigure}[h]{0.31\columnwidth}
		\includegraphics[width=\textwidth]{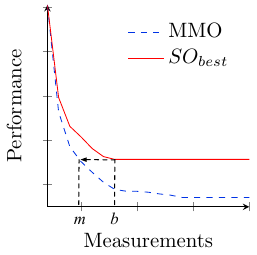}
		\subcaption{\footnotesize $s={b \over m} > 1$}
	\end{subfigure}
	~
	\begin{subfigure}[h]{0.31\columnwidth}
		\includegraphics[width=\textwidth]{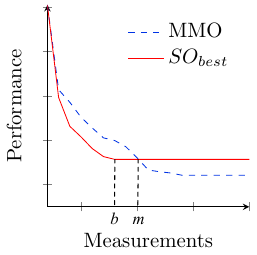}
		\subcaption{\footnotesize $s={b \over m} < 1$}
	\end{subfigure}
	~
	\begin{subfigure}[h]{0.31\columnwidth}
			\includegraphics[width=\textwidth]{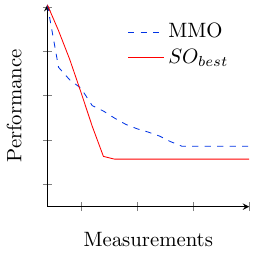}
	 	\subcaption{\footnotesize Failed}
	\end{subfigure}

	\caption{Illustrating the calculations of speedup $s$.}
	\label{fig:rq2-exp}
\end{figure}

From the example in Figure~\ref{fig:rq2-exp}, we see that:

\begin{itemize}

\item In Figure~\ref{fig:rq2-exp}a, $s>1$ means that the approach reaches the converging performance of SO$_{best}$ by using less measurements, hence is more efficient when achieving the best of what can be produced by the baseline.

\item In Figure~\ref{fig:rq2-exp}b, $s<1$ suggests that the approach reaches the converging performance of SO$_{best}$ by using more measurements, hence is less efficient even though it can lead to better results when the full budget is exhausted.

\item In Figure~\ref{fig:rq2-exp}c, we use $s<0$ to denote the case where the approach has never been able to reach the converging performance of SO$_{best}$, denoted as ``failed''.

\end{itemize}

Of course, when $s=1$, both approaches reach the same performance at exactly the same number of measurements.

Clearly, the greater the $s$, the better speedup, and hence more resources can be saved against that consumed by SO$_{best}$. In particular, 
if the MMO is resource-efficient, then we would expect at least $s = 1$ and ideally $s > 1$. Since in our context, the resource is the number of measurements, it reflects the time and computation required by a model. Again, we use the same SO$_{best}$ and MMO-FSE$_{best}$ from \textbf{RQ1}.


\begin{figure}[t!]
\centering
\includegraphics[width=\columnwidth]{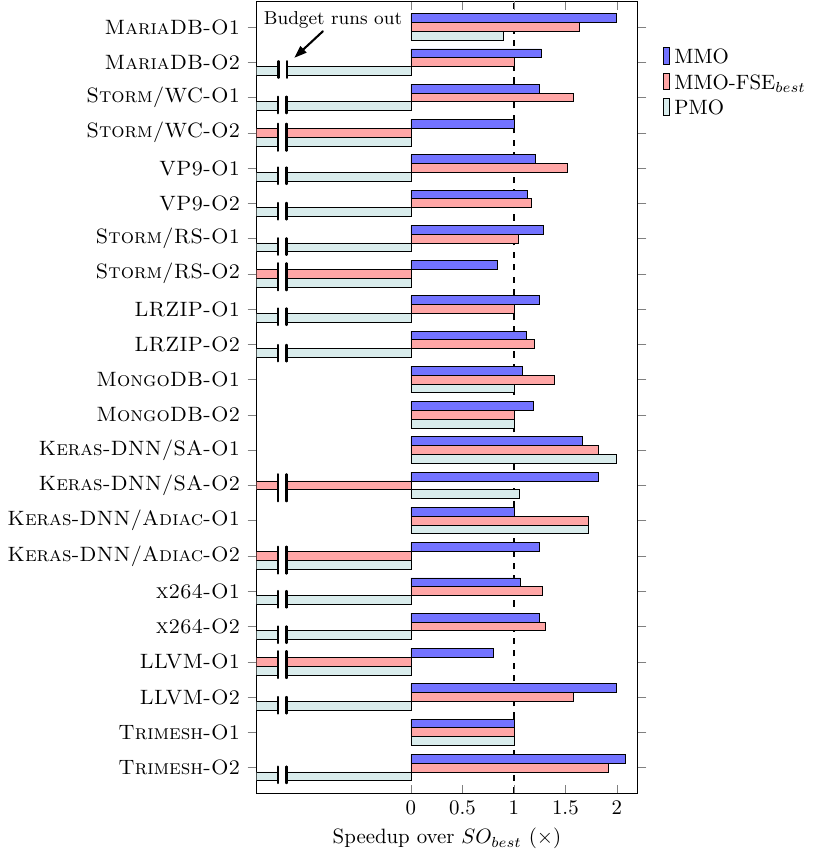}
\caption{Speedup (denoted as $s$) for MMO, MMO-FSE with the best weight (MMO$_{best}$), and the PMO for converging to the best value (the average over 50 runs) of performance objective, $T$, by the best single-objective counterpart (SO$_{best}$), using its budget consumption as the baseline (the dashed line at speedup $1\times$). $0<s<1$ indicating that the method is even slower (by using more measurements) than SO$_{best}$ to reach its best result, suggesting an inefficient utilization of resources. The broken bars mean $T$ has not been reached when the tunning terminates, i.e., $s < 0$.}
\label{fig:resource}
\end{figure}

\begin{figure*}[t!]
	\centering
	\begin{subfigure}[h]{0.2\textwidth}
		\includegraphics[width=\textwidth]{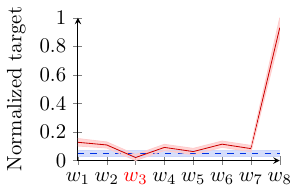}
		\subcaption{\scriptsize \textsc{MariaDB-O1}}
	\end{subfigure}
	~\hspace{-0.3cm}
	\begin{subfigure}[h]{0.2\textwidth}
		\includegraphics[width=\textwidth]{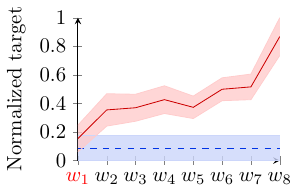}
	\subcaption{\scriptsize \textsc{MariaDB-O2}}
	\end{subfigure}
	~\hspace{-0.3cm}
	\begin{subfigure}[h]{0.2\textwidth}
		\includegraphics[width=\textwidth]{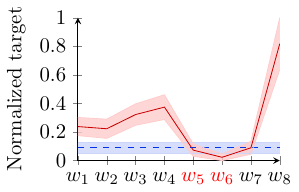}
	\subcaption{\scriptsize \textsc{Storm/WC-O1}}
	\end{subfigure}
		~\hspace{-0.3cm}
	\begin{subfigure}[h]{0.2\textwidth}
		\includegraphics[width=\textwidth]{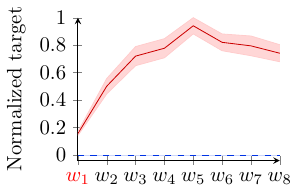}
	\subcaption{\scriptsize \textsc{Storm/WC-O2}}
	\end{subfigure}
		~\hspace{-0.3cm}
	\begin{subfigure}[h]{0.2\textwidth}
		\includegraphics[width=\textwidth]{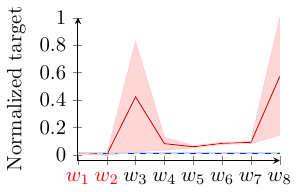}
	\subcaption{\scriptsize \textsc{VP9-O1}}
	\end{subfigure}
	
		\begin{subfigure}[h]{0.2\textwidth}
		\includegraphics[width=\textwidth]{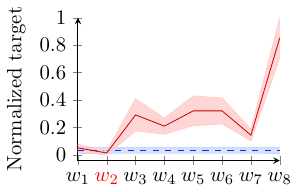}
		\subcaption{\scriptsize \textsc{VP9-O2}}
	\end{subfigure}
	~\hspace{-0.3cm}
	\begin{subfigure}[h]{0.2\textwidth}
		\includegraphics[width=\textwidth]{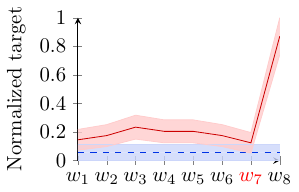}
	\subcaption{\scriptsize \textsc{Storm/RS-O1}}
	\end{subfigure}
	~\hspace{-0.3cm}
	\begin{subfigure}[h]{0.2\textwidth}
		\includegraphics[width=\textwidth]{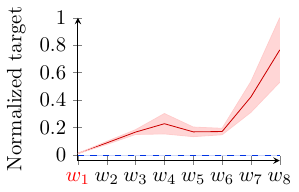}
	\subcaption{\scriptsize \textsc{Storm/RS-O2}}
	\end{subfigure}
		~\hspace{-0.3cm}
	\begin{subfigure}[h]{0.2\textwidth}
		\includegraphics[width=\textwidth]{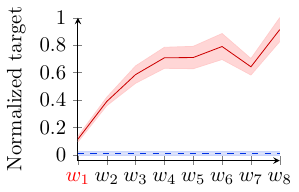}
	\subcaption{\scriptsize \textsc{Lrzip-O1}}
	\end{subfigure}
		~\hspace{-0.3cm}
	\begin{subfigure}[h]{0.2\textwidth}
		\includegraphics[width=\textwidth]{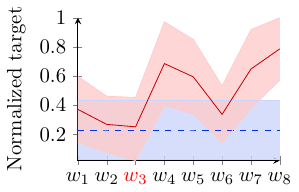}
	\subcaption{\scriptsize \textsc{Lrzip-O2}}
	\end{subfigure}

		\begin{subfigure}[h]{0.2\textwidth}
		\includegraphics[width=\textwidth]{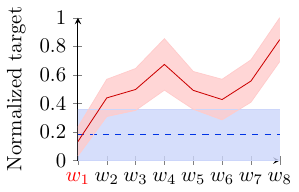}
		\subcaption{\scriptsize \textsc{MongoDB-O1}}
	\end{subfigure}
	~\hspace{-0.3cm}
	\begin{subfigure}[h]{0.2\textwidth}
		\includegraphics[width=\textwidth]{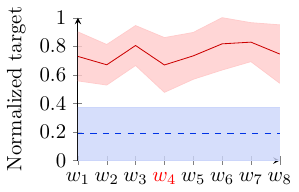}
	\subcaption{\scriptsize \textsc{MongoDB-O2}}
	\end{subfigure}
	~\hspace{-0.3cm}
	\begin{subfigure}[h]{0.2\textwidth}
		\includegraphics[width=\textwidth]{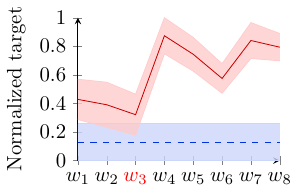}
	\subcaption{\scriptsize \textsc{Keras-DNN/SA-O1}}
	\end{subfigure}
		~\hspace{-0.3cm}
	\begin{subfigure}[h]{0.2\textwidth}
		\includegraphics[width=\textwidth]{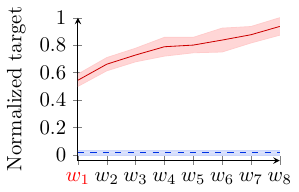}
	\subcaption{\scriptsize \textsc{Keras-DNN/SA-O2}}
	\end{subfigure}
		~\hspace{-0.3cm}
	\begin{subfigure}[h]{0.2\textwidth}
		\includegraphics[width=\textwidth]{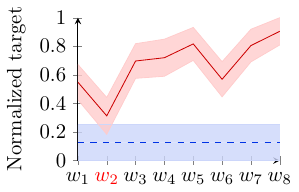}
	\subcaption{\scriptsize \textsc{Keras-DNN/Adiac-O1}}
	\end{subfigure}
	
		\begin{subfigure}[h]{0.2\textwidth}
		\includegraphics[width=\textwidth]{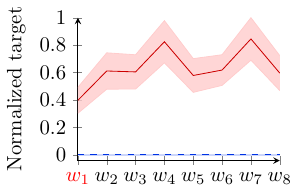}
		\subcaption{\scriptsize \textsc{Keras-DNN/Adiac-O2}}
	\end{subfigure}
	~\hspace{-0.3cm}
	\begin{subfigure}[h]{0.2\textwidth}
		\includegraphics[width=\textwidth]{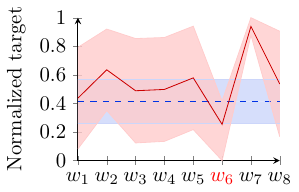}
	\subcaption{\scriptsize \textsc{x264-O1}}
	\end{subfigure}
	~\hspace{-0.3cm}
	\begin{subfigure}[h]{0.2\textwidth}
		\includegraphics[width=\textwidth]{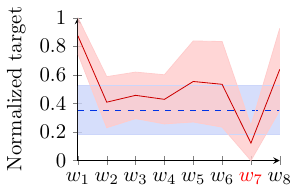}
	\subcaption{\scriptsize \textsc{x264-O2}}
	\end{subfigure}
		~\hspace{-0.3cm}
	\begin{subfigure}[h]{0.2\textwidth}
		\includegraphics[width=\textwidth]{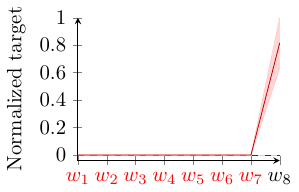}
	\subcaption{\scriptsize \textsc{LLVM-O1}}
	\end{subfigure}
		~\hspace{-0.3cm}
	\begin{subfigure}[h]{0.2\textwidth}
		\includegraphics[width=\textwidth]{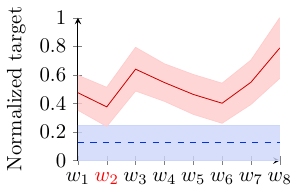}
	\subcaption{\scriptsize \textsc{LLVM-O2}}
	\end{subfigure}
	
		\begin{subfigure}[h]{0.2\textwidth}
		\includegraphics[width=\textwidth]{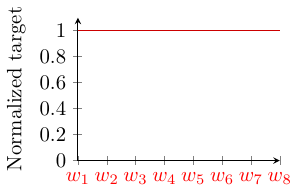}
		\subcaption{\scriptsize \textsc{Trimesh-O1}}
	\end{subfigure}
	~\hspace{-0.3cm}
	\begin{subfigure}[h]{0.2\textwidth}
		\includegraphics[width=\textwidth]{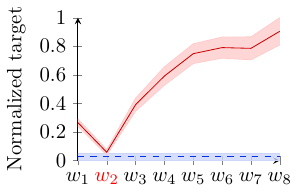}
	\subcaption{\scriptsize \textsc{Trimesh-O2}}
	\end{subfigure}
	~\hspace{-0.3cm}
		\begin{subfigure}[h]{0.2\textwidth}
		\includegraphics[width=0.8\textwidth]{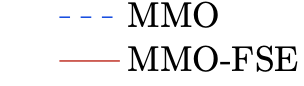}
	\end{subfigure}
		~\hspace{-0.3cm}
	\begin{subfigure}[h]{0.2\textwidth}
		\includegraphics[width=\textwidth]{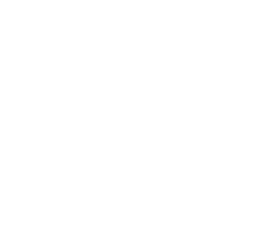}
	\end{subfigure}
		~\hspace{-0.3cm}
	\begin{subfigure}[h]{0.2\textwidth}
		\includegraphics[width=\textwidth]{figures/rq3/blank.pdf}
	\end{subfigure}
	
		\caption{Comparing MMO and MMO-FSE with different weights on the normalized target performance objective (mean and standard error) over 50 runs; the smaller, the better. $w_1, w_2, w_3, w_4, w_5, w_6, w_7, w_8$ denote weight setting of $\{0.01,0.1,0.3,0.5,0.7,0.9,1,10\}$, respectively. For each case, the promising weight(s) of MMO-FSE, i.e., the one(s) that has better (or identical) mean than MMO or that has the best mean if no weight outperforms MMO, is highlighted.}
	\label{fig:rq3}
\end{figure*}

\subsubsection{Findings}

As can be seen from Figure~\ref{fig:resource}, despite a very small number of cases 
where the MMO uses more resources to reach the performance level achieved by the best single-objective counterpart,
most commonly it uses less number of measurements than, or at least identical to, the baseline to find the same or better results, e.g., it obtains a speedup up to $2.09 \times$. In particular, the MMO achieves 17 cases of $s > 1$; 3 cases of $s = 1$; and 2 cases of $0 < s < 1$. Remarkably, there is no case where it fails to reach the performance level  (the divided bars, denoted as $s < 0$). This indicates that the MMO overcomes local optima better and more efficiently---a key attraction to software configuration tuning due to its expensive measurements. The MMO-FSE$_{best}$ does show competitive results with respect to MMO: it has 13 cases of $s > 1$ and 4 cases of $s = 1$, but there are 5 cases of $s < 0$ due to the issues discussed in Section~\ref{sec:why-new}, which could be undesirable on certain domains. Again, the above is due to MMO covering the key properties of software configuration tuning (Section~\ref{sec:method}) while making it much less sensitive to the weight parameter.

In contrast, the PMO exhibits the worst resource efficiency in terms of the speedup over $SO_{best}$, as it has 3 cases of $s > 1$, together with 3 cases of $s = 1$; 1 case of $0 < s < 1$, respectively, while the remaining 16 cases are $s < 0$. This is a clear sign that PMO is generally resource-hungry as discussed in Section~\ref{sec:method}.

\textcolor{black}{Interestingly, however, we see that PMO is considerably resource-efficient in 3 cases (\textsc{Keras-DNN/SA-O1}, \textsc{Keras-DNN/SA-O1}, and \textsc{Keras-DNN/Adaic-O1}). Despite being rare, this is indeed possible, because it can be severely affected by the relationship between the target and auxiliary performance objective. When there is a conflicting or weak relationship, then certainly optimizing the auxiliary performance objective would be harmful to the target one, as the valuable budget is wasted on something meaningfulless. However, when the relationship is harmonic, optimizing one can in fact beneficial to the other. In such a case, the drawback of PMO we discussed in Section~\ref{sec:method} would become blurred. We observed that the inference time and AUC on the three cases tend to be harmonic objectives. Note that the relationship between the two objectives need not be symmetric. For example, improving inference time can also help to significantly improve AUC, but finding configurations with better AUC may slightly improve inference time. In other words, the extent of interaction can be different. This is why PMO tends to be efficient on \textsc{Keras-DNN/Adaic-O1} but not on \textsc{Keras-DNN/Adaic-O2}.}

As the conclusion for \textbf{RQ2}, we say:

\begin{quotebox}
   \noindent
      \textit{The MMO is resource-efficient as we found that}
      \begin{itemize}[leftmargin=0.5cm]
   \item \textit{it saves generally more resources than the best single-objective counterpart to reach the same or better results (for 17 out of 22 cases with up to $2.09 \times$ speedup).}
   \item \textit{it leads to very competitive resource-saving and fewer cases of ``failed'' when compared with MMO-FSE$_{best}$.}
   \item \textit{the PMO, in contrast, is much more resource-hungry.}
         \end{itemize}
\end{quotebox}

\subsection{RQ3: Benefits over MMO-FSE}
\label{sec:rq3}

\subsubsection{Method}

In \textbf{RQ3}, we seek to verify the benefits provided by the weight-free design in MMO are indeed meaningful over MMO-FSE. Particularly, on each of the 22 systems/environments, we examine how MMO performs against the MMO-FSE under different weights using the full-scale experiment. Indeed, if the differences between MMO and MMO-FSE over different weights are small, then perhaps using the MMO-FSE can be sufficient. Our goal is to confirm if there are some promising weights that often lead to good enough results. As such, over 50 runs, we say a weight value is promising in MMO-FSE if:

\begin{itemize}
    \item It leads to a result that is generally better than (or identical to) that of MMO;
    \item or when there are no weight values in MMO-FSE can outperform MMO overall, it is the one with the best mean result. 
\end{itemize}


The other aspect we are interested in is how much extra resource would be required in order to identify at least one promising weight when using MMO-FSE. This makes sense as if the effort to find some good weights in MMO-FSE is trivial, then one would merely need to find such weight in a case-by-case manner. To investigate such, we use the following procedure in each of the 22 systems/environments:

\begin{enumerate}

\item Run MMO-FSE under all weights studied with an incremental search budget that is proportional to that of the full-scale experiment, i.e., 10\%, 20\%, $...$, 100\%. The experiment under each proportion of the budget is repeated 50 runs.

\item Find the smallest proportion of the search budget, $p$, which discovers at least one of the promising weights as that identified previously under the full-scale experiment.


\item The $p$ is then reported.

\end{enumerate}

\begin{figure}[t!]
\centering
\begin{tikzpicture}
\begin{axis}[
height=8cm, width=12cm,
grid style = {dashed},
legend pos=outer north east,
ymin=0, ymax=100,
enlarge x limits=0.05,
ybar=0pt,
bar width=9pt,
xtick={1,2,3,4,5,6,7,8,9,10,11,12,13,14,15,16,17,18,19,20,21,22},
xticklabels={\textsc{MongoDB-O1},\textsc{MongoDB-O2},\textsc{Storm/RS-O1},\textsc{Keras-DNN/SA-O1},\textsc{x264-O2},\textsc{x264-O1},\textsc{LLVM-O2},\textsc{MariasDB-O2},\textsc{Keras-DNN/Adiac-O2},\textsc{LRZIP-O2},\textsc{Storm/WC-O1},\textsc{VP9-O1},\textsc{LRZIP-O1},\textsc{MariaDB-O1},\textsc{VP9-O2},\textsc{Keras-DNN/SA-O2},\textsc{Storm/WC-O2},\textsc{Storm/RS-O2},\textsc{Keras-DNN/Adiac-O1},\textsc{LLVM-O1},\textsc{Trimesh-O1},\textsc{Trimesh-O2}},
ylabel=\% of full-scale measurements,
xticklabel style={rotate=90},
legend columns=1,
legend cell align={left},
  legend style={nodes={scale=1, transform shape},draw=none,fill=none,at={(0.7,0.95)}},
extra y ticks = 50,
        extra y tick labels={},
        extra y tick style={grid=major,major grid style={thick,draw=black}}
]

\addplot[fill=steel!55,draw=black]
coordinates { 
(1,100)
(2,100)
(3,100)
(4,80)
(5,80)
(6,70)
(7,70)
(8,60)
(9,60)
(10,60)
(11,60)
(12,60)
(13,50)
(14,30)
(15,30)
(16,20)
(17,10)
(18,10)
(19,10)
(20,10)
(21,10)
(22,10)
};


\end{axis}
\end{tikzpicture}
\caption{The necessary resource consumed $p$ (in terms of \% on the full-scale experiments' search budget) for the MMO-FSE to find the promising weight(s) as that identified under the full-scale experiments; these are the resources that would have been saved by using MMO. As a reference, the dashed line highlights the 50\% threshold of the budget. All cases are sorted in descending order.}
\label{fig:mmo-resource}
\end{figure}

\subsubsection{Findings}
Figure~\ref{fig:rq3} shows the mean on MMO and MMO-FSE under different weights; the promising ones have been highlighted. As can be seen, we observe that:

\begin{itemize}
    \item In the majority of cases, MMO can outperform MMO-FSE over all weight settings.
    \item For all the cases, the performance of MMO-FSE deviates significantly under different weights.
    \item The promising weights often achieve considerably superior results than most of the others, i.e., up to $10\times$ better.
    \item It is difficult to conclude a generally promising weight over the cases. Most commonly, the promising weight can be radically diverse, e.g., it is $w=0.01$ for \textsc{MariaDB-O2} while $w=0.9$ for \textsc{x264-O1}---a $90\times$ difference.
    
\end{itemize}


\begin{algorithm}[t]
    \DontPrintSemicolon
    \footnotesize

    \caption{\textsc{Flash}$_\textsc{MMO}$}
    \label{alg:flash-mmo}
    \KwIn{Configuration space $\mathcal{V}$; the system $\mathcal{F}$}

    \KwOut{${s}_{best}$ the best configuration on $f_t(\vect{x})$}
        \kwDeclare{vector of surrogates $\mathcal{M}$ (one for each performance objective)}
    Randomly initialize a size of $k$ configurations $\mathcal{P}$\\
  \textsc{measure($\mathcal{P},\mathcal{F}$)}\\
    $\mathcal{V}\leftarrow$$\mathcal{V} - \mathcal{P}$\\
    \While{The search budget is not exhausted}
    {  
    $\mathcal{M}=$ \textsc{trainCARTs($\mathcal{P}$)}\\
       \tcc{searching an estimated-best configuration in the transformed meta-objective space defined by the MMO (with NSGA-II), as shown in Algorithm~\ref{alg:mmo-new}}
          \textcolor{red!80!black}{\sout{${o}=$ \textsc{findBestConfiguration($\mathcal{V},\mathcal{M}$)}}}\\
     \textcolor{green!50!black}{${o}=$ \textsc{MMOonNSGA-II($\mathcal{V},\mathcal{M}$)}}\\
             \textsc{measure(${o},\mathcal{F}$)}\\
    $\mathcal{V}\leftarrow$$\mathcal{V} - {o}$\\
          $\mathcal{P}\leftarrow$$\mathcal{P} + {o}$\\
          \If{${o}$ is measured to be better than ${s}_{best}$ on $f_t(\vect{x})$} 
          {
          ${s}_{best}={o}$\\
          
          }
    }

        \Return ${s}_{best}$
      
\end{algorithm}

\begin{algorithm}[t]
    \DontPrintSemicolon
    \footnotesize

    \caption{\textsc{BOCA}$_\textsc{MMO}$}
    \label{alg:boca-mmo}
  \KwIn{Configuration space $\mathcal{V}$; the system $\mathcal{F}$}

    \KwOut{${s}_{best}$ the best configuration on $f_t(\vect{x})$}
        \kwDeclare{vector of surrogates $\mathcal{M}$}
    Randomly initialize a size of $k$ configurations $\mathcal{P}$\\
  \textsc{measure($\mathcal{P},\mathcal{F}$)}\\
    \While{The search budget is not exhausted}
    {  
    $\mathcal{M}=$ \textsc{trainRandomForest($\mathcal{P}$)}\\
    \textcolor{red!80!black}{\sout{$\mathcal{I}=$\textsc{getAllSettingsOnImportantOptions($\mathcal{M}, K$)}}}\\
    \For{\textcolor{red!80!black}{\sout{$i \in \mathcal{I}$}}}{
    \textcolor{red!80!black}{\sout{${c}=$\textsc{decay($j$)}}}\\
    \textcolor{red!80!black}{\sout{$\mathcal{I}=$\textsc{getUnimportantSettings($c$)}}}\\
    \textcolor{red!80!black}{\sout{$\mathcal{P'}\leftarrow$\textsc{combinedSamples($i,\mathcal{U}$)}}}
    }
    
          \tcc{searching an estimated-best configuration in the transformed meta-objective space defined by the MMO (with NSGA-II), as shown in Algorithm~\ref{alg:mmo-new}}
    \textcolor{red!80!black}{\sout{${o}=$ \textsc{findBestConfiguration($\mathcal{P'}$)}}}\\
     \textcolor{green!50!black}{${o}=$ \textsc{MMOonNSGA-II($\mathcal{V},\mathcal{M}$)}}\\
             \textsc{measure(${o},\mathcal{F}$)}\\
          $\mathcal{P}\leftarrow$$\mathcal{P} + {o}$\\
          \If{${o}$ is measured to be better than ${s}_{best}$ on $f_t(\vect{x})$} 
          {
          ${s}_{best}={o}$\\
          
          }
    }

        \Return ${s}_{best}$
      
\end{algorithm}

To understand how much extra resource is required to tune the weight in MMO-FSE in order to find a promising one, Figure~\ref{fig:mmo-resource} illustrates the results. Clearly, we see that for 13 out of the 22 cases, it needs 50\% or more of the full-scale search budget to identify a promising weight, which may not be acceptable when using the MMO-FSE under a case; or otherwise, the quality of tuning could be compromised. Even for the 9 cases where the extra resources required are between 10\% and 30\%, it may still be undesirable since some systems, such as \textsc{VP9}, can take up to 190 seconds to measure a single configuration.

\begin{table*}[t!]
\caption{Improvements on \textsc{Flash} and \textsc{BOCA} using MMO over 50 runs. The format is the same as Table~\ref{tb:rq1}.}
    \label{tb:rq4}
    \footnotesize
  \begin{center}
    \begin{adjustbox}{max width = 1\textwidth}

    \begin{tabular}{cc@{}cc@{}cc@{}cc}
        \begin{tabular}{lcc}
            \cellcolor[gray]{1}\textbf{}  & \cellcolor[gray]{1}\textbf{Mean/SE} & \cellcolor[gray]{1}\textbf{$\hat{A}_{12}$ ($p$ value)}  \\
            \hline
 
\textsc{Flash}&  \quart{46.94}{53.06}{73.47}{100} &\cellcolor{steel!30}\textbf{.62 S ($<$.001)}\\ 
\textsc{Flash}$_\textsc{MMO}$&  \bquart{0.00}{41.92}{20.96}{100} &-\\ 
\cdashlinelr{1-3}
\textsc{BOCA}&  \quart{42.66}{57.34}{71.33}{100} &\cellcolor{steel!30}\textbf{.59 S ($<$.001)}\\ 
\textsc{BOCA}$_\textsc{MMO}$&  \bquart{0.00}{46.30}{23.15}{100} &-\\

        \end{tabular} & 
        &
       \begin{tabular}{lcccl}
                 \cellcolor[gray]{1}\textbf{}  & \cellcolor[gray]{1}\textbf{Mean/SE} & \cellcolor[gray]{1}\textbf{$\hat{A}_{12}$ ($p$ value)}  \\
            \hline
&  \bquart{0.00}{62.86}{31.43}{100} &\cellcolor{red!10}{.48 T ($=$.101)}\\ 
&  \quart{45.68}{54.32}{72.84}{100} &-\\ 
\cdashlinelr{1-3}
&  \quart{16.90}{83.10}{58.45}{100} &\cellcolor{steel!30}{.52 T ($=$.018)}\\ 
&  \bquart{0.00}{81.17}{40.58}{100} &-\\ 
        \end{tabular} &
        &
       \begin{tabular}{lcccl}
                \cellcolor[gray]{1}\textbf{}  & \cellcolor[gray]{1}\textbf{Mean/SE} & \cellcolor[gray]{1}\textbf{$\hat{A}_{12}$ ($p$ value)}  \\
            \hline
&  \bquart{0.00}{34.91}{17.45}{100} &\cellcolor{red!10}\textbf{.33 M ($<$.001)}\\ 
&  \quart{52.36}{47.64}{76.18}{100} &-\\ 
\cdashlinelr{1-3}
&  \bquart{0.00}{99.06}{49.53}{100} &\cellcolor{red!10}{.46 T ($=$.016)}\\ 
&  \quart{5.94}{94.06}{52.97}{100} &-\\ 
        \end{tabular} &
        &
       \begin{tabular}{lcccl}
               \cellcolor[gray]{1}\textbf{}  & \cellcolor[gray]{1}\textbf{Mean/SE} & \cellcolor[gray]{1}\textbf{$\hat{A}_{12}$ ($p$ value)}  \\
            \hline
&  \quart{23.88}{76.12}{61.94}{100} &\cellcolor{steel!30}{.53 T ($<$.001)}\\ 
&  \bquart{0.00}{81.20}{40.60}{100} &-\\ 
\cdashlinelr{1-3}
&  \quart{42.03}{57.97}{71.01}{100} &\cellcolor{steel!30}\textbf{.57 S ($<$.001)}\\ 
&  \bquart{0.00}{53.97}{26.99}{100} &-\\ 
        \end{tabular}  \\
            (a). \textsc{MariaDB-O1}&& (b). \textsc{MariaDB-O2} && (c). \textsc{Storm/WC-O1}&& (d). \textsc{Storm/WC-O2}

        \\
        \\
           \begin{tabular}{lcccl}
          \cellcolor[gray]{1}\textbf{}  & \cellcolor[gray]{1}\textbf{Mean/SE} & \cellcolor[gray]{1}\textbf{$\hat{A}_{12}$ ($p$ value)}  \\
            \hline

\textsc{Flash}&  \bquart{0.00}{46.36}{23.18}{100} &\cellcolor{red!10}{.49 T ($=$.530)}\\ 
\textsc{Flash}$_\textsc{MMO}$&  \quart{26.03}{73.97}{63.02}{100} &-\\ 
\cdashlinelr{1-3}
\textsc{BOCA}&  \quart{48.34}{51.66}{74.17}{100} &\cellcolor{steel!30}\textbf{.98 L ($<$.001)}\\ 
\textsc{BOCA}$_\textsc{MMO}$&  \bquart{0.00}{4.38}{2.19}{100} &-\\ 
        \end{tabular} & 
        &
       \begin{tabular}{lcccl}
                     \cellcolor[gray]{1}\textbf{}  & \cellcolor[gray]{1}\textbf{Mean/SE} & \cellcolor[gray]{1}\textbf{$\hat{A}_{12}$ ($p$ value)}  \\
            \hline
&  \quart{87.35}{12.65}{93.68}{100} &\cellcolor{steel!30}\textbf{.91 L ($<$.001)}\\ 
&  \bquart{0.00}{17.50}{8.75}{100} &-\\ 
\cdashlinelr{1-3}
&  \quart{89.33}{10.67}{94.66}{100} &\cellcolor{steel!30}\textbf{.95 L ($<$.001)}\\ 
&  \bquart{0.00}{15.56}{7.78}{100} &-\\ 
        \end{tabular} &
        &
       \begin{tabular}{lcccl}
            \cellcolor[gray]{1}\textbf{}  & \cellcolor[gray]{1}\textbf{Mean/SE} & \cellcolor[gray]{1}\textbf{$\hat{A}_{12}$ ($p$ value)}  \\
            \hline
&  \bquart{0.00}{26.99}{13.50}{100} &\cellcolor{red!10}\textbf{.31 M ($<$.001)}\\ 
&  \quart{48.79}{51.21}{74.40}{100} &-\\ 
\cdashlinelr{1-3}
&  \quart{80.86}{19.14}{90.43}{100} &\cellcolor{steel!30}\textbf{.88 L ($<$.001)}\\ 
&  \bquart{0.00}{15.78}{7.89}{100} &-\\ 

        \end{tabular} &
        &
       \begin{tabular}{lcccl}
               \cellcolor[gray]{1}\textbf{}  & \cellcolor[gray]{1}\textbf{Mean/SE} & \cellcolor[gray]{1}\textbf{$\hat{A}_{12}$ ($p$ value)}  \\
            \hline

&  \quart{5.69}{94.31}{52.85}{100} &\cellcolor{steel!30}\textbf{.56 S ($=$.040)}\\ 
&  \bquart{0.00}{13.04}{6.52}{100} &-\\ 
\cdashlinelr{1-3}
&  \quart{22.15}{77.85}{61.07}{100} &{.50 T ($=$.015)}\\ 
&  \bquart{0.00}{7.45}{3.73}{100} &-\\

        \end{tabular}  \\ 
                (e). \textsc{VP9-O1}&& (f). \textsc{VP9-O2} && (g). \textsc{Storm/RS-O1}&& (h). \textsc{Storm/RS-O2}
        \\
        
               \\
           \begin{tabular}{lcccl}
             \cellcolor[gray]{1}\textbf{}  & \cellcolor[gray]{1}\textbf{Mean/SE} & \cellcolor[gray]{1}\textbf{$\hat{A}_{12}$ ($p$ value)}  \\
            \hline
\textsc{Flash}&  \quart{0.00}{100.00}{50.00}{100} &\cellcolor{steel!30}\textbf{.58 S ($=$.044)}\\ 
\textsc{Flash}$_\textsc{MMO}$&  \bquart{3.93}{93.02}{50.44}{100} &-\\ 
\cdashlinelr{1-3}
\textsc{BOCA}&  \bquart{0.00}{35.06}{17.53}{100} &\cellcolor{red!10}\textbf{.38 S ($<$.001)}\\ 
\textsc{BOCA}$_\textsc{MMO}$&  \quart{41.80}{58.20}{70.90}{100} &-\\ 

        \end{tabular} & 
        &
       \begin{tabular}{lcccl}
              \cellcolor[gray]{1}\textbf{}  & \cellcolor[gray]{1}\textbf{Mean/SE} & \cellcolor[gray]{1}\textbf{$\hat{A}_{12}$ ($p$ value)}  \\
            \hline
&  \quart{42.76}{57.24}{71.38}{100} &\cellcolor{steel!30}\textbf{.58 S ($=$.015)}\\ 
&  \bquart{0.00}{59.71}{29.85}{100} &-\\ 
\cdashlinelr{1-3}
&  \quart{82.83}{17.17}{91.42}{100} &\cellcolor{steel!30}\textbf{.97 L ($<$.001)}\\ 
&  \bquart{0.00}{9.87}{4.93}{100} &-\\ 

        \end{tabular} &
        &
       \begin{tabular}{lcccl}
              \cellcolor[gray]{1}\textbf{}  & \cellcolor[gray]{1}\textbf{Mean/SE} & \cellcolor[gray]{1}\textbf{$\hat{A}_{12}$ ($p$ value)}  \\
            \hline
&  \quart{65.36}{34.64}{82.68}{100} &\cellcolor{steel!30}\textbf{.67 M ($=$.008)}\\ 
&  \bquart{0.00}{44.56}{22.28}{100} &-\\  
\cdashlinelr{1-3}
&  \quart{80.48}{19.52}{90.24}{100} &\cellcolor{steel!30}\textbf{.94 L ($<$.001)}\\ 
&  \bquart{0.00}{21.43}{10.71}{100} &-\\ 
        \end{tabular} &
        &
       \begin{tabular}{lcccl}
                \cellcolor[gray]{1}\textbf{}  & \cellcolor[gray]{1}\textbf{Mean/SE} & \cellcolor[gray]{1}\textbf{$\hat{A}_{12}$ ($p$ value)}  \\
            \hline
&  \quart{58.20}{41.80}{79.10}{100} &\cellcolor{steel!30}\textbf{.73 L ($<$.001)}\\ 
&  \bquart{0.00}{40.34}{20.17}{100} &-\\ 
\cdashlinelr{1-3}
&  \quart{78.16}{21.84}{89.08}{100} &\cellcolor{steel!30}\textbf{.95 L ($<$.001)}\\ 
&  \bquart{0.00}{16.79}{8.39}{100} &-\\ 

        \end{tabular}  \\
            (i). \textsc{LRZIP-O1}&& (j). \textsc{LRZIP-O2} && (k). \textsc{MongoDB-O1}&& (l). \textsc{MongoDB-O2}
        \\
        
               \\
           \begin{tabular}{lcccl}
               \cellcolor[gray]{1}\textbf{}  & \cellcolor[gray]{1}\textbf{Mean/SE} & \cellcolor[gray]{1}\textbf{$\hat{A}_{12}$ ($p$ value)}  \\
            \hline
\textsc{Flash}&  \quart{61.70}{38.30}{80.85}{100} &\cellcolor{steel!30}\textbf{.63 S ($=$.011)}\\ 
\textsc{Flash}$_\textsc{MMO}$&  \bquart{0.00}{41.24}{20.62}{100} &-\\ 
\cdashlinelr{1-3}
\textsc{BOCA}&  \bquart{32.11}{66.76}{65.49}{100} &\cellcolor{red!10}{.48 T ($<$.001)}\\ 
\textsc{BOCA}$_\textsc{MMO}$&  \quart{0.00}{100.00}{50.00}{100} &-\\

        \end{tabular} & 
        &
       \begin{tabular}{lcccl}
             \cellcolor[gray]{1}\textbf{}  & \cellcolor[gray]{1}\textbf{Mean/SE} & \cellcolor[gray]{1}\textbf{$\hat{A}_{12}$ ($p$ value)}  \\
            \hline
&  \quart{60.75}{39.25}{80.38}{100} &\cellcolor{steel!30}\textbf{.78 L ($<$.001)}\\ 
&  \bquart{0.00}{3.55}{1.77}{100} &-\\ 
\cdashlinelr{1-3}
&  \bquart{0.00}{8.07}{4.04}{100} &\cellcolor{red!10}\textbf{.03 L ($<$.001)}\\ 
&  \quart{73.24}{26.76}{86.62}{100} &-\\ 
        \end{tabular} &
        &
       \begin{tabular}{lcccl}
            \cellcolor[gray]{1}\textbf{}  & \cellcolor[gray]{1}\textbf{Mean/SE} & \cellcolor[gray]{1}\textbf{$\hat{A}_{12}$ ($p$ value)}  \\
            \hline
&  \quart{88.76}{11.24}{94.38}{100} &\cellcolor{steel!30}\textbf{.61 S ($=$.005)}\\ 
&  \bquart{0.00}{48.84}{24.42}{100} &-\\ 
\cdashlinelr{1-3}
&  \bquart{0.00}{13.23}{6.61}{100} &\cellcolor{red!10}\textbf{.05 L ($<$.001)}\\ 
&  \quart{88.32}{11.68}{94.16}{100} &-\\ 

        \end{tabular} &
        &
       \begin{tabular}{lcccl}
             \cellcolor[gray]{1}\textbf{}  & \cellcolor[gray]{1}\textbf{Mean/SE} & \cellcolor[gray]{1}\textbf{$\hat{A}_{12}$ ($p$ value)}  \\
            \hline
&  \quart{25.02}{62.67}{56.35}{100} &\cellcolor{steel!30}\textbf{.67 M ($=$.002)}\\ 
&  \bquart{0.00}{100.00}{50.00}{100} &-\\ 
\cdashlinelr{1-3}
&  \bquart{0.00}{5.62}{2.81}{100} &\cellcolor{red!10}\textbf{.28 L ($<$.001)}\\ 
&  \quart{31.63}{68.37}{65.81}{100} &-\\ 

        \end{tabular}  \\
             (n). \textsc{Keras-DNN/SA-O1}&& (m). \textsc{Keras-DNN/SA-O2} && (o). \textsc{Keras-DNN/Adiac-O1}&& (p). \textsc{Keras-DNN/Adiac-O2}
        \\
        
               \\
           \begin{tabular}{lcccl}
               \cellcolor[gray]{1}\textbf{}  & \cellcolor[gray]{1}\textbf{Mean/SE} & \cellcolor[gray]{1}\textbf{$\hat{A}_{12}$ ($p$ value)}  \\
            \hline

\textsc{Flash}&  \bquart{0.00}{48.62}{24.31}{100} &\cellcolor{red!10}\textbf{.26 L ($<$.001)}\\ 
\textsc{Flash}$_\textsc{MMO}$&  \quart{55.12}{44.88}{77.56}{100} &-\\ 
\cdashlinelr{1-3}
\textsc{BOCA}&  \quart{27.69}{72.31}{63.85}{100} &\cellcolor{steel!30}\textbf{.73 L ($<$.001)}\\ 
\textsc{BOCA}$_\textsc{MMO}$&  \bquart{0.00}{65.64}{32.82}{100} &-\\ 
        \end{tabular} & 
        &
       \begin{tabular}{lcccl}
              \cellcolor[gray]{1}\textbf{}  & \cellcolor[gray]{1}\textbf{Mean/SE} & \cellcolor[gray]{1}\textbf{$\hat{A}_{12}$ ($p$ value)}  \\
            \hline
&  \bquart{0.00}{58.25}{29.12}{100} &\cellcolor{red!10}\textbf{.36 M ($<$.001)}\\ 
&  \quart{36.74}{63.26}{68.37}{100} &-\\
\cdashlinelr{1-3}
&  \quart{8.08}{91.92}{54.04}{100} &\cellcolor{steel!30}\textbf{.59 S ($<$.001)}\\ 
&  \bquart{0.00}{88.54}{44.27}{100} &-\\ 
        \end{tabular} &
        &
       \begin{tabular}{lcccl}
             \cellcolor[gray]{1}\textbf{}  & \cellcolor[gray]{1}\textbf{Mean/SE} & \cellcolor[gray]{1}\textbf{$\hat{A}_{12}$ ($p$ value)}  \\
            \hline
&  \quart{62.77}{37.23}{81.38}{100} &\cellcolor{steel!30}\textbf{.78 L ($<$.001)}\\ 
&  \bquart{0.00}{25.54}{12.77}{100} &-\\ 
\cdashlinelr{1-3}
&  \quart{69.73}{30.27}{84.87}{100} &\cellcolor{steel!30}\textbf{.75 L ($<$.001)}\\ 
&  \bquart{0.00}{23.48}{11.74}{100} &-\\ 
        \end{tabular} &
        &
       \begin{tabular}{lcccl}
            \cellcolor[gray]{1}\textbf{}  & \cellcolor[gray]{1}\textbf{Mean/SE} & \cellcolor[gray]{1}\textbf{$\hat{A}_{12}$ ($p$ value)}  \\
            \hline
&  \bquart{0.00}{91.53}{45.77}{100} &\cellcolor{red!10}{.47 T ($<$.001)}\\ 
&  \quart{5.80}{94.20}{52.90}{100} &-\\ 
\cdashlinelr{1-3}
&  \quart{12.12}{87.88}{56.06}{100} &\cellcolor{steel!30}{.55 T ($<$.001)}\\ 
&  \bquart{0.00}{85.10}{42.55}{100} &-\\ 
        \end{tabular}  \\
             (q). \textsc{x264-O1}&& (r). \textsc{x264-O2} && (s). \textsc{LLVM-O1}&& (t). \textsc{LLVM-O2}
        \\
        
               \\
           \begin{tabular}{lcccl}
                 \cellcolor[gray]{1}\textbf{}  & \cellcolor[gray]{1}\textbf{Mean/SE} & \cellcolor[gray]{1}\textbf{$\hat{A}_{12}$ ($p$ value)}  \\
            \hline

\textsc{Flash}&  \quart{18.26}{81.74}{59.13}{100} &\cellcolor{steel!30}{.51 T ($<$.001)}\\ 
\textsc{Flash}$_\textsc{MMO}$&  \bquart{0.00}{71.91}{35.95}{100} &-\\ 
\cdashlinelr{1-3}
\textsc{BOCA}&  \quart{60.84}{39.16}{80.42}{100} &\cellcolor{steel!30}\textbf{.63 S ($<$.001)}\\ 
\textsc{BOCA}$_\textsc{MMO}$&  \bquart{0.00}{0.00}{0.00}{100} &-\\ 
        \end{tabular} & 
        &
       \begin{tabular}{lcccl}
                 \cellcolor[gray]{1}\textbf{}  & \cellcolor[gray]{1}\textbf{Mean/SE} & \cellcolor[gray]{1}\textbf{$\hat{A}_{12}$ ($p$ value)}  \\
            \hline

&  \quart{40.27}{59.73}{70.13}{100} &\cellcolor{steel!30}{.54 T ($=$.103)}\\ 
&  \bquart{0.00}{74.02}{37.01}{100} &-\\ 
\cdashlinelr{1-3}
&  \quart{52.88}{47.12}{76.44}{100} &\cellcolor{steel!30}\textbf{.67 M ($<$.001)}\\ 
&  \bquart{0.00}{46.53}{23.26}{100} &-\\ 
        \end{tabular} &
        
           \begin{minipage}{0.001\textwidth}
         \begin{tabular}{p{5.05cm}ccc}
              &   \textbf{\% Win}& \textbf{\% Lose} & \textbf{\% Tie}   \\
            \hline
\textsc{Flash}$_\textsc{MMO}$  vs. \textsc{Flash}&68\%&32\%&0\%\\ 
&&&\\ 
\textsc{BOCA}$_\textsc{MMO}$  vs. \textsc{BOCA}&68\%&27\%&5\%\\ 
&&&\\ 
        \end{tabular}    
     \end{minipage}

        &
     
        &
       \\
            (u). \textsc{Trimesh-O1}&& (v). \textsc{Trimesh-O2}  & \begin{minipage}{0.001\textwidth}
                        \begin{tabular}{c}
           (w). Overall \% win/loss/tie for using MMO in \textsc{Flash}/\textsc{BOCA} based on $\hat{A}_{12}$
        \end{tabular} 
                 \end{minipage}&&&
        \\
    \end{tabular}
  \end{adjustbox}
   \end{center}
\end{table*}

Therefore, for \textbf{RQ3}, we say:

\begin{quotebox}
   \noindent
   \textit{Thanks to the new normalization, the weight-free feature is meaningful in MMO as we found that}
   \begin{itemize}[leftmargin=0.5cm]

\item   \textit{the MMO-FSE can be highly sensitive to the weight setting---there exist some rather diverse, case-dependent, and promising weights that perform significantly better than the others, which can only be discovered via pre-tuning.}
\item   \textit{the effort required by the MMO-FSE to identify a promising weight is non-trivial, i.e., it takes 50\% or more of the full-scale search budget for 13 out of 22 cases, which would otherwise be saved by using MMO instead without compromising the quality.}

\end{itemize}

\end{quotebox}

\subsection{RQ4: Consolidating Model-based Tuning}

\subsubsection{Method}

For \textbf{RQ4}, we extended \textsc{Flash} and \textsc{BOCA} with our MMO, denoted as \textsc{Flash}$_\textsc{MMO}$ and \textsc{BOCA}$_\textsc{MMO}$, respectively. As shown in Algorithm~\ref{alg:flash-mmo} and~\ref{alg:boca-mmo}, the change is highlighted in colors, from which we see that the amendment is merely a single line of code which changes the original search strategy that solely optimizes the target performance objective to searching over the space of MMO (working with NSGA-II). In this way, the search is conducted in the transformed meta-objective space of the surrogate-predicted objectives, in which the system $\mathcal{F}$ is replaced by the surrogates $\mathcal{M}$. For all optimizers, we allow 1,000 evaluations, including redundant ones, on the surrogate (50 population size and 20 generations in \textsc{Flash}$_\textsc{MMO}$ and \textsc{BOCA}$_\textsc{MMO}$) as from existing work~\cite{DBLP:journals/asc/Bingul07,DBLP:conf/emo/KnowlesH05,DBLP:conf/gecco/PreussRW10}.

Similar to the previous sections, the statistical test\footnote{We use the Wilcoxon signed-rank test here since the comparisons are paired, i.e., on each run, all optimizers use the same set of randomly sampled training data for building the surrogate.} and effect size are reported for every pairwise comparison between the \textsc{Flash}$_\textsc{MMO}$ and \textsc{Flash} (\textsc{BOCA}$_\textsc{MMO}$ and \textsc{BOCA}) over 50 runs.

\subsubsection{Findings}

From Table~\ref{tb:rq4}, it is clear that \textsc{Flash}$_\textsc{MMO}$ obtains better results than \textsc{Flash} in general: it wins 15 out of 22 cases while loses 7 others. In particular, in those cases where \textsc{Flash}$_\textsc{MMO}$ wins, 12 of them are statistically significant. In contrast, only 4 of those that it loses have $p<0.05$ and non-trivial effect size. The relative magnitude of gains has also been significant, e.g., for \textsc{VP9}-O2 and \textsc{Storm/RS}-O2. Similar results have also been registered for comparing \textsc{BOCA}$_\textsc{MMO}$ and \textsc{BOCA}. This means that, even when searching within the surrogate-predicted space, our MMO can bring considerable improvement on model-based tuning methods like \textsc{Flash} and \textsc{BOCA}. 

\begin{figure}[t!]
\centering

	\begin{subfigure}[h]{0.5\columnwidth}
		\includegraphics[width=\textwidth]{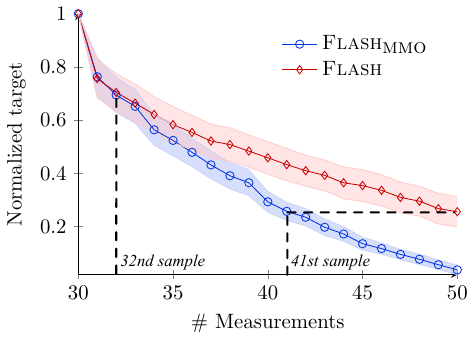}
		\subcaption{on \textsc{Flash}}
	\end{subfigure}
	~\hspace{-0.3cm}
	\begin{subfigure}[h]{0.5\columnwidth}
		\includegraphics[width=\textwidth]{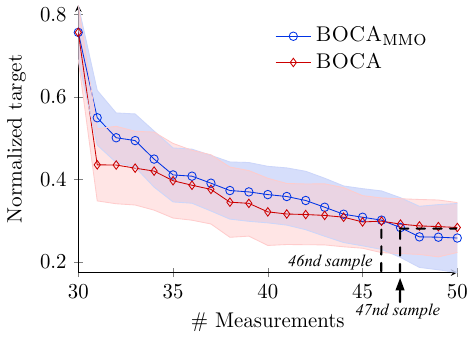}
	\subcaption{on \textsc{BOCA}}
	\end{subfigure}
\caption{The search trajectories on the model-based tuning methods extended by MMO and the original ones. Each point shows the average and the corresponding standard error over all cases and 50 runs. The target performance objectives are normalized and converted as minimizing objectives, if needed; hence the smaller, the better.}
\label{fig:flash-resource}
\end{figure}

To take a closer investigation, Figure~\ref{fig:flash-resource} shows the overall search trajectories for the 20 measurements that are actually spent on tuning. Clearly, \textsc{Flash}$_\textsc{MMO}$ produces a trajectory with a steeper slope than that of \textsc{Flash} over all cases and runs. The standard error of the average performance is also smaller, implying that MMO can also consolidate the stability of outcomes. Of particular interesting points are at the 32\textit{th} and 41\textit{st} measurement: the former means that \textsc{Flash}$_\textsc{MMO}$ improves the results at as little as the 2\textit{nd} measurement into the tuning (as the first 30 are for pre-training the surrogate); while the latter reflects that the MMO helps to improve resource efficiency, achieving a ${50\over41}= 1.22 \times$ speedup over \textsc{Flash} when reaching its best outcome under the search budget. When comparing \textsc{BOCA}$_\textsc{MMO}$ with \textsc{BOCA}, the improvement on efficiency is less obvious since \textsc{BOCA} leverages the information of important options. However, we see that at the 46\textit{th} measurement, \textsc{BOCA}$_\textsc{MMO}$ starts to become superior to its counterpart while achieving the best of \textsc{BOCA} at the 47\textit{th} measurement, enabling a ${50\over47}= 1.06 \times$ speedup.

\textcolor{black}{Although the use of a surrogate can transform the original configuration landscape into a different one according to the estimated value, the issue of local optima remains present~\cite{DBLP:conf/sigsoft/JChen2023}, providing that the accuracy is sufficient. Therefore, the reason that improves \textsc{BOCA} can be attributed to the fact that MMO relieves the issue of local optima trap, as the search strategy in \textsc{BOCA} is restricted to a certain region of the search space with respect to the important options. As for \textsc{Flash}, which is resilient to local optima due to the random search nature, the improvement is the result of MMO being able to preserve the tendency towards the best of target performance objectives, providing better guidance in the tuning.}


Overall, for \textbf{RQ4}, we have:

\begin{quotebox}
   \noindent
   \textit{The MMO can significantly consolidate model-based tuning methods because}
   \begin{itemize}[leftmargin=0.5cm]

\item   \textit{it improves the results on both \textsc{Flash} and \textsc{BOCA} for 15 out of 22 cases (68\%), within which 12 and 13 cases are statistically significant, respectively.}
\item   \textit{it enables a $1.22 \times$ and $1.06 \times$ speedup over \textsc{Flash} and \textsc{BOCA}, respectively, with gradually more stable outcomes across all the cases overall.}

\end{itemize}

\end{quotebox}

\section{Using MMO in Practice}
\label{sec:practice}

Using MMO in practical scenarios for any new system is straightforward. In what follows, we describe the basic steps for the application of MMO in practice:

\begin{enumerate}

\item Build the benchmark under which the configurable software system needs to be tuned. This is often selected from some well-known ones (e.g., \textsc{WordCount} for \textsc{Storm}) or emulated depending on the software engineers' understanding of the software's running environment.  

\item Identify the key configuration options and their possible values. In practice, one can either derive these from previous studies (as we have done in this work) or based on experience and domain understanding.

\item Confirm the target performance objective and an arbitrarily chosen auxiliary performance objective, as well as how they are measured. It is important to ensure that both objectives can be influenced by different configurations. For example, when \textit{runtime} is an important target for a system, then the auxiliary performance objective may be chosen from \textit{CPU load}, \textit{memory consumption}, and \textit{energy usage}, etc.

\item Define a tuning budget in terms of the number of measurements. The parameters of the underlying NSGA-II of MMO can be chosen from some widely used ones. At this point, one can also decide on whether to use MMO as a measurement-based optimizer or a model-based one (e.g., by pairing MMO with optimizers like \textsc{Flash} or \textsc{BOCA}). In general, the model-based version is recommended if the budget is small.

\item Run MMO on the system. 


\end{enumerate}

\section{Threats to Validity}
\label{sec:discussion}

Threats to \textbf{internal validity} can be related to the search budget. To tackle this, we have set a budget that reaches a reasonable convergence for all the optimizers compared---a typical setup when the study aims to examine the effectiveness of mitigating local optima~\cite{DBLP:conf/icst/PanichellaKT15}. The other parameter settings follow what has been pragmatically used from the literature~\cite{Chen2018FEMOSAA,DBLP:conf/sigsoft/ShahbazianKBM20,DBLP:journals/infsof/ChenLY19,DBLP:journals/asc/Bingul07,DBLP:conf/emo/KnowlesH05,DBLP:conf/gecco/PreussRW10}. or tuned through preliminary runs. However, we acknowledge that examining alternative parameters can be an interesting topic and we leave this for future work. To mitigate bias, we repeated 50 experiment runs under each case.

The metrics and evaluation used may pose threats to \textbf{construct validity}. Since there is only a single performance concern, 
there is no need to consider metrics with respect to multi-objective optimization~\cite{Li2020}.
We conduct the comparison based on the target performance objective and to the resource (number of measurements) required to converge to the same result. 
Both of these are common metrics in software configuration tuning~\cite{nair2018finding,DBLP:conf/icse/GaoZ0LY21}. To verify statistical significance and effect size, we use the Wilcoxon test (including both the paired and non-paired versions depending on the research questions) and $\hat{A}_{12}$ to examine the results. While the MMO model is optimizer-agnostic, we examine mainly on NSGA-II in this work; using alternative multi-objective optimizers is unlikely to invalidate our conclusion but we admit the usefulness of evaluating over a wide range of optimizers with MMO, which can be part of the future work.

Threats to \textbf{external validity} can be raised from the subjects studied. We mitigated this by using 11 systems/environments that are of different domains, scales, and performance attributes, as used by prior work~\cite{nair2018finding,DBLP:conf/mascots/JamshidiC16,DBLP:conf/mascots/MendesCRG20,DBLP:conf/sigsoft/JamshidiVKS18,DBLP:journals/corr/abs-2106-02716}. We also compared the proposed MMO (under the new normalization) with four state-of-the-art single-objective counterparts for software configuration tuning, PMO model, and the MMO-FSE. Further, we examine how it can help to consolidate \textsc{Flash}, a recent model-based tuning method from the software engineering community. Nonetheless, we agree that studying additional systems and optimizers may prove fruitful.

%

\section{Related Work}
\label{sec:related}


Broadly, optimizers for software configuration tuning can be either \textit{measurement-based} or \textit{model-based}.



\subsection{Measurement-based Tuning}

In measurement-based tuning methods, the optimizer is guided by directly measuring the configuration of the software systems. Despite the expensiveness, the measurements can accurately reflect the good or bad of a configuration (and the extents thereof). A wide range of optimizers have been studied, such as random search~\cite{DBLP:journals/jmlr/BergstraB12,DBLP:conf/sigmetrics/YeK03,DBLP:conf/sigsoft/OhBMS17,DBLP:conf/icac/RamirezKCM09}, hill climbing~\cite{DBLP:conf/www/XiLRXZ04,DBLP:conf/hpdc/LiZMTZBF14,DBLP:conf/icpads/DingLQ15}, single-objective genetic algorithm~\cite{DBLP:conf/sc/BehzadLHBPAKS13,DBLP:conf/sigsoft/ShahbazianKBM20,DBLP:conf/ssbse/SinhaCC20} and simulated annealing~\cite{garvin2009improved,guo2010evaluating}, to name a few. 

Under such a single-objective model, a key difference for those optimizers lies in the tricks that attempt to overcome the issues of local optima. For example, some extend the random search to consider a wider neighboring radius of the configuration structure, hence it is more likely to jump out from the local optima~\cite{DBLP:conf/sigsoft/OhBMS17}. Others rely on restarting from a different point, such as in restarted hill climbing, hence increasing the chance to find the ``right'' path from local optima to the global optimum~\cite{DBLP:conf/cloud/ZhuLGBMLSY17,DBLP:conf/www/XiLRXZ04}. More recently, Krishna \textit{et al.}~\cite{9134972} has relied on probabilistically accepting worse configurations to jump out of local optima---a typical feature of the simulated annealing~\cite{DBLP:conf/icpads/DingLQ15,guo2010evaluating}.

Our MMO differs from all the above as it lies in a higher level of abstraction---the optimization model---as opposed to the level of optimization method. In particular, with the new normalization, the purposely-crafted Pareto relation in MMO has been shown to be able to better overcome the local optima for software configuration tuning. 


\subsection{Model-based Tuning}

Instead of solely using the measurements of software systems, the model-based tuning methods apply a surrogate (analytical~\cite{DBLP:conf/icse/Kumar0BB20,DBLP:conf/gecco/0001LY18,DBLP:journals/infsof/ChenLY19} or machine learning based~\cite{nair2018finding,DBLP:conf/mascots/JamshidiC16,DBLP:conf/msr/GongC22,DBLP:conf/fse/GongC24}) to evaluate configurations, which guides the search in an optimizer. The intention is to speed up the exploration of configurations as the model evaluation is much cheaper. Yet, it has been shown that the model accuracy and the availability of initial data can become an issue~\cite{DBLP:conf/cloud/ZhuLGBMLSY17}.

Studies on model-based tuning for software systems differ mainly on the way of building surrogates and the choice of acquisition function. Among others, Jamshidi and Casale~\cite{DBLP:conf/mascots/JamshidiC16} use Bayesian optimization to tune software configuration, wherein the search is guided by the Gaussian Process based surrogate trained from the data collected. Nair \textit{et al.}~\cite{nair2018finding} follow a similar idea to propose \textsc{Flash}, but the CART is used instead as the surrogate. More recently, Chen \textit{et al.}~\cite{DBLP:conf/icse/0003XC021} also follow BO and CART to propose \textsc{BOCA}, but they additionally identify the ``important configuration options'' from the Random Forest model. Such information would then inform the optimization of the acquisition function in determining what to sample next.

Since MMO lies in the level of optimization model, it is complementary to the model-based methods in which the MMO would take the surrogate values as inputs instead of the real measurements. This, as we have shown using \textsc{Flash} as a case, can better consolidate the tuning results.


\subsection{General Parameter Tuning}


Optimizers proposed for the parameter tuning of general algorithms can also be relevant~\cite{DBLP:journals/ec/BlotMJH19,DBLP:journals/tec/HuangLY20,DBLP:series/sci/Bezerra0S20,DBLP:conf/ppsn/PushakH18}, including IRace~\cite{lopez2016irace}, ParamILS~\cite{DBLP:journals/jair/HutterHLS09}, SMAC~\cite{DBLP:conf/lion/HutterHL11}, GGA$++$~\cite{DBLP:conf/ijcai/AnsoteguiMSST15}, as well as their multi-objective variants, such as MO-ParamILS~\cite{DBLP:conf/lion/BlotHJKT16} and SPRINT-Race~\cite{DBLP:conf/gecco/ZhangGA15}. To examine a few examples, ParamILS~\cite{DBLP:journals/jair/HutterHLS09} relies on iterative local search---a search procedure that may jump out of local optima using strategies similar to that of SA and SHC-r. Further, a key contribution is the capping strategy, which helps to reduce the need to measure an algorithm under some problem instances, hence saving computational resources. This is one of the goals that we seek to achieve too. Similar to Nair \textit{et al.}~\cite{nair2018finding}, SMAC~\cite{DBLP:conf/lion/HutterHL11} uses Bayesian optimization but relies on a Random Forest model, which additionally considers the performance of an algorithm over a set of instances.

However, their work differs from ours in two aspects. 
	Firstly, general algorithm configuration requires working on a set of problem instances, 
	each coming with different features. 
	The software configuration tuning, in contrast, is often concerned with tuning software systems under a given benchmark (i.e., one instance)~\cite{nair2018finding,DBLP:conf/mascots/JamshidiC16,DBLP:conf/cloud/ZhuLGBMLSY17,Chen2018FEMOSAA}. Therefore, most of their designs for saving resources (such as the capping in ParamILS) were proposed to reduce the number of instances measured. Of course, it is possible to generalize the problem to consider multiple benchmarks at the same time, yet this is outside the scope of this paper. 
	Secondly, none of them works on the level of optimization model, and therefore, similar to the case of \textsc{Flash} and \textsc{BOCA}, our MMO is still complementary to their optimizers.



\subsection{Multi-objectivization in SBSE}


Multi-objectivization, which is the notion behind our MMO model, has been applied in other SBSE problems~\cite{DBLP:journals/tse/YuanB20,derakhshanfar2020good,DBLP:conf/ssbse/MkaouerKBC14,DBLP:conf/ssbse/SoltaniDPDZD18}. For example, to reproduce a crash based on the crash report, one can purposely design a new auxiliary objective, which measures how widely a test case covers the code, to be optimized alongside with the target crash distance~\cite{derakhshanfar2020good}. A multi-objective optimizer, e.g., NSGA-II, is directly used thereafter. A similar case can be found also for the code refactoring problem~\cite{DBLP:conf/ssbse/MkaouerKBC14}. However, during the tuning process, such a model, i.e., PMO in this paper, can result in poor resource efficiency as it wastes a significant amount of resources in optimizing the auxiliary objective, 
which is of no interest. 
This is a particularly unwelcome issue for software configuration tuning where the measurement is expensive. As we have shown in Section~\ref{sec:result}, PMO performs even worse than the classic single-objective model in most of the cases.






\section{Conclusion and Future Work}
\label{sec:con}

To mitigate the local optima issue in software configuration tuning, this paper takes a different perspective---multi-objectivizing the single objective optimization scenario. 
We do this by proposing a meta multi-objective model (MMO), 
at the level of optimization model (external part), 
as opposed to existing work that focuses on developing an effective single-objective optimizer (internal part). This work provides a sound analysis to interpret the principle behind MMO and explain what causes its limitation, namely eliminating the need for the weight parameter in the MMO model. Deriving on the theoretical understanding of the root cause, we then overcome such a limitation by proposing a simple yet effective new normalization. 


We compare MMO under the new normalization with four state-of-the-art single-objective optimizers, the plain multi-objectivization model, and the MMO model with the old normalization from our prior FSE work over 22 cases that are of diverse performance attributes, systems, and environments. 
The results reveal that 
the MMO model:


\begin{itemize}

\item can generally be more effective in overcoming local optima with better results;
\item and does so by utilizing resources more efficiently (better speedup) in most cases;
\item saves a considerable amount of extra resources that would otherwise be required for identifying the best weight.
\end{itemize}

Furthermore, we use MMO as part of \textsc{Flash} and \textsc{BOCA}, two recent model-based efforts from the software engineering community for configuration tuning, and revealing that it can:

\begin{itemize}
\item considerably consolidate the results;
\item while enabling good speedup overall.
\end{itemize}

Future directions of this work are exciting and fruitful, as it paves a new way of thinking about the resolution for mitigating local optima in software configuration and perhaps in a wider context of SBSE---multi-objectivizing at the level of optimization model instead of working at the level of an optimizer/algorithm. Specifically, the most immediate next steps include extending MMO beyond two meta-objectives 
(e.g., through considering multiple auxiliary objectives if any)
and exploring the possibility of designing a tailored multi-objectivization model for other SBSE problems. At the same time, through the analysis discussed in this work, we hope that there will be subsequent studies that take the unique characteristics of MMO into account, which can further advance software configuration tuning and beyond.

\ifCLASSOPTIONcompsoc
 \section*{Acknowledgments}
\else
 \section*{Acknowledgment}
\fi

This work was supported by a UKRI Grant (10054084) and a NSFC Grant (62372084). The authors wish to thank the anonymous reviewers for their highly constructive and insightful comments. Tao Chen would like to thank Yvonne and Savion for their assistance.
%
%

\bibliographystyle{IEEEtranS}
\bibliography{IEEEabrv,reference}

\ifCLASSOPTIONcaptionsoff
  \newpage
\fi

\end{document}